\documentclass[seceq]{ptptex}

\usepackage{graphicx}
\usepackage{wrapft}

\newcommand{\be}{\begin{equation}}
\newcommand{\ee}{\end{equation}\noindent}
\newcommand{\bea}{\begin{eqnarray}}
\newcommand{\eea}{\end{eqnarray}}
\newcommand{\Tr}{\mbox{Tr}}
\newcommand{\mD}{\mathcal{D}}
\newcommand{\psibar}{\bar{\psi}}

\newcommand{\1}{{\bf 1}}
\newcommand{\maprightb}[1]{\smash{\mathop{
\hbox to 1cm{\rightarrowfill}}\limits_{#1}}}
\title{
Lattice QCD at Finite Density
\footnote{based on talks at Workshop ``Thermal Field Theory'',
Yukawa Institute, Aug. 9. 2002 and Symposium ``Towards
understanding of finite density systems'', JPS meeting, Sept. 14, 2002}%
}

\subtitle{An introductory review}    

\author{
Shin \textsc{Muroya$^{1)}$}, Atsushi \textsc{Nakamura$^{2)}$}, 
Chiho \textsc{Nonaka$^{3)}$} and 
Tetsuya \textsc{Takaishi$^{4)}$ }%
}

\inst{
\it $^{1)}$Tokuyama Women's College, Shunan 745-8511, Japan \\
\it $^{2)}$RIISE, Hiroshima University,  Higashi-Hiroshima 739-8521, Japan \\
\it $^{3)}$Dept.\ of Physics, Duke University,  Durham, NC 27708-0305, USA\\
\it $^{4)}$Hiroshima University of Economics, Hiroshima 731-0192, Japan
}

\abst{
This is a pedagogical review of the lattice study of finite density QCD, which 
is intended to provide the minimum necessary contents, so that the paper may 
be used as the first reading for a newcomer to the field and also for those 
working in nonlattice communities.  After a brief introduction to argue why 
finite density QCD can be a new attractive subject, we describe fundamental 
formulae which are necessary for the following sections.

Then we survey lattice QCD simulations in small chemical potential regions,
where several prominent works have been reported recently.  Next, two-color 
QCD calculations are discussed, where we have a chance to glance at many new 
features of finite density QCD, and indeed recent simulations indicated quark 
pair condensation and the in-medium effect.  Tables of SU(3) and SU(2) lattice 
simulations at finite baryon density are given. 

In the next section, we make a survey of several related works which may be a 
starting point of the new development in the future, although some works do 
not attract much attention now.  Materials are described in a pedagogical 
manner.  Starting from a simple 2-d model, we briefly discuss a lattice 
analysis of NJL model.  We describe a non-perturbative analytical approach, 
i.e., strong coupling approximation method and some results.   Canonical 
ensemble approach instead of usual grand canonical ensample may be another 
route to reach high density.  We examine the density of state method and show 
this old idea includes recently proposed factorization method. An alternative 
method, complex Langevin equation and an interesting model, finite isospin 
model, are also discussed.

In the Appendix, we give several technical points which are useful in 
practical calculations.
}

\begin{document}

\maketitle
\tableofcontents
\section{Introduction}\label{sec-intro}

\begin{figure}[hbt]
\vspace{-3mm}
\begin{center}
\includegraphics[width=.6 \linewidth]{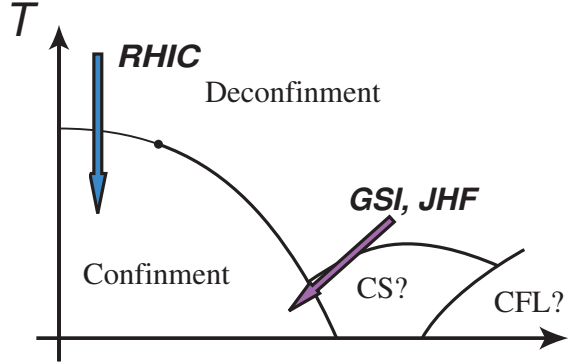}
\end{center}
\caption{Schematic phase diagram of QCD in $(T,\mu)$ plane.}
\label{fig1-1}
\end{figure}

Finite density QCD (Quantum Chromodynamics) has attracted considerable
attention in  high energy physics, nuclear physics and astrophysics.
Many theorists now believe that QCD has a very rich structure
when we study it in temperature and density parameter space,
and some experimentalists want to reveal it.   
Fig.\ref{fig1-1} shows a schematical phase diagram in $(T,\mu)$
plane.
Lattice QCD has been expected to provide fundamental information
about QCD in nonperturbative regions, and 
recently many promising works have appeared in this field. 

In this paper, we discuss these activities together with 
a pedagogic introduction and related works.
This is not a textbook-type comprehensive review.
Minimum necessary knowledge and possible hints for future
are given. 
Materials are taken from our notes, which we have been
preparing during our research in this field.
We hope this paper is a useful starting point for
those who wish to join the field.
For further reading, we cite several reviews. 
\cite{Barbour98, Hands01, KogutReviw02, Lombardo02, KarschLaermann03}
~

\begin{figure}[hbt]
\begin{center}
\includegraphics[width=.6 \linewidth]{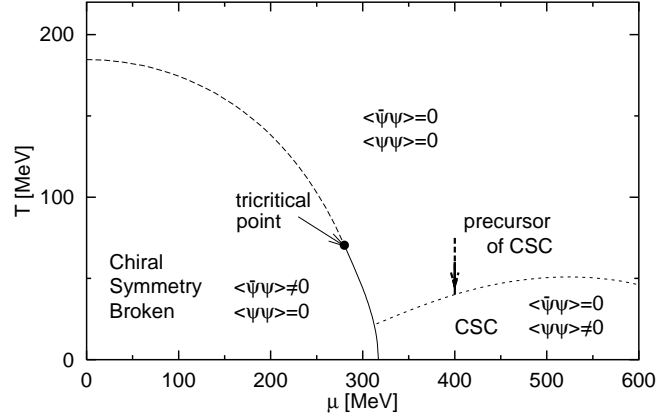}
\end{center}
\caption{ Phase diagram $(T,\mu)$ plane by NJL model.
(Ref.~\citen{NJL01}).
}
\label{fig1-2}
\end{figure}

The most attractive possibility of finite density QCD
is probably the color super conductivity (CSC).
This was revealed first by one gluon exchange type calculation:
Barrois, Bailin and Love, and Iwasaki and Iwado pointed out
that an attractive force near the Fermi surface creates a Cooper pair, 
and results in CSC in the case of QCD at very low temperature and 
finite density \cite{CS1}.  
In the late nineties, using the instanton type modeling of the attractive
force, Alford, Rajagopal and Wilczek \cite{CS2a} and
Rapp, Schaefer, Shuryak and Velkovsky \cite{CS2b}
argued that the gap energy is of the order of 100 MeV, 
and therefore the transition temperature, which should be
around the gap energy, is relatively high.
In these arguments, two-flavor, i.e., $u$ and $d$ quark, color 
superconductivity is considered and termed 2CS.  
Fig.\ref{fig1-2} is a prediction of the phase by NJL model. \cite{NJL01}

Since these works, the field has been actively explored and there
have been many interesting observations, e.g., 
at sufficiently large chemical potential, a state with a special
combination of three flavors and color is stable and called color flavor
locking (CFL) \cite{CFL}; a new phase LOFF (Larkin, Ovchinnikov, Fulde 
and Ferrell), where pairing with different chemical potentials results
in nonzero total momentum, may be favored \cite{AlfordBowersRaja00}; 
the gap energy is $\sim \exp(-c/g)$ instead of $\sim \exp(-c/g^2)$ 
\cite{Son} and so on.
See Ref.~\citen{RW}. Introduction of Ref.~\citen{PR} also gives a good
overview of the field.

\begin{figure}[hbt]
\begin{center}
\includegraphics[width=.8 \linewidth]{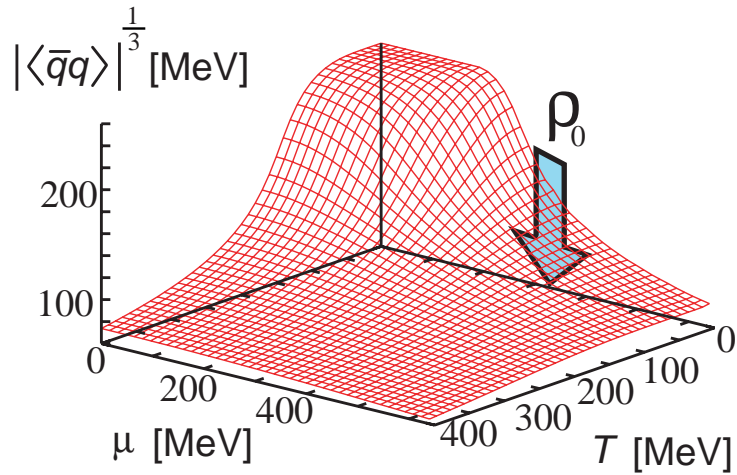}
\end{center}
\caption{
$\langle \bar{q}q \rangle $ as a function of $(T,\mu)$. 
We solve gap equations at finite temperature and density
derived in Ref.~\citen{LutzKlimtWeise92}.
We employ all parameter values given in \citen{LutzKlimtWeise92},
except t$G_D$, which is set to be zero, i.e., we drop the
s-quark contribution for simplicity.
The arrow indicates the value of $\mu$ corresponding to 
the normal nuclear density $\rho_0$.
}
\label{fig-NJL2}
\end{figure}

Another interesting, but probably less well known possibility is 
chiral restoration in the nuclear medium. In the normal nuclear
matter, the baryonic density is $\rho_0 \sim 0.17/{\mbox {fm}}^3$.
In a Nambu-Jona-Lashino (NJL) type model, the quark condensation,
$\langle \bar{q}q \rangle $ at $\rho_0$ is less than that of the vacuum.
We show the behavior of $\langle\bar{\psi}\psi\rangle$ as a function of
$\mu$ and $T$ in a simple NJL model calculation in Fig.\ref{fig-NJL2}.
The quark condensation is an important parameter and 
causes many effects, such as vector meson masses.
There are several recent experimental data which may yield information
regarding in-medium hadrons,
\begin{itemize}
\item
CERES: 
Large low mass $e^+ e^-$ pair enhancement was observed in CERN 
SPS in Pb+Au collisions at 158 A GeV/c.
This non trivial enhancement is also found
at 40 AGeV/c where it is even larger, and the data may only be reproduced
if the properties of the intermediate $\rho $ in the hot and dense medium
are modified.
\cite{CERES}
\item
KEK:
At KEK, invariant mass spectra of electron-positron pairs were measured
in the region below the $\omega$ meson mass for the p+C and p+Cu collisions.
The result was interpreted as being a signature of the $\rho/\omega$
modification at a normal nuclear-matter density.
\cite{Enyo}
\item
STAR:
The invariant mass of $\rho$ meson decays in Star experiment at $\sqrt{s}=
200$ GeV Au+Au collisions at RHIC shows 60-70 MeV downward shift of
the peak from its vacuum value.  This suggests the modification of
the spectral function at finite $T$ and $\mu$.
\cite{Star}
\item
Pionic 1s states of Sn:
Precision spectroscopy of pionic 1s states of Sn nuclei suggests
$f^{*}_{\pi} (\rho_0)^2/f_{\pi}^2 =0.65\pm0.05$, at the normal nuclear 
density. \cite{Pionic1s}

\end{itemize}
See Ref.~\citen{Kunihiro} for more detailed arguments and
related theoretical and experimental works. 

Hagedorn noted \cite{Hagedorn} 
that if the hadron spectrum increases exponentially,
\begin{equation}
\rho (m) = C {\rm e}^{m/T_0},
\end{equation}
then the partition function diverges at some temperature $T_c$,
called Hagedorn temperature,
\begin{equation}
\int dm \rho (m) {\rm e}^{-m/T} = \infty \hspace{0.3cm} 
{\mbox {\rm if }} T > T_0.
\end{equation}
Cabbibo and Parisi pointed out that the singularity of the partition 
function does not necessary mean the upper limit of the temperature,
but the phase transition \cite{C-P}.

In 1981, McLerran and Svetitsky \cite{M-S}, and Kuti, Polonyi and 
Szlachanyi \cite{K-P-S}, demonstrated that when the temperature
increases, a QCD matter encounters a phase transition from the confinement to 
the deconfinement phase
by calculating Polyakov lines for two-color QCD.
Since these pioneering works, lattice simulations have provided
many detailed analyses of QCD at finite temperature.
Still the order of the phase transition, i.e., the cross-over, the first
or the second order, with the physical {\it u}, {\it d} and {\it s} 
quark masses is not
yet finally determined. 
Lattice QCD is the first principle calculation, and
can be considered as the most reliable starting point.
See Ref.~\citen{Kanaya02} for recent lattice investigations 
at finite temperature.

Then how about lattice QCD at finite density ?
A lattice simulation of color SU(2) at finite temperature and
density was carried out in 1984
and the transition from the confinement to the deconfinement
state was observed \cite{Nakamura84}.  
Since then, to our knowledge, few lattice calculations were performed 
until 1999 \cite{MH99,HKLM99} except algorithmic study.

The reason is that when the chemical potential is introduced, 
the fermion determinant, $\det\Delta(\mu)$ becomes complex, 
which  appears in the Euclidean path integral measure,
\begin{equation}
Z = \mbox{Tr}\, e^{-\frac{1}{kT}(H-\mu N)}
  = \int \mD U \mD\bar{\psi}\mD\psi
    \, e^{-\frac{1}{kT} S_G - \bar{\psi}\Delta\psi}
 = \int \mD U \det\Delta \, e^{-\frac{1}{kT} S_G} .
\label{pathinte}
\end{equation}
Here $\Delta$ is diagonal in flavor, 
$\Delta=\mbox{diag}(\Delta_u,\Delta_d,\Delta_s,\cdots)$.

\begin{figure}[hbt]
\begin{center}
\includegraphics[width=.6 \linewidth]{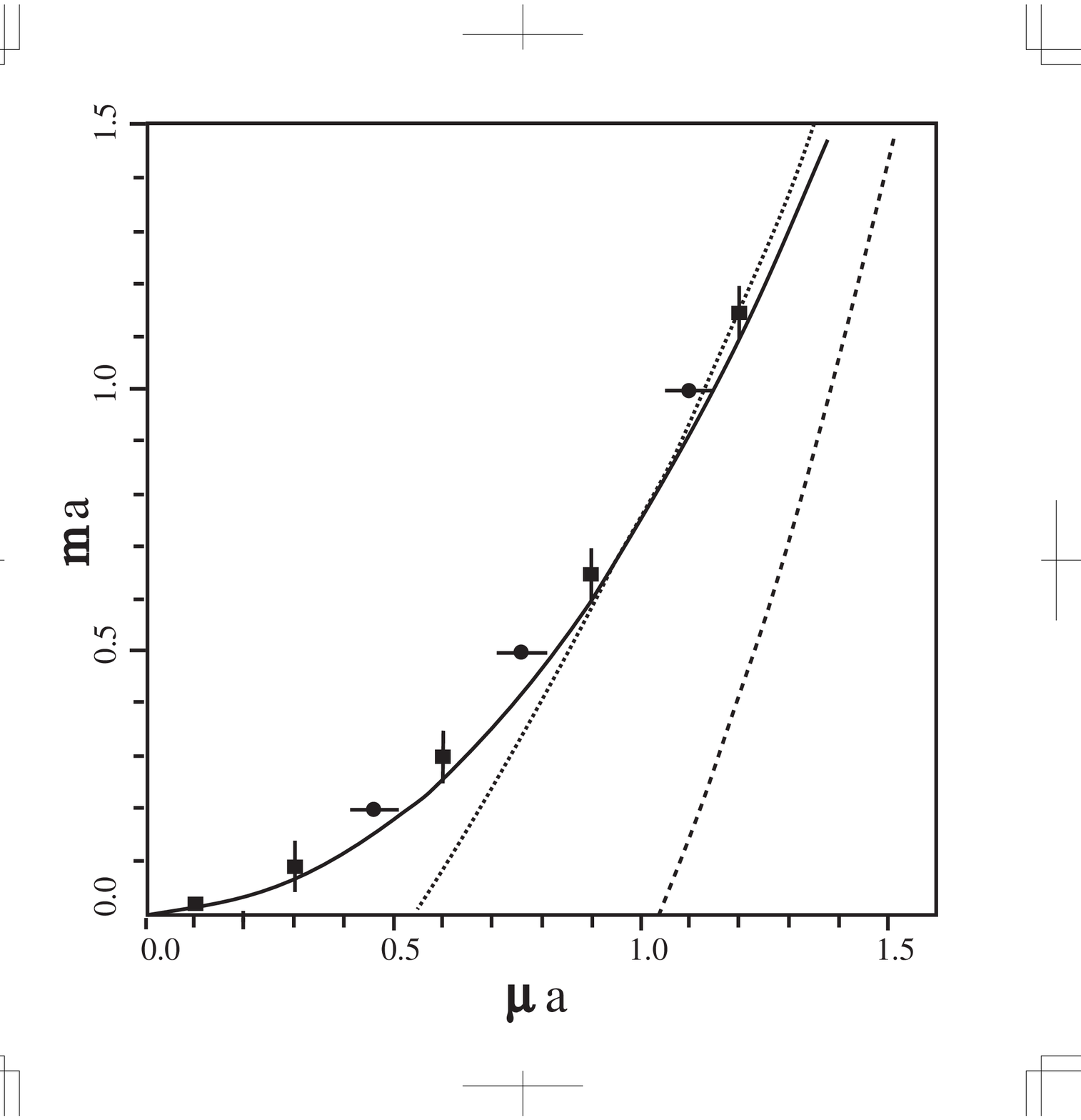}
\end{center}
\caption{
Phase diagram in the quark mass-chemical potential plane of 
the SU(3) gauge theory with staggered fermions at $\beta=0$ in
the quench approximation. \cite{quench}
The data points show the threshold value where physical
observables start deviation from their $\mu=0$ values.  
The dotted line the expected critical line corresponding to the
baryon threshold and the curve is the pion threshold which seems
to describe the Monte Carlo data.  The dotted curve is the result of
the mean field calculation.
}
\label{quench-pi}
\end{figure}

The simplest way to avoid the problem is to use the quench approximation,
where we discard $\det\Delta(\mu)$.  This approximation works
well in many lattice calculations such as spectroscopy.
However at the finite density it was found that  
the onset of chiral symmetry restoration occurs at a chemical
potential of half the pion mass,
\begin{equation}
\mu_c = \frac{m_\pi}{2} 
\end{equation}
i.e., at the chiral limit, $\mu_c=0$.  \cite{quench}
Fig.\ref{quench-pi} shows data points in the quark
mass-chemical potential plane at which physical observables start
deviating from their $\mu=0$ values in quench simulations at $\beta=0$.
The curve (dotted line) indicates the expected critical line correpoinding 
to pion (baryon) threshold.

Stephanov pointed out that quenched QCD is not the $N_f \to 0$ limit
of QCD, but the limit of a theory with $2N_f$ quarks \cite{Stephanov}. 
Let us consider the resolvent of the Dirac operator $\Delta$ for zero
mass case,
\begin{equation}
G(x,y) = \langle \Tr \frac{1}{z-\Delta} \rangle 
= \int dx' dy' \rho(x',y')\frac{1}{z-z'},
\end{equation}
where $z = x + iy$ is a complex variable.  Banks-Casher relation
is $\langle\bar{\psi}\psi\rangle=\pi\rho(0)$.
$\rho$ can be inversely obtained as
\begin{equation}
\rho = \frac{1}{\pi}\frac{\partial G}{\partial z^*} .
\end{equation}
Using the formula, $\det A = \exp\Tr \log A$, we can write $G$ as
\begin{equation}
G = -\frac{\partial V}{\partial z},
\end{equation}
where 
\begin{equation}
V = -\langle \log \det(z-\Delta)\rangle.
\end{equation}
If the standard replica trick works, $V$ can be obtained as 
$\lim_{n\rightarrow 0} V_n$, where
\begin{equation}
V_n = -\frac{1}{n} \log\langle {\det}^n (z-\Delta) \rangle.
\end{equation}

Taking $\psi_{L,R} \equiv (1\pm\gamma_5)/2\psi$ as bases, we can write
$\Delta$ as
\begin{equation}
\Delta = \left( 
\begin{array}{cc}
 0                & iX+\mu \\
 iX^{\dagger}+\mu & 0
\end{array} \right )
\end{equation}
where we employ the chiral representation of the Dirac matrices.
Using the random matrix model where $X$ is treated as complex Gaussian random
variables,
Stephanov found that the above $V_n$ does not lead to the natural
result, and we should start from $V_n$ which includes not only $z$
but also $z^*$,
\begin{equation}
V_n = -\frac{1}{n} \log\langle {\det}^n (z-\Delta)(z^*-\Delta^{\dagger}) \rangle.
\end{equation}
This corresponds to a system composed of
$n$ quarks with original action and $n$ quarks with the conjugate action.
Therefore to understand QCD at finite density, the quenched approximation 
is not appropriate and one needs dynamical simulations.

We have learned :
\begin{itemize}
\item
QCD at finite density is an interesting attractive
field of physics, and lattice QCD should play an important
role here. 
\item
Quench approximation of lattice QCD is not appropriate for
the finite density system.
\item
Due to the complex nature of the fermion determinant $\det\Delta$,
the standard Monte Carlo simulation is very difficult to
obtain physical quantities.
\end{itemize}
In the following we continue our discussion, assuming the reader
accepts these points. 

In Sec.2, we give the fundamental formulae in lattice QCD at finite
baryon density.
In Sec.3, we describe recent prominent activities in small chemical
potential and high temperature regions.
In Sec.4, several numerical simulations of two-color QCD are reported,
which suggest new interesting features at large $\mu$.
In Sec.5, we collect several ideas and calculations which may give
us a hint for the next step. 
We summarize and conclude our discussions in Sec. 6.

\section{Formulation}\label{sec-formulation}

A thermodynamical system is described by the partition function,
\begin{eqnarray}
Z &=& {\rm{Tr}} ({\rm e }^{-\frac{1}{kT} ({H} - \mu {N})})
\nonumber \\
  &=& \int \mD U \mD 
\bar{\psi}\mD\psi{\rm e}^{-\int_0^\beta d \tau \int d^3x 
(L + \mu n)}
\end{eqnarray}
where $\beta=1/T$
and we impose the periodic (anti-periodic) boundary condition for
$U$ ($\psi$) along the temporal direction \cite{G-P-Y} . 
The chemical potential is introduced in fermion action as
$\mu\bar{\psi}\gamma_0\psi$.  Therefore the fermion determinant
has the form,
\begin{equation}
\Delta(\mu) = D_\nu\gamma_\nu + m + \mu\gamma_0 .
\label{det1}
\end{equation}

Adopting anti-hermite $D$ and hermite $\gamma$ (final result does
not depend on the representation.), we can easily show 
\begin{equation}
\Delta(\mu)^\dagger = -D_\nu\gamma_\nu + m + \mu\gamma_0
= \gamma_5 \Delta(-\mu) \gamma_5
\label{det2}
\end{equation}
and then
\begin{equation}
(\det\Delta(\mu))^*
= \det\Delta(\mu)^\dagger
= \det\Delta(-\mu) .
\label{det3}
\end{equation}
At $\mu=0$, this relationship guarantees that $\det\Delta$ is real.

In the continuum field theory, the chemical potential is introduced
by the substitution, 
\begin{equation}
p_4 \rightarrow p_4 - i\mu
\end{equation} 
in the fermion propagators.
On the lattice with the lattice spacing $a$,
the $\bar{\psi}\partial_\mu\psi$ term leads to $\bar{\psi}e^{ip_\mu a}\psi$.
For example, a free propagator of Wilson fermions is given as,
\begin{equation}
\frac{1}{ 1
 - \kappa \sum_{\mu=1}^{4} \left\{
   (1-\gamma_\mu) e^{i p_\mu a} + (1+\gamma_\mu) e^{-i p_\mu a} \right\}
  } .
\label{FreeWfermion}
\end{equation}
Therefore, we can naturally include the chemical potential exponentiated
in the fermion matrix as,
\begin{eqnarray}
\Delta(x,x') = \delta_{x,x'}
 - \kappa \sum_{i=1}^{3} \left\{
        (1-\gamma_i) U_i(x) \delta_{x',x+\hat{i}}
      + (1+\gamma_i) U_i^{\dagger}(x') \delta_{x',x-\hat{i}} \right\}
\nonumber
\\
  - \kappa \left\{
        e^{+\mu}(1-\gamma_4) U_4(x) \delta_{x',x+\hat{4}}
      + e^{-\mu}(1+\gamma_4) U_4^{\dagger}(x') \delta_{x',x-\hat{4}}
\right\} .
\label{Wfermion}
\end{eqnarray}
For staggered (Kogut-Susskind) fermions,
\begin{eqnarray}
\Delta(x,x') = m \delta_{x,x'} 
  + \frac{1}{2} \sum_{i=1}^{3} \eta_i(x) \left\{ 
    U_i(x) \delta_{x',x+\hat{i}} -  U_i^{\dagger}(x') \delta_{x',x-\hat{i}}
    \right\}
\nonumber
\\
  + \frac{1}{2} \eta_4(x) \left\{ e^{+\mu} U_4(x) \delta_{x',x+\hat{4}} 
                   -  e^{-\mu} U_4^{\dagger}(x') \delta_{x',x-\hat{4}}
\right\} ,
\label{KSfermion}
\end{eqnarray}
where 
\begin{equation}
\eta_{\mu}(x) = 1\ (\mu=1), \ \ 
       (-1)^{x_1}\ (\mu=2), \ \ 
       (-1)^{x_1+x_2}\ (\mu=3), \ \ 
       (-1)^{x_1+x_2+x_3}\ (\mu=4) .
\end{equation}
In other words,
the chemical potential is introduced as,
\begin{eqnarray}
U_t(x)         &\to& e^\mu U_t(x),             \nonumber \\
U_t^\dagger(x) &\to& e^{-\mu} U_t^\dagger(x),
\label{ChePot}
\end{eqnarray}
where $U_\mu(x) = \exp(iaA_\mu(x))$.
Hasenfratz and Karsch have shown that the formula (\ref{ChePot}) avoids
nonphysical divergence of the free energy of quarks \cite{HK83}. 
Gavai considered more general form than $\exp(\pm\mu)$ 
in Eq.(\ref{ChePot}). \cite{Gavai85}
Creutz discussed how the chemical potential appears in lattice fermion
formulation. \cite{Creutz99}

The relation (\ref{det2}) holds for Wilson fermions, while in staggered
fermion case, it is easy to check 
\begin{equation}
\Delta(\mu)^\dagger = \Gamma_5 \Delta(-\mu) \Gamma_5 , 
\end{equation}
where
\begin{equation}
\Gamma_5(x,x') = (-1)^{x_1+ x_2+ x_3+ x_4} \delta_{x,x'} .
\end{equation}

In the relativistic formulation, we expect the invariance under
$\mu \rightarrow -\mu$.  If we expand the partition function as
\begin{equation}
Z(\mu) = \sum_k \frac{\mu^k}{k!} \frac{\partial^k}{\partial\mu^k}
\int \mD U \det\Delta e^{-\beta S_G}|_{\mu=0} ,
\end{equation}
the terms with odd power of $\mu$ vanish.
Let us check $k=1$ case explicitly;
At $\mu=0$,
\begin{eqnarray}
\frac{\partial}{\partial\mu}\det\Delta
 =  \Tr [ \Delta^{-1}\frac{\partial\Delta}{\partial\mu} ] \det\Delta
 =  \Tr [ \gamma_5\Delta^{-1}\gamma_5
        \gamma_5\frac{\partial\Delta}{\partial\mu}\gamma_5 ] 
       \det\gamma_5\Delta\gamma_5
\nonumber \\
 =  \Tr [ \Delta^{\dagger-1}
        (- \frac{\partial\Delta^{\dagger}}{\partial\mu}) ] 
       \det\Delta^{\dagger}
= -  \Tr [ \Delta(U^{\dagger})^{-1}
         \frac{\partial\Delta(U^{\dagger})}{\partial\mu} ]  
       \det\Delta(U^{\dagger}) ,
\end{eqnarray}
where we use Eq.(\ref{det2}).
The integration measure $\mD U$ and the gluon action $S_G$ are 
invariant under the change $U \rightarrow U^{\dagger}$, we get
$\partial Z/\partial\mu=-\partial Z/\partial\mu = 0$ at $\mu = 0$.

\begin{figure}[hbt]
\begin{center}
\begin{minipage}{ 0.45\linewidth}
\includegraphics[width=0.95\linewidth]{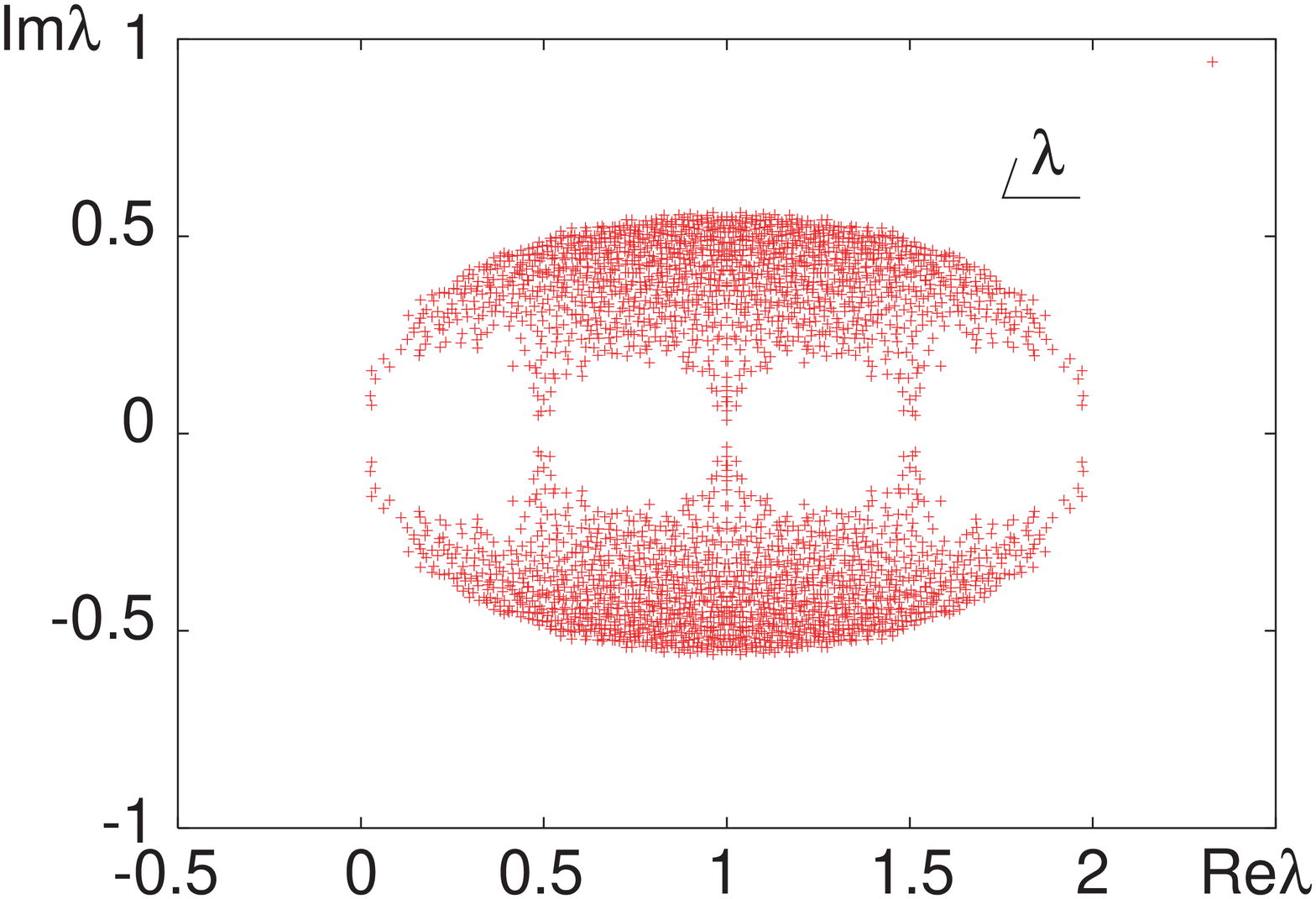}
\end{minipage}
\hspace{1mm}
\begin{minipage}{ 0.45\linewidth}
\includegraphics[width=0.90\linewidth]{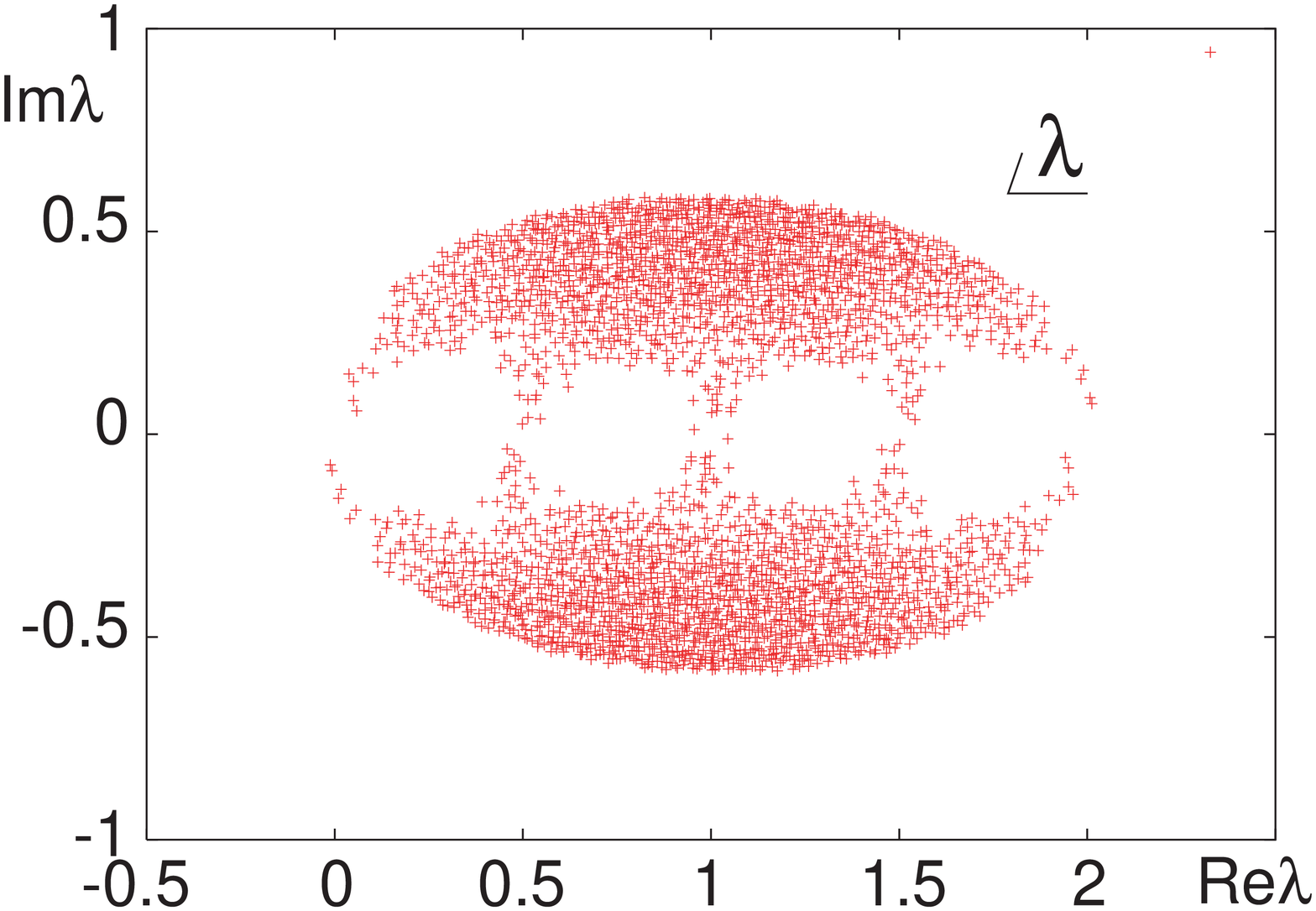}
\end{minipage}
\end{center}
\begin{center}
\begin{minipage}{ 0.45\linewidth}
\includegraphics[width=0.95\linewidth]{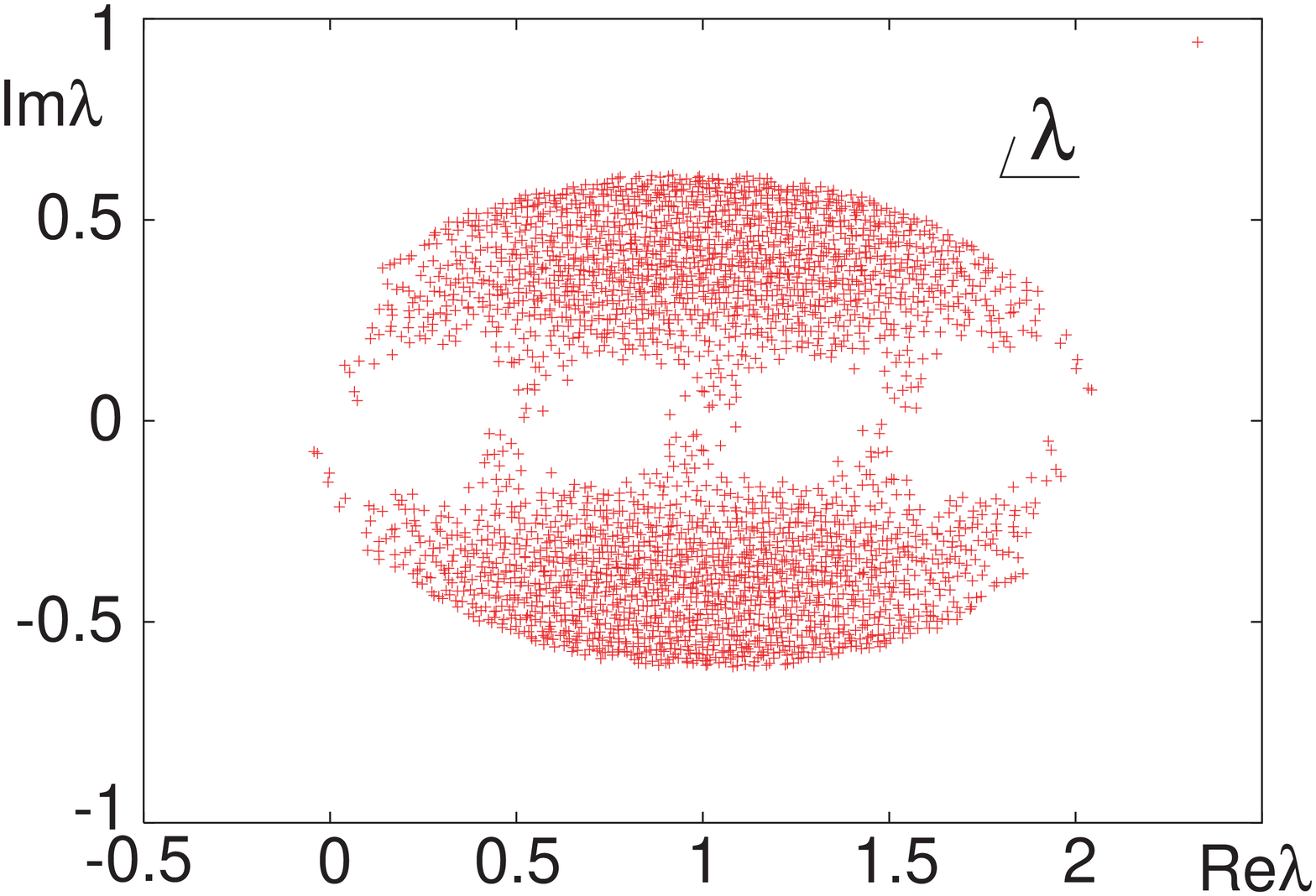}
\end{minipage}
\hspace{1mm}
\begin{minipage}{ 0.45\linewidth}
\includegraphics[width=0.90\linewidth]{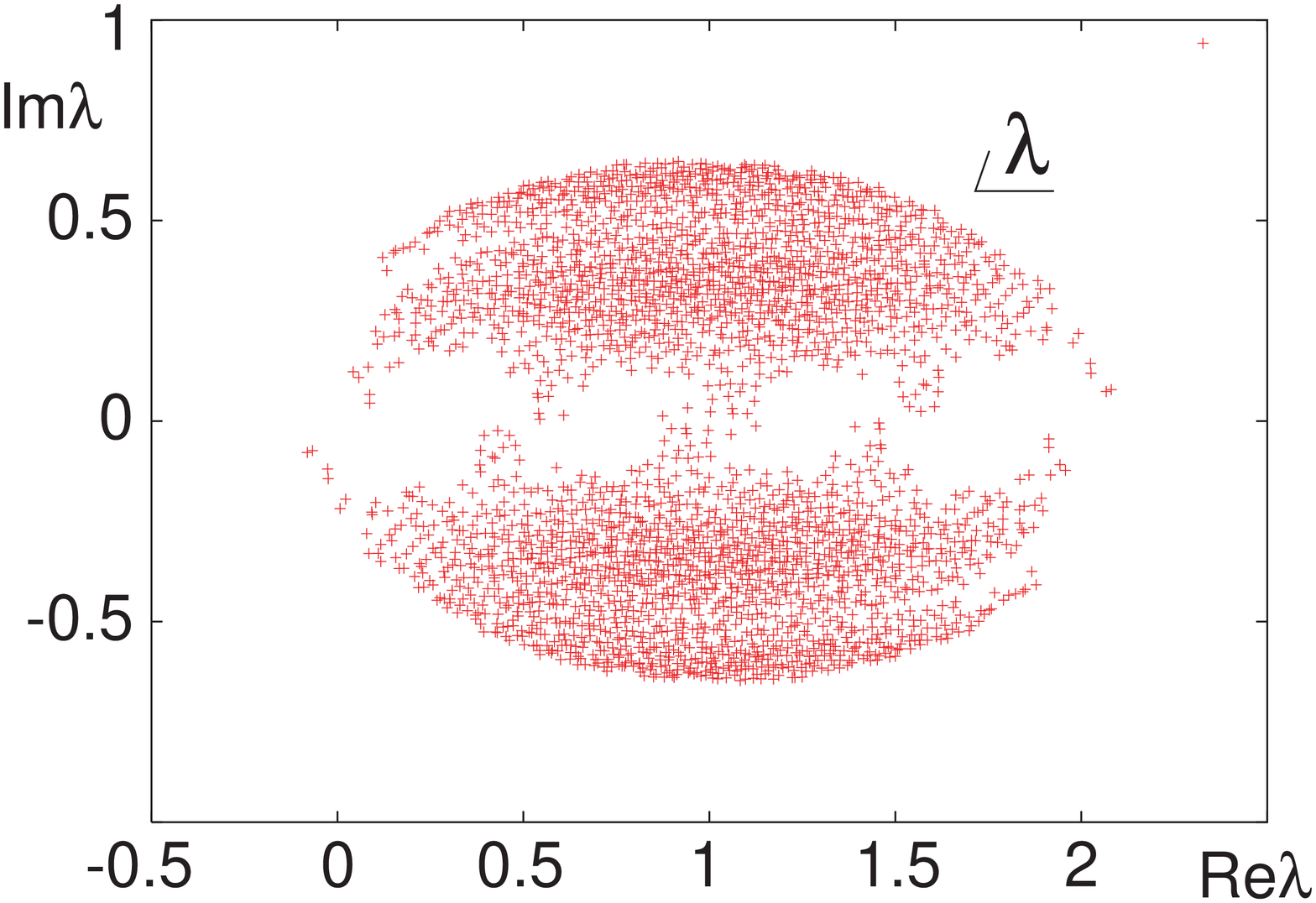}
\end{minipage}
\end{center}
\caption{
Eigen value distribution.  Wilson fermions.
$4^4$ lattice. $ma=0.1$. Quench $\beta=5.7$.
}
\label{fig2-1}
\end{figure}

\begin{figure}[hbt]
\begin{center}
\begin{minipage}{ 0.45\linewidth}
\includegraphics[width=0.95\linewidth]{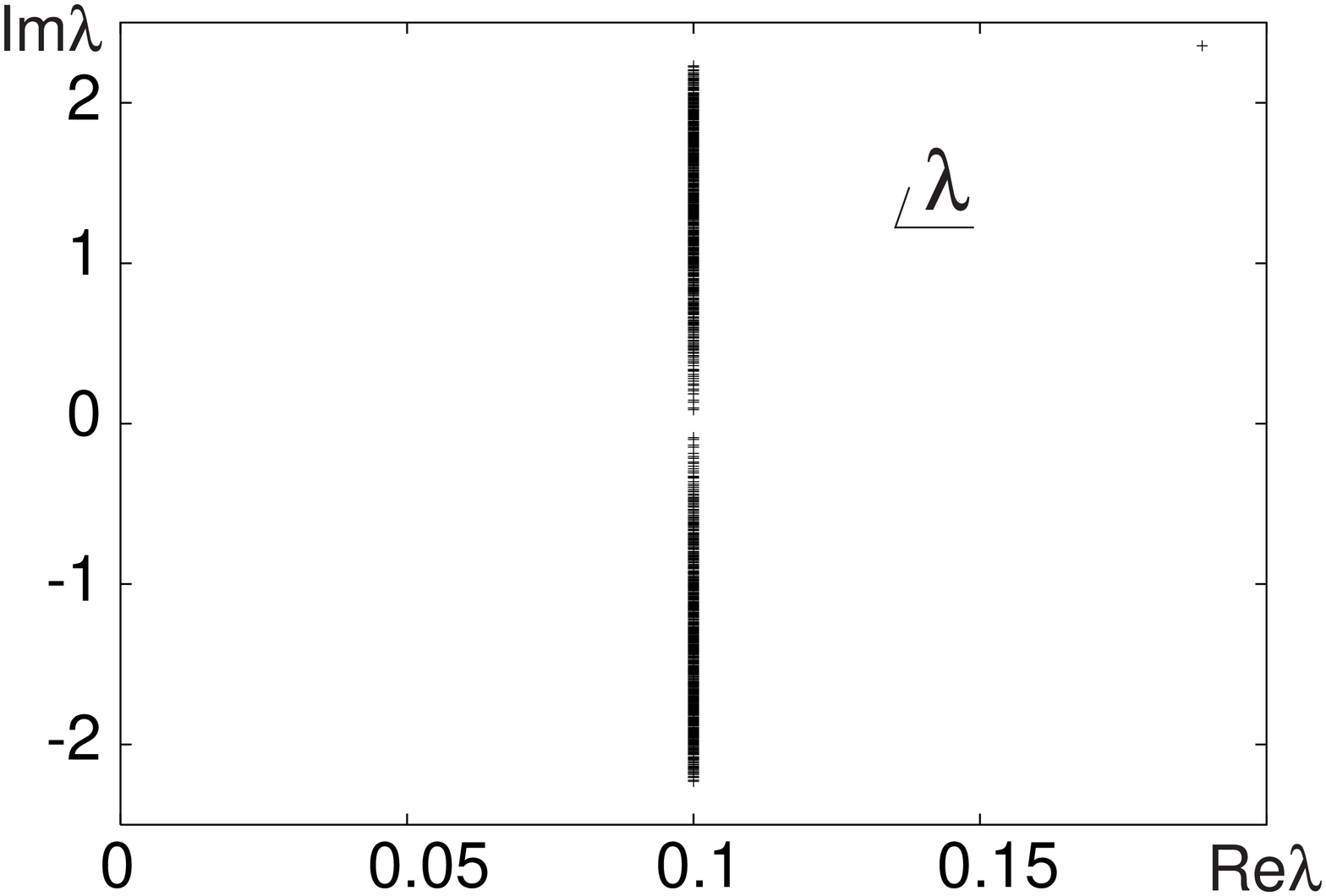}
\end{minipage}
\hspace{1mm}
\begin{minipage}{ 0.45\linewidth}
\includegraphics[width=0.90\linewidth]{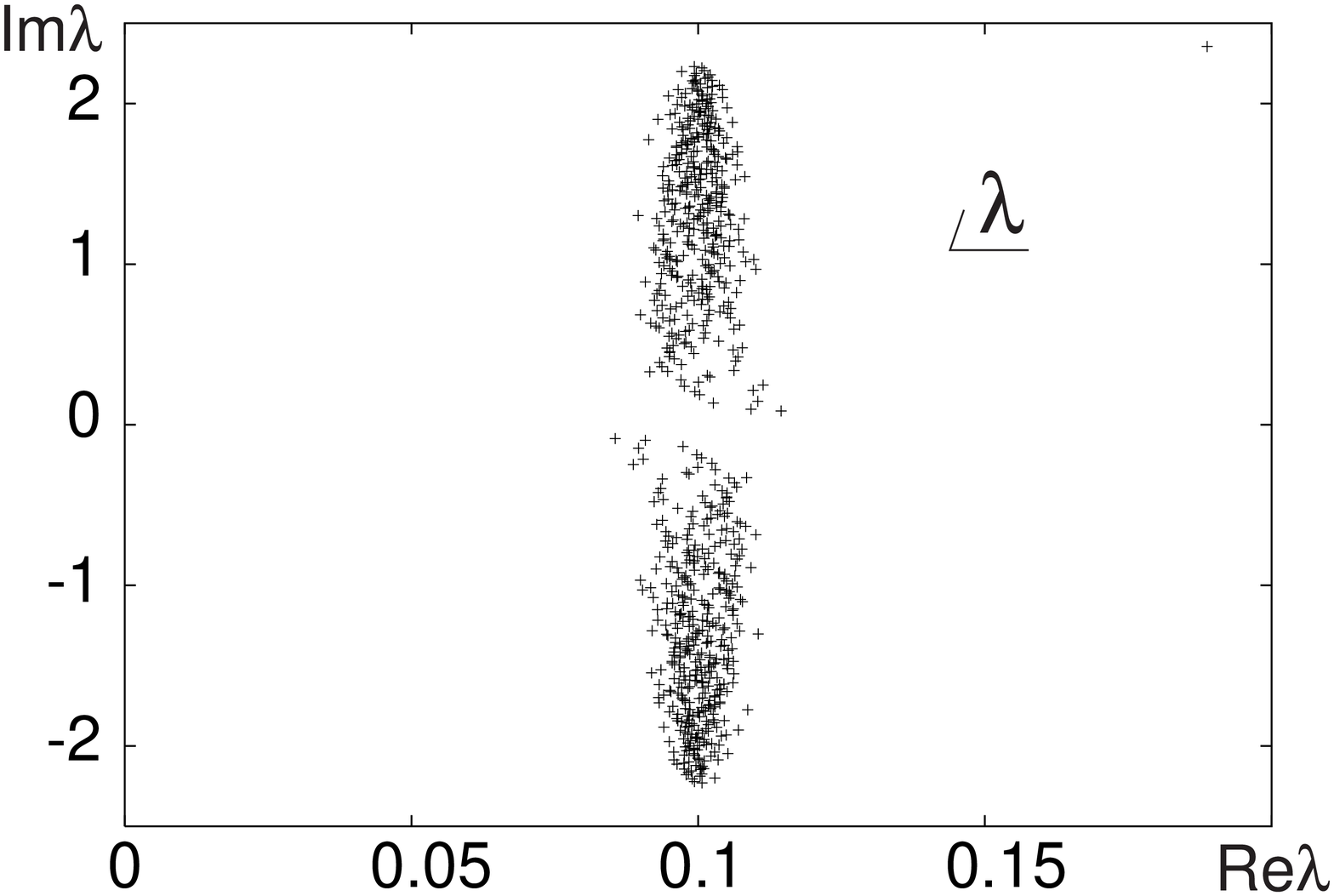}
\end{minipage}
\end{center}
\begin{center}
\begin{minipage}{ 0.45\linewidth}
\includegraphics[width=0.95\linewidth]{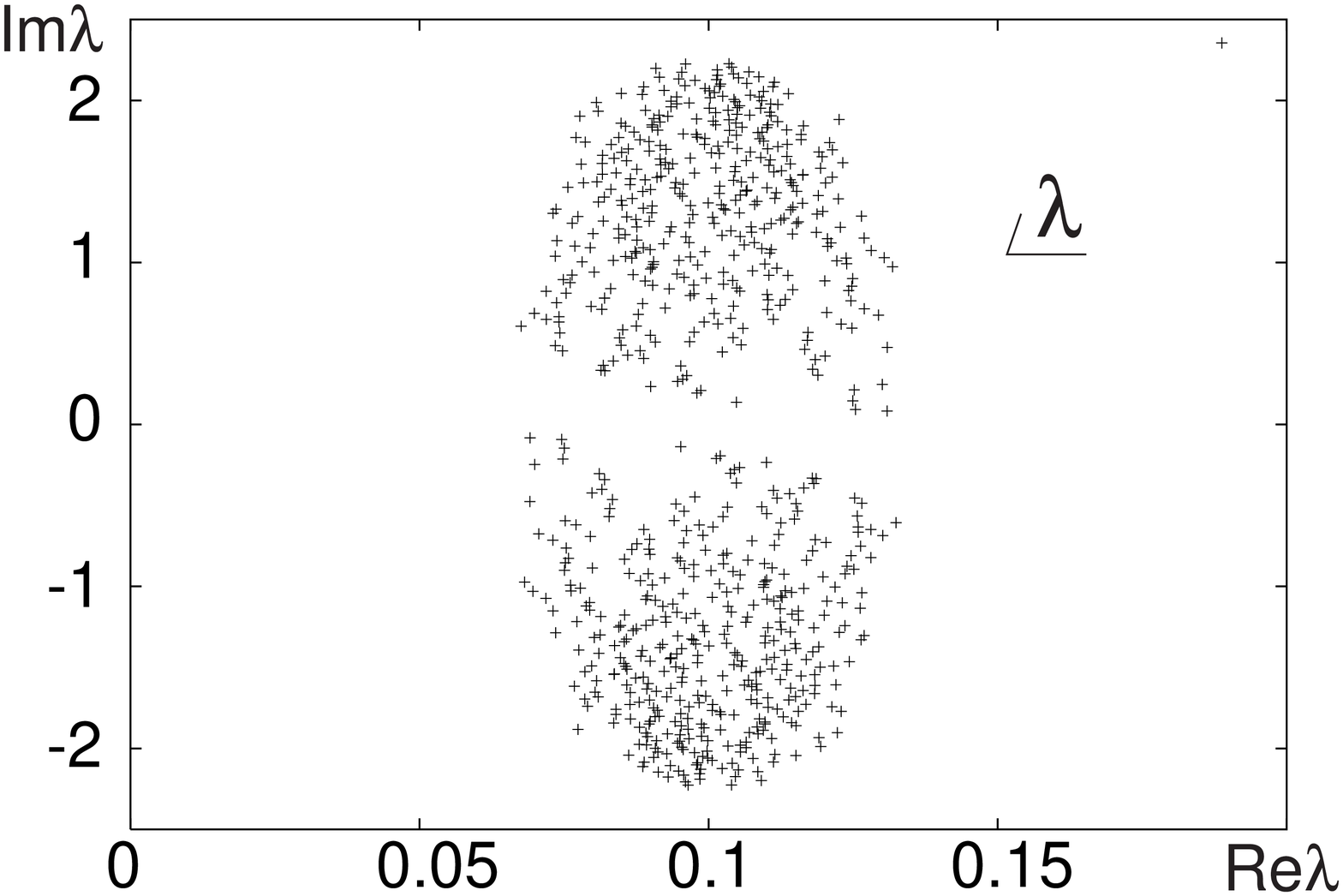}
\end{minipage}
\hspace{1mm}
\begin{minipage}{ 0.45\linewidth}
\includegraphics[width=0.90\linewidth]{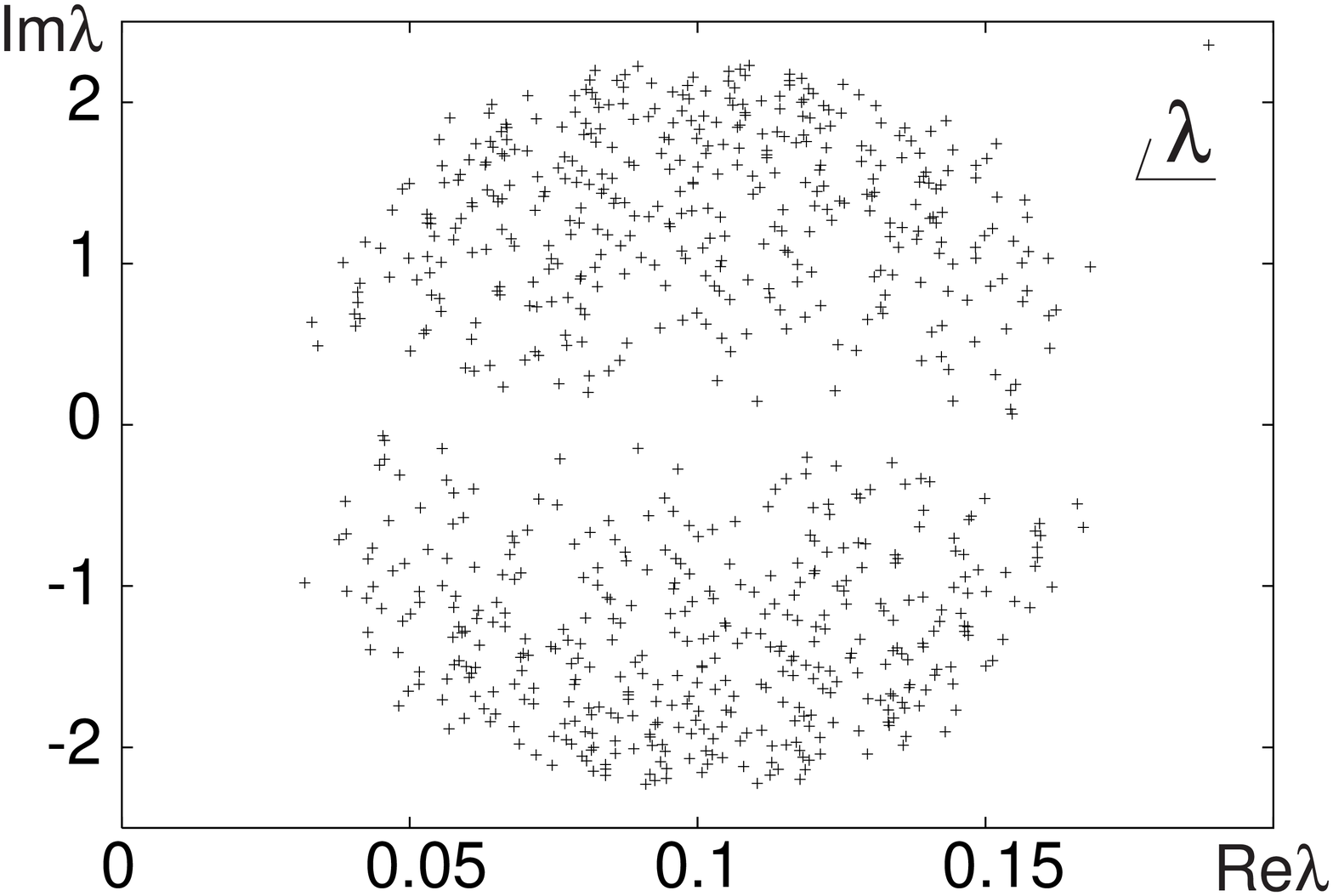}
\end{minipage}
\end{center}
\caption{
Eigen value distribution.  Staggered fermions.
$4^4$ lattice. $ma=0.1$. Quench $\beta=5.7$.
}
\label{fig2-2}
\end{figure}

We show in Figs.\ref{fig2-1} and \ref{fig2-2} typical eigen value distributions
for Wilson fermions (Fig.\ref{fig2-1}) and staggered 
fermions (Fig.\ref{fig2-2}).
Here configurations are generated in the quench approximation,
since so far there is no algorithm which enables us to calculate
 the path integral
(\ref{pathinte}) at finite $\mu$.

It is sometimes convenient to change variables, $\psi$ and $\bar{\psi}$
as
\begin{eqnarray}
\psi(\vec{x},t_i) \rightarrow e^{-t_i \mu} \psi(\vec{x},t_i)
\nonumber \\
\bar{\psi}(\vec{x},t_i) \rightarrow e^{+t_i \mu} \bar{\psi}(\vec{x},t_i)
\end{eqnarray}
where $t_i=1,2,\cdots,N_t$.
Then
\begin{equation}
\bar{\psi}(\vec{x},t_i) e^{\mu} U_t(\vec{x},t_i) \psi(\vec{x},t_{i+1})
\rightarrow
\bar{\psi}(\vec{x},t_i)  U_t(\vec{x},t_i) \psi(\vec{x},t_{i+1})
\label{muatEdge1}
\end{equation}
i.e., we do not need to put a factor $e^{\pm \mu}$ 
to the link variables, $U_t$ and $U^\dagger_t$, 
except at the temporal edge of the lattice where
\begin{eqnarray}
e^{N_t\mu} U_t(\vec{x},N_t) &=& e^{\mu/T} U_t(\vec{x},N_t)
\nonumber \\
e^{-N_t\mu} U_t^\dagger(\vec{x},N_t) &=& e^{-\mu/T} U_t^\dagger(\vec{x},N_t) .
\label{muatEdge2}
\end{eqnarray}
The partition function should be a function of $N_t\mu=\beta\mu=\mu/T$ 
(see Eq.(\ref{pathinte})),
therefore the above expressions appear to be natural.

\section{Three Color}\label{sec-3color}

Even though the fermion determinant $\det\Delta$ is complex,
one may perform Monte Carlo simulations by taking its modulus
as a measure,
\begin{eqnarray}
\langle O \rangle  &=& \frac{1}{Z}  \int \mD U \,O\,\det\Delta \, e^{-\beta S_G}
 = \frac{
 \int \mD U \,O\,|\det\Delta|e^{i\theta} \, e^{-\beta S_G}
 }
 {
 \int \mD U |\det\Delta|e^{i\theta} \, e^{-\beta S_G}
 }
\nonumber \\
 &=& \frac{
 \int \mD U \,O\,|\det\Delta|e^{i\theta} \, e^{-\beta S_G}
 }
 {
  \int \mD U |\det\Delta| \, e^{-\beta S_G}
 } \displaystyle{ /} 
 \frac
 {
 \int \mD U |\det\Delta|e^{i\theta} \, e^{-\beta S_G}
 }
 {
  \int \mD U |\det\Delta| \, e^{-\beta S_G}
 }
\nonumber \\
&=& \frac{
\langle Oe^{i\theta} \rangle _0
}{
\langle e^{i\theta} \rangle _0
}
\label{mod-det}
\end{eqnarray}
where $\langle \cdots \rangle _0$ is the expectation value with $|\det\Delta|$
as the measure, i.e., phase quenching measure.
The direct calculations were pursued on small lattices, 
but a large phase fluctuation hinders us 
from obtaining a  meaningful signal 
in low temperature and large chemical potential regions,
i.e., the numerator and denominator of the last term of Eq.(\ref{mod-det})
are very small \cite{Nakamura90,Toussaint,Hasen-To}.
Lattice QCD at finite chemical potential suffers from a severe sign problem.
In spite of this difficulty, there have been many challenging efforts, which
we will survey in this section.  Table \ref{Tab-su3} is a compilation of
numerical simulations of three color system.

\begin{table}[htbp]
	\begin{center}
	\caption{Three Color}
		\begin{tabular}{|l|p{50mm}|p{45mm}|l|} \hline \hline
		\hline
		Action & parameters & comments & Ref. \\ \hline
          Plaq.     + KS &
           $N_f=2$, $m_q=0.00625, 0.0125$,     
           $16^3 \times 8$  
          &
          quark number susceptibility     &
        \citen{Gott} \\ \hline
		Plaq. + KS & 
		$N_F=2$, $m_q = 0.0125, 0.0170, 0.0250$,  
		$16\times 8^2 \times 4$ 
		& 
		response of observables  at $\mu = 0$& 
        \citen{QCD-Taro02}
		\\ \hline
		Plaq. + KS &
		$N_f = 2+1$, $m_{u,d}=0.025$, $m_s=0.2$,  
		$N_s^3 \times 4$, $N_s=$ 8,10,12
		& 
		phase diagram, 
		q method 
		 &
        \citen{FodorKatz}\\ \hline
		Imp. + Imp. KS & 
		$N_f=2$, $m_q=0.1, 0.2$,  
		$16^3 \times 4$ 
		&
		 Taylor expansion, 
		 reweighting method 
		              & 
        \citen{Allton}
		\\ \hline
               Imp. + Imp. KS &
               $N_f=2$, $m/T=0.1$, 0.4, 
               $16^3 \times 4$ 
               &
               Taylor expansion, analytic frame work,  
               equation of state
               &
        \citen{Allton03}
                \\ \hline

        Plaq. + KS &  
		$N_f=4$, $m_q=0.05$,  
		$16^3 \times 4$ 
		&  
        imaginary chemical potential &
  		\citen{DElia}\\ \hline
		Plaq. + KS & 
		$N_f=2$, $m_q=0.025$,  
		$8^3\times 4$, $6^3 \times 4$
		& 
		QCD phase diagram, 
		imaginary chemical potential 
		 &
		\citen{FP}
	    \\ \hline
        Plaq. + KS  &
		$N_f = 2+1$, $m_{u,d}=0.025$, $m_s=0.2$,  
		$N_s^3 \times 4, N_s=$ 8,10,12
		& 
		equation of state    &
        \citen{FodorKatz2}\\ \hline
            \end{tabular}
	\end{center}
	\label{Tab-su3}
\end{table}


\subsection{Response of observables with respect to $\mu$ at $\mu=0$ }

Although the direct simulations at finite $\mu$ is very hard, one may
measure the effect of
the chemical potential through the response of physical observables with
respect to the chemical
potential at $\mu=0$. Such an attempt was first pursued by Gottlieb et.
al.\cite{Gott}.

\begin{figure}[hbt]
\begin{center}
\includegraphics[width=.6 \linewidth]{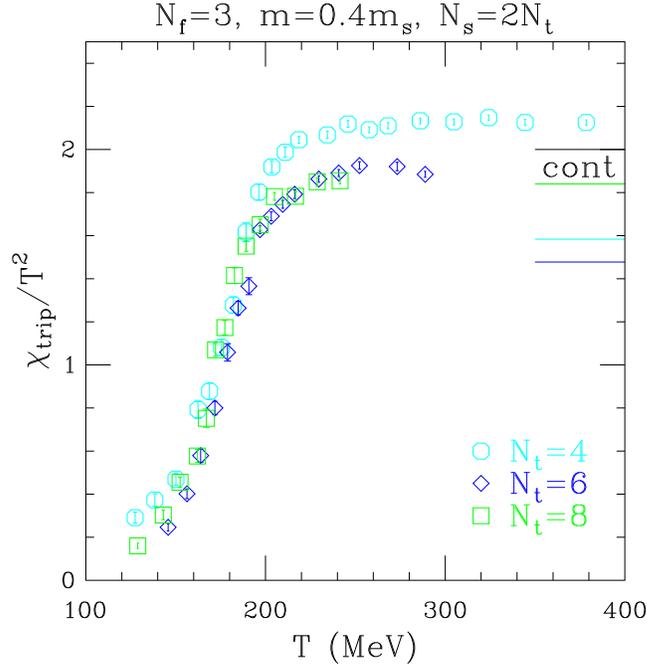}
\end{center}
\caption{The triplet quark number susceptibility fo $N_f =3$
on $8^3 \times 4$, $12^3\times 6$ and $16^3\times 8$ lattices.
}
\label{MILC1}
\end{figure}

\begin{figure}[hbt]
\begin{center}
\includegraphics[width=.6 \linewidth]{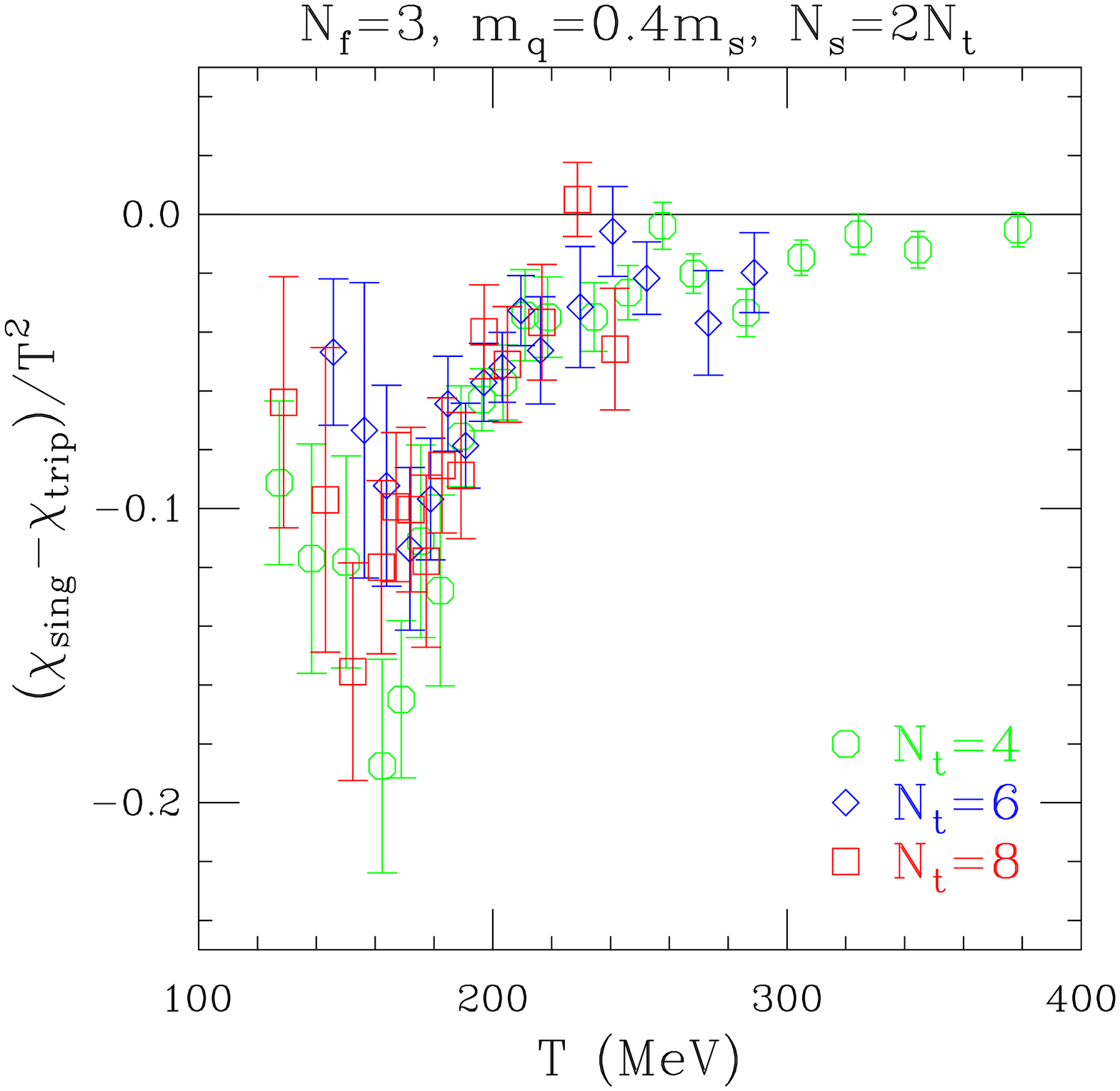}
\end{center}
\caption{The difference between singlet and triplet quark number
susceptibility for
$N_f=3$ on $8^3 \times 4$, $12^3\times 6$ and $16^3\times 8$ lattices.
}
\label{MILC2}
\end{figure}

They calculated the singlet and non-singlet quark number susceptibilities,
$\chi_S(+)$ and $\chi_{NS}(-)$,
\begin{equation}
\chi_{S,NS} = \left(\frac{\partial}{\partial \mu_u} \pm \frac{\partial}{\partial
\mu_d}\right)
(n_u \pm n_d)
\end{equation}
where $n_i$ for $i=u, d$ is the quark number densities given by
\begin{equation}
n_i = \frac{T}{V}\frac{\partial \log Z}{\partial \mu_i}
\end{equation}
These susceptibilities are found to be very small in the low temperature
phase
and to increase abruptly at the transition temperature.
The two susceptibilities are almost same, e.g. $\chi_S \approx \chi_{NS}$
which indicates that fluctuations in the $u$ and $d$ quark densities
are uncorrelated.
These susceptibilities are important since they are related to event-by-event
fluctuations and diffusion
in relativistic heavy-ion collisions\cite{EVENT}.
%
Further studies in the susceptibilities for quenched and 2 flavors QCD can
be found in
\cite{GOTTLIEB,GAVAI,GavaiGupta,GGM}.
MILC (MIMD Lattice Computation) collaboration recently calculated the
susceptibilities for
three flavors with improved staggered fermions.\cite{BERNARD}
Fig.\ref{MILC1} shows the triplet susceptibility $\chi_{trip} (=\chi_{NS}$
here) on $8^3\times 4$,
$12^3\times 6$ and $16^3\times 8$ lattices and indicates the excellent
scaling properties
between the $N_t =6$ and $N_t=8$ results.
Fig.\ref{MILC2} shows the difference between singlet and triplet
susceptibility.
One can see the clear difference between the two susceptibilities around the
phase transition which occurs at about 180 MeV.
%
%


QCD-TARO (Thousand-cell ARray processors for Omni-purposes) collaboration
has extended the idea of Ref.\cite{Gott} further and
calculated derivatives of meson screening masses and the quark condensation
with respect to the chemical potential at $\mu=0$.\cite{QCD-Taro02}
Consider a meson propagator $C(z)$ in $z$ direction consisting of $u$ and $d$ quarks as an example,
\begin{eqnarray}
C(z) &=& A ( e^{-M z} + e^{M(L_z-z)} )
\label{meson-propa1}
\\
&=&
\frac{1}{Z}  \int \mD U \,H(z)H^\dagger(0)\, \,
e^{-\beta S_G-\bar{\psi}\Delta\psi}
\nonumber \\
&=&
\frac{1}{Z}  \int \mD U G(z)
\det\Delta \, e^{-\beta S_G} ,
\label{meson-propa2}
\end{eqnarray}
where $H(x)=\bar{\psi}(x)\Gamma\psi(x)$ is a meson operator
and
\begin{equation}
G(z) \equiv \Tr \Gamma\Delta_u^{-1}(z,0) \Gamma\Delta_d^{-1}(0,z) .
\end{equation}
We neglect possible higher states for simplicity.

The first derivative of $C(z)$ with respect to
the chemical potential leads to
\begin{equation}
\frac{1}{C(z)} \frac{dC(z)}{d\mu}
 = \frac{1}{A} \frac{d A}{d\mu}
 + \frac{dM}{d\mu}
  \left\{ \left( z-\frac{L_z}{2} \right) \tanh
     \left[ M \left( z-\frac{L_z}{2} \right) \right]
    - \frac{L_z}{2} \right\} .
\label{deri-1}
\end{equation}
Similarly the second derivative is given as
\begin{eqnarray}
\frac{1}{C(z)} \frac{d^2C(z)}{d\mu^2}
 &=& \frac{1}{A} \frac{d^2 A}{d\mu^2}\left( \frac2{A}\frac{d A}{d\mu}
\frac{dM}{d\mu} +\frac{d^2M}{d\mu^2}\right)
\times \left\{ \left( z-\frac{L_z}{2} \right) \tanh
     \left[ M \left( z-\frac{L_z}{2} \right) \right]
    - \frac{L_z}{2} \right\} \nonumber \\
 & + & \left(\frac{dM}{d\mu}\right)^2 \left\{\left( z-\frac{L_z}{2} \right)^2
+ \frac{L_z^2}{4} -L_z \left( z-\frac{L_z}{2} \right) \tanh
     \left[ M \left( z-\frac{L_z}{2} \right) \right]
\right\}.
\label{deri-2}
\end{eqnarray}

The numerical data of the derivative of $C(z)$ to $\mu$ is given as follows.
\begin{equation}
\frac{dC(z)}{d\mu}=\frac{d<G(z)>}{d\mu}
= \left \langle \dot{G}+G\frac{ \dot{D} }{ D } \right \rangle
  -  \langle G \rangle  \left< \frac{\dot{D} }{ D } \right \rangle  ,
\label{deri4}
\end{equation}

\begin{equation}
\frac{d^2<G(z)>}{d\mu^2}
= \left \langle \ddot{G}+2\dot{G}\frac{ \dot{D} }{ D } +G\frac{ \ddot{D} }{ D }\right \rangle
 -2\left \langle \dot{G}+G\frac{ \dot{D} }{ D } \right \rangle \left< \frac{\dot{D} }{ D } \right \rangle
  -  \langle G \rangle  \left\{ \left< \frac{\ddot{D} }{ D } \right >
  -2 \left \langle \frac{\dot{D} }{ D } \right \rangle^2 \right\} ,
\label{deri5}
\end{equation}
where $D$ stands for $\det \Delta$ and 
the dotted $\dot{O}$ and $\ddot{O}$ denote the first and 
the second derivatives of an operator $O$ with respect to $\mu$.
Fitting the numerical data of Eqs.(\ref{deri4}) and (\ref{deri5}) 
to Eqs.(\ref{deri-1}) and (\ref{deri-2})
we can determine the derivatives of the residue $A$ and the meson mass $M$.
Thus with the derivatives of $M$, we can obtain the behavior of
a hadron mass in the vicinity of $\mu=0$ as a function of the chemical potential,
\begin{equation}
M(\mu) = M(0) + \mu \left .\frac{\partial M}{\partial\mu} \right |_{\mu=0}
+\left . \frac{1}{2}\mu^2 \frac{\partial^2 M}{\partial\mu^2} \right |_{\mu=0}
+ \cdots
\end{equation}

\begin{figure}[hbt]
\begin{center}
\includegraphics[width=.6 \linewidth]{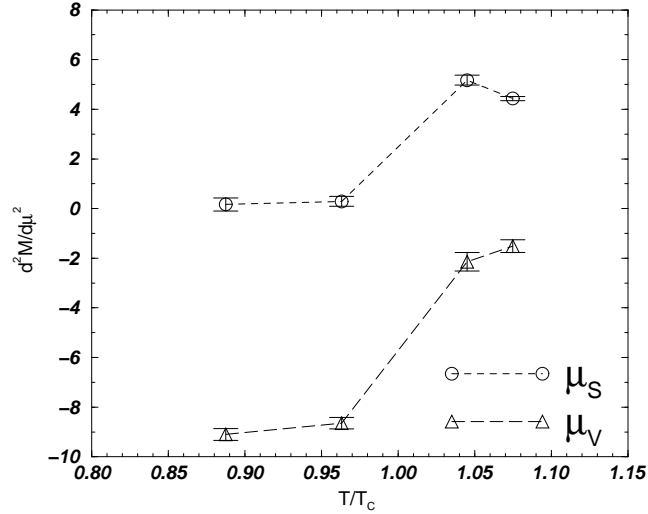}
\end{center}
\caption{
The second derivative of the pseudo scalar mass
$\left .\frac{d^2M}{d\mu^2}\right |_{\mu=0}$ for isoscalar and isovector
chemical potentials, $\mu_S$ and $\mu_V$.
(Ref.~\citen{QCD-Taro02}).
}
\label{fig3-1}
\end{figure}

The first derivative of the pseudo-scalar (PS) mass turned out to be consistent with zero $\mu=0$.
In Fig.\ref{fig3-1}, we show the second derivative of the PS
mass where iso-scalar and iso-vector
type chemical potentials, $\mu_S$ and $\mu_V$, are
introduced,
\begin{eqnarray}
\mu_S &\equiv& \mu_u = \mu_d ,
\nonumber \\
\mu_V &\equiv& \mu_u = - \mu_d .
\end{eqnarray}
In the low temperature phase the second derivative of the PS mass $M$
with respect to $\mu_S$ is small. 
This is expected since below the critical temperature and in the vicinity 
of zero $\mu_S$, the PS meson still keeps a property as a Goldstone boson 
and persists its behavior.
On the other hand, above $T_c$, the PS meson is no longer a Goldstone boson
and the second derivative of $M$ seems to remain finite.
In contrast to the case of $\mu_S$ the second derivative of $M$ with respect to
$\mu_V$ is significantly large in the low temperature phase and decreases
in magnitude above $T_c$.

As we will see later, the fermion determinant with the isovector
chemical potential is equivalent to that of a two-flavor system without
the phase.  
Large nonzero values of the second derivative of the pseudo-scalar
meson mass may suggest some phase structure in finite but small
chemical potential regions. 
Son and Stephanov have shed a new light on the model as a finite isospin
density system. See section \ref{sec-FiniteIsospin}. 

\begin{figure}[hbt]
\begin{center}
\includegraphics[width=.6 \linewidth]{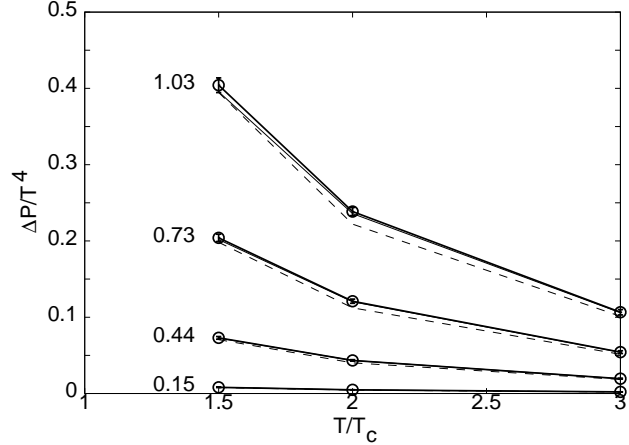}
\end{center}
\caption{$\displaystyle \frac{\Delta P}{T^4}$ as a function of $T/T_c$ for
the values of $\mu /T_c$.}
\label{GG1}
\end{figure}

Gavai and Gupta calculated the depedence of the pressure $P$ on $\mu$
through
a Taylor expansion at $\mu=0$.\cite{GAVAIG} For a
homogeneous system,
the pressure $P$ is obtained by $P=-F/V$, where $F=-T \log Z$.
Therefore the Taylor expansion of $P$ at $\mu_i=0$ ($i=u$,$d$,...) is given
by
\begin{eqnarray}
P(\mu_u,\mu_d,...) & = & P(0) -\frac1{V} \sum_i \frac{\partial F}{\partial
\mu_i}
-\frac1{2V}\sum_{i,j} \frac{\partial^2 F}{\partial \mu_i \partial \mu_j}
+ ...   \\
& = & P(0) + \sum_i \chi_i + \sum_{i,j}\chi_{ij} + ...
\end{eqnarray}
The second term $\chi_i$ corresponds to the quark number density which is
zero at $\mu=0$ and
the third therm $\chi_{ij}$, the quark number susceptibility.
One can consider the higher order derivative terms and all the odd
derivatives vanish by CP symmetry.
The Taylor series of $P$ up to the fourth order for $\mu_u=\mu_d=\mu$ is
given by
\begin{equation}
\frac{\Delta P}{T^4} =
\left(\frac{\chi_{uu}}{T^2}\right)\left(\frac{\mu}{T}\right)^2
\left[1+\left(\frac{\mu/T}{\mu_*/T}\right)^2\right]
\end{equation}
where $\Delta P = P(\mu) -P(0)$, and
\begin{equation}
\frac{\mu_*}{T} =\sqrt{\frac{12\chi_{uu}/T^2}{|\chi_{uuuu}|}}
\end{equation}
$\chi_{uudd}$ is neglected since it turns out to be numerically smaller than
the errors in $\chi_{uuuu}$.
Fig.\ref{GG1} shows $\frac{\Delta P}{T^4}$ obtained in quenched QCD.
For $\mu/T_c=0.15$, the effects of $\mu$ on $P$ are very small, which
implies that
at RHIC region where $\mu/T_c\sim 0.06$, the effects are also small.
As $\mu/T_c$ increases, the effects become significant.
One could see the effects at SPS region where  $\mu/T_c\sim 0.45$.

\subsection{Glasgow approach}

The Glasgow group has been developing the way towards simulations of 
lattice QCD at finite density. See Ref.~\citen{barbour98} and references 
therein.

Consider the partition function at $\mu$ normalized by that of $\mu=0$.
\begin{eqnarray}
Z &=& 
\frac{\int \mD U \det\Delta(\mu) \, e^{-\beta S_G}}
{\int \mD U \det\Delta(0) \, e^{-\beta S_G}}
=
\frac{\int \mD U 
\frac{\det\Delta(\mu)}{\det\Delta(0)} 
\det\Delta(0)\, e^{-\beta S_G}}
{\int \mD U \det\Delta(0) \, e^{-\beta S_G}}
\nonumber \\
&=&
\left \langle \frac{\det\Delta(\mu)}{\det\Delta(0)} \right \rangle _{\mu=0}
=
\sum_{n=-3N_s^3}^{+3N_s^3} \langle b_{|n|} \rangle _{\mu=0} \, e^{n\mu/T}
\end{eqnarray}
Taking the derivative of $Z$ with respect to $T$ and $\mu$,
we obtain the energy and the baryon number density, respectively.  
By investigating zeros in the complex $\mu$-plane,  
\begin{equation}
Z(\mu) = 0
\end{equation}
we have Lee-Yang zero.
It was very difficult to obtain reliable values of $ \langle b_{|n|} \rangle $ numerically
at low temperature when $\mu$ increases.

\subsection{Reweighting by Fodor and Katz}

\begin{figure}[hbt]
\begin{center}
\includegraphics[width=.6 \linewidth]{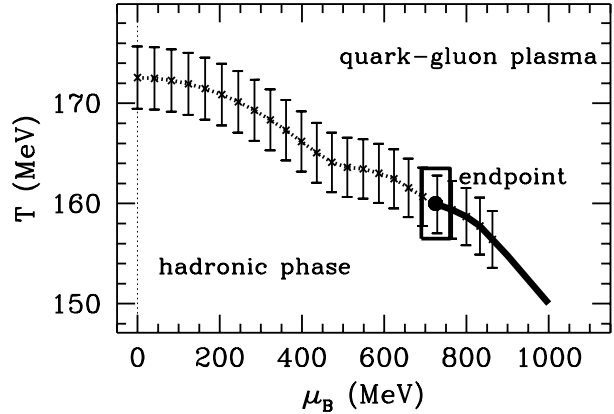}
\end{center}
\caption{
Phase diagram obtained by Fodor and Katz.
(Ref.~\citen{FodorKatz})
}
\label{fig3-3}
\end{figure}

Fodor and Katz \cite{FodorKatz} have achieved a great step in lattice QCD 
simulation at finite density, i.e., they have for the first time succeeded 
in estimating the phase transition line at finite $\mu$ in $(T,\mu)$
parameter plane. See Fig.\ref{fig3-3}.

\begin{figure}[hbt]
\begin{center}
\includegraphics[width=.6 \linewidth]{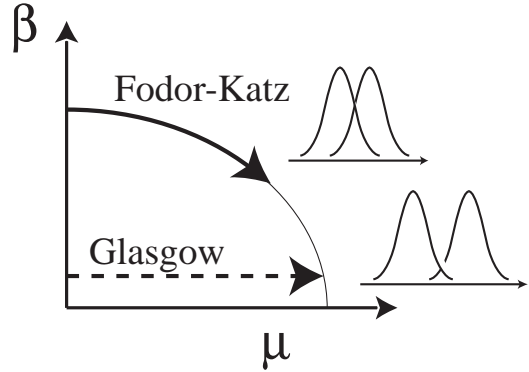}
\end{center}
\caption{
Schematic diagram of multi-parameter reweighting.
}
\label{fig3-2}
\end{figure}

They proposed using the multi-parameter reweighting method.
The essential idea is to use the gauge coupling constant $\beta$ as
a controlling parameter,
\begin{eqnarray}
\langle O \rangle  &=& \frac{1}{Z(\mu)} \int \mD U \, O \, \det\Delta(\mu) \, e^{-\beta S_G} 
\nonumber \\
&=&
\frac{1}{Z(\mu)}
\int \mD U \, O \, \frac{\det\Delta(\mu)}{\det\Delta(0)}
\, e^{-(\beta-\beta_0) S_G}
\det\Delta(0) 
\, e^{-\beta_0 S_G}
\nonumber \\
&=&
\frac{ \langle O \frac{\det\Delta(\mu)}{\det\Delta(0)}e^{-\Delta\beta S_G} \rangle _0}
{ \langle \frac{\det\Delta(\mu)}{\det\Delta(0)}e^{-\Delta\beta S_G} \rangle _0}
\label{multi-para-rew}
\end{eqnarray}
and choose $\beta_0$ so that
\begin{equation}
\left|
\frac{\det\Delta(\mu)}{\det\Delta(0)}
\right|
e^{-\Delta\beta S_G}
\label{multi-para-fac}
\end{equation}
becomes as large as possible, i.e., the overlap should be large.
In Eq.(\ref{multi-para-rew}) , $ \langle \cdots \rangle _0$ 
stands for the evaluation
at $(\beta_0,\mu=0)$.
The coupling constant $\beta_0$ is taken at the confinement/deconfinement
transition point, and physical quantities near the phase transition line
are calculated by the formula (\ref{multi-para-rew}), i.e., quantities on the
phase transition line in $(T,\mu)$ plane are evaluated at $(T_c,\mu=0)$
with the reweighting factor.  A possible interpretation for this 
procedure is
that points on the phase transition line have similar nature.  
(Fig.\ref{fig3-2})

If the reweighting factor, (\ref{multi-para-fac}), is large, we can
get better signal. Further if in some regions of $(\mu,\beta)$ parameter
space, this value changes little, we have similar reweighting effect
there.  Ejiri investigated the problem, and observed that the chemical
potential effect through the phase of the fermion matrix does not
contribute to the $\beta$ dependence of the gauge action, and
$\mu$ in the amplitude, $|\det\Delta(\mu)|$, correlates strongly
with $e^{-\Delta\beta S_G}$. \cite{Ejiri} 

In this calculation we need the ratio of the determinant,
$\det\Delta(\mu)/\Delta(0)$. Gibbs obtained a useful formula, \cite{Gibbs}
\begin{equation}
\det\Delta(\mu) = e^{N_t V_s \mu}\prod_{i=1}^{6V_s} 
(\lambda_i + e^{-N_t \mu}) ,
\label{Eq-Gibbs}
\end{equation}
where $\lambda_i$ are eigen values of a matrix constructed from
the spatial and mass terms of fermion matrix, which does not
depend on the chemical potential $\mu$.  Therefore once we
obtain the eigenvalues $\lambda_i$, we can calculate the 
determinant at any $\mu$ for a fixed gauge field. 
See Appendix A.

\begin{figure}[hbt]
\begin{center}
\includegraphics[width=.6 \linewidth]{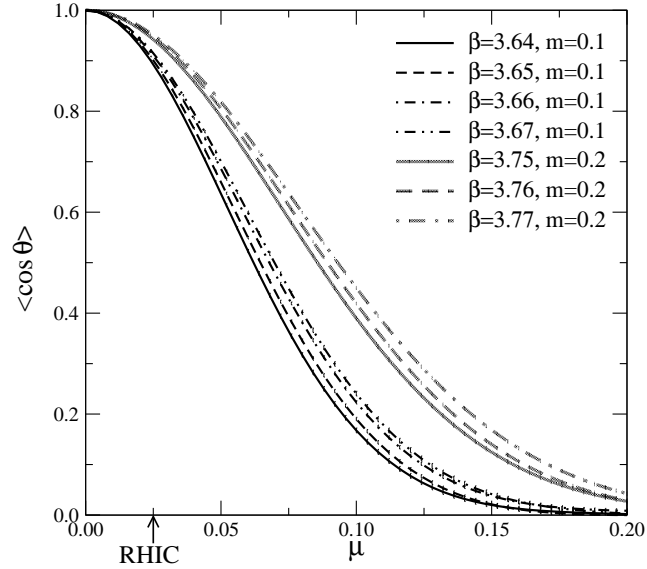}
\end{center}
\caption{
The phase of complex determinant calculated by Bielefeld-Swansea
collaboration.
(Ref.~\citen{Allton}).
}
\label{figa-1}
\end{figure}

Note that, though Fodor and Katz obtained great success near the 
phase transition line, they did not 
solve the sign problem of the finite density
lattice QCD.
It still remains difficult to study a low-temperature region where
the phase fluctuation coming from the complex fermion determinant
is most probably large.
Fig.\ref{figa-1} shows $\langle \cos\theta \rangle$ as a function
of $\mu$ obtained by Bielefeld-Swansea group for various $beta$
and quark masses.  At large $\mu$, the complex phase of the fermion
determinant becomes large, i.e., the nominator and denominator
in Eq.\ref{mod-det} are small.

\subsection{Taylor expansion}

Fodor and Katz calculated determinants in Eq.(\ref{multi-para-rew}),
but it is difficult to study large lattices by the present computers.
In any case, we cannot study large chemical potentials because of
the sign problem.

Bielefeld-Swansee collaboration\cite{Allton} proposed using Taylor expansion
at $\mu=0$,
\begin{equation}
\ln\left(\frac{\det\Delta(\mu)}{\det\Delta(0)}\right)
=
\sum_{n=1}^{\infty} 
\frac{\mu^n}{n!}
\frac{\partial^n\ln\det\Delta}{\partial\mu^n}
\end{equation}
At the lowest order, the phase of the determinant is obtained as 
$\displaystyle \mu {\rm Im} {\rm Tr} M^{-1}\frac{\partial M}{\partial \mu}$
which is easier to evaluate than the direct calculation 
from the determinant itself\footnote{$\displaystyle \rm{Tr} M^{-1}\frac{\partial M}{\partial \mu}$
is usually estimated by the noise method which uses noise vectors but this term might need
a number of vectors to obtain its value with accuracy\cite{FKT}}.

\begin{figure}[hbt]
\begin{center}
\includegraphics[width=.6 \linewidth]{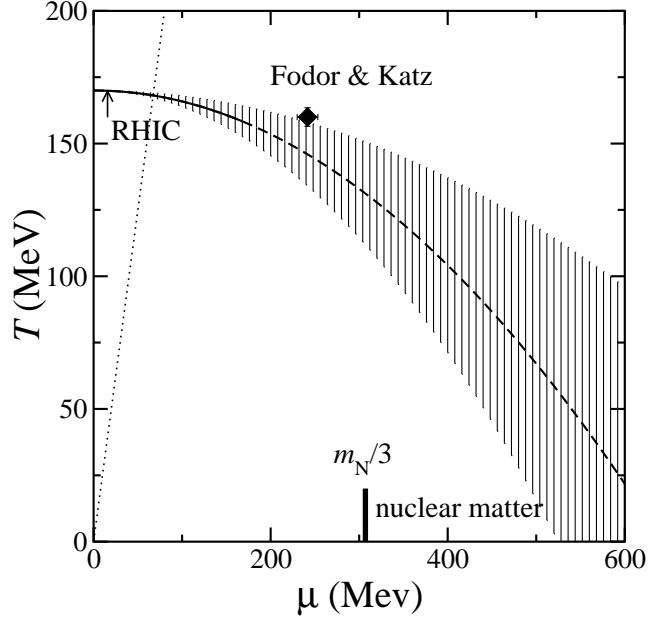}
\end{center}
\caption{
Phase diagram obtained by Bielefeld-Swansea collaboration.
(Ref.~\citen{Allton}).
}
\label{3-4}
\end{figure}

\begin{figure}[hbt]
\begin{center}
\includegraphics[width=.6 \linewidth]{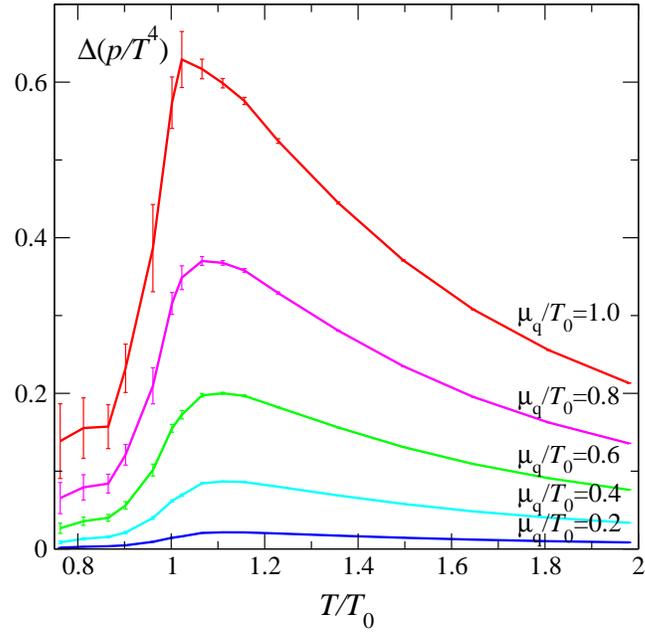}
\end{center}
\caption{
The equation of state correction $\Delta(p/T^4)$ vs. $T/T_{c0}$
for various $\mu/T_{c0}$ by Bielefeld-Swansea collaboration.
(Ref.~\citen{Allton03}).
}
\vspace{5mm}
\label{3-4b}
\end{figure}
Employing the formula, they have performed thorough studies in small
$\mu$ regions which cover RHIC experiments
and obtained the phase diagram similar to that of Fodor and Katz.

Recently they study the thermodynamical grand potential up to
the fourth order of the derivative with respect to the chemical
potential, and calculate the quation of state. \citen{Allton03}
Fig.\ref{3-4b} shows $\Delta(p/T^4)$ which is defined as
\begin{equation}
\Delta\left( \frac{p}{T^4}\right) \equiv 
 \left.\frac{p}{T^4}\right\|_{T,\mu} - \left.\frac{p}{T^4}\right\|_{T,0}.
\end{equation}
The correction of the pressure is large for $0.9\le T/T_{0}\le 1.3$, 
$\mu/T_{0}\ge.5$, but will decrease as $T$ rises further.
Fodor, Katz and Szabo reported similar result. \citen{FodorKatzSzabo}


\subsection{Imaginary chemical potential}

If the chemical potential is pure imaginary, i.e.,
$\mu=i\mu_I$, then the fermion matrix $\Delta$ becomes, 
\begin{equation}
\Delta = D_\nu\gamma_\nu + m + i\mu_I\gamma_0
\label{det4}
\end{equation}
and 
\begin{equation}
\Delta^\dagger = \gamma_5 \Delta \gamma_5.
\end{equation}
Thus $\det\Delta$ with a pure imaginary chemical potential is real.

If we use the expressions \ref{muatEdge2} in section 2,
the link variable at $t=N_t$ has the form,
\begin{equation}
e^{i\frac{\mu_I}{T}} U_t(\vec{x},N_t) .
\end{equation}
\noindent
As is well known, the gauge action $S_G$ has $Z_3$ symmetry, i.e.,
$S_G$ is invariant when link variables on a time slice change as
\begin{equation}
U_t \rightarrow z U_t
\end{equation}
where $z$ is an element of $Z_3$ group,
\begin{equation}
z = e^{i\phi}, \qquad \phi=0, \frac{2\pi}{3}, \frac{4\pi}{3}.
\end{equation}
But the fermion action is not invariant under the transformation.

In the case of the imaginary chemical potential, the $Z_3$ transformation
of the link variables can be absorbed into $\mu_I$ in the fermion action.
Conversely, the shift of the imaginary chemical potential,
\begin{equation}
\mu_I \rightarrow \mu_I + \phi
\end{equation}
can be managed as the change of the link variables, $U_t \to z U_t$,
which can be regarded as the simple change of the integration
variables $U_t$ to $zU_t$. 
Therefore, the path integral gives the same result,
\begin{equation}
Z(\mu_I+\phi) = Z(\mu_I) .
\end{equation}

\begin{figure}[hbt]
\begin{center}
\begin{minipage}{ 0.48\linewidth}
\includegraphics[width=1.0 \linewidth]{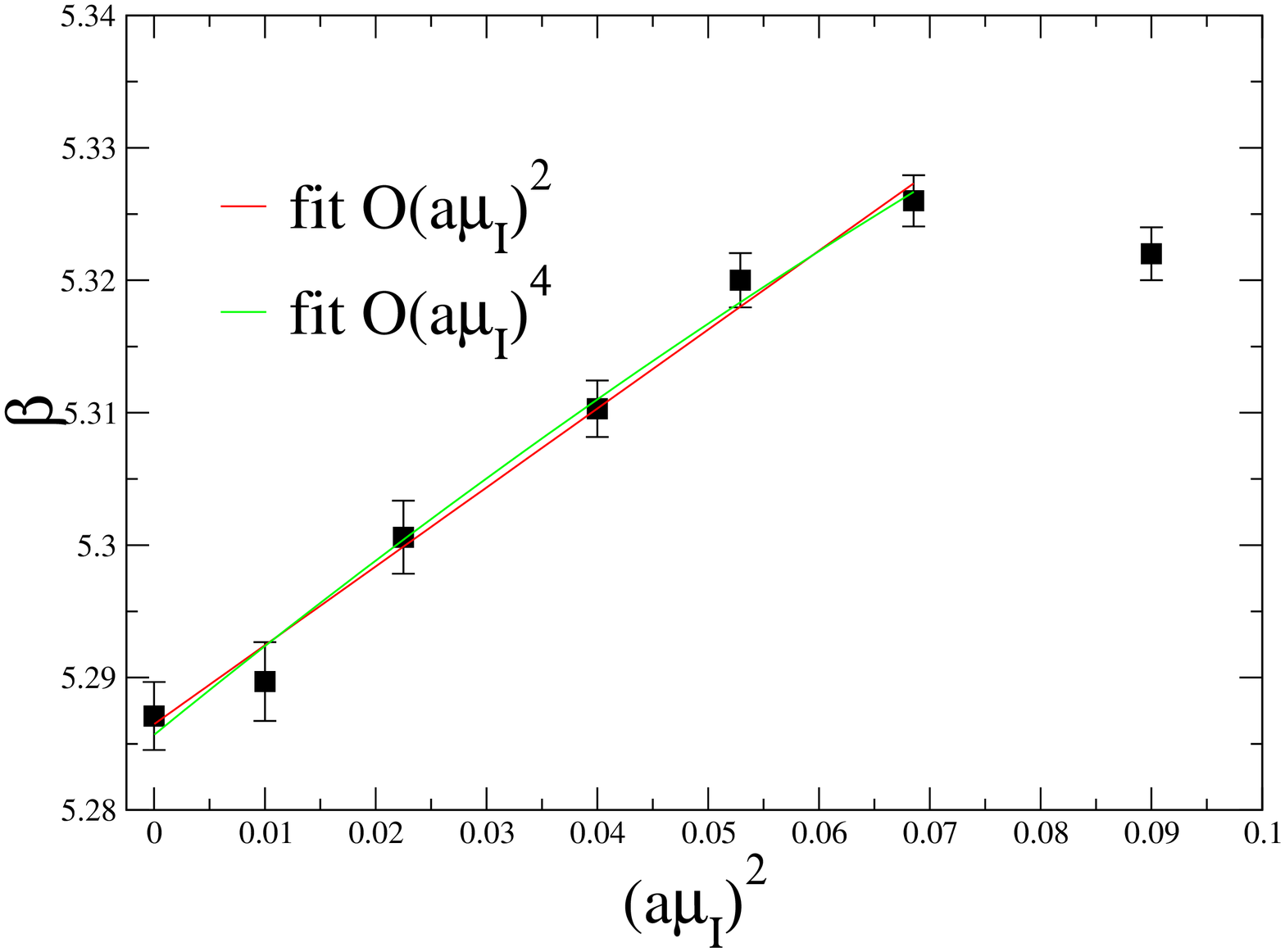}
\end{minipage}
\hspace{1mm}
\begin{minipage}{ 0.48\linewidth}
\includegraphics[width=1.1\linewidth]{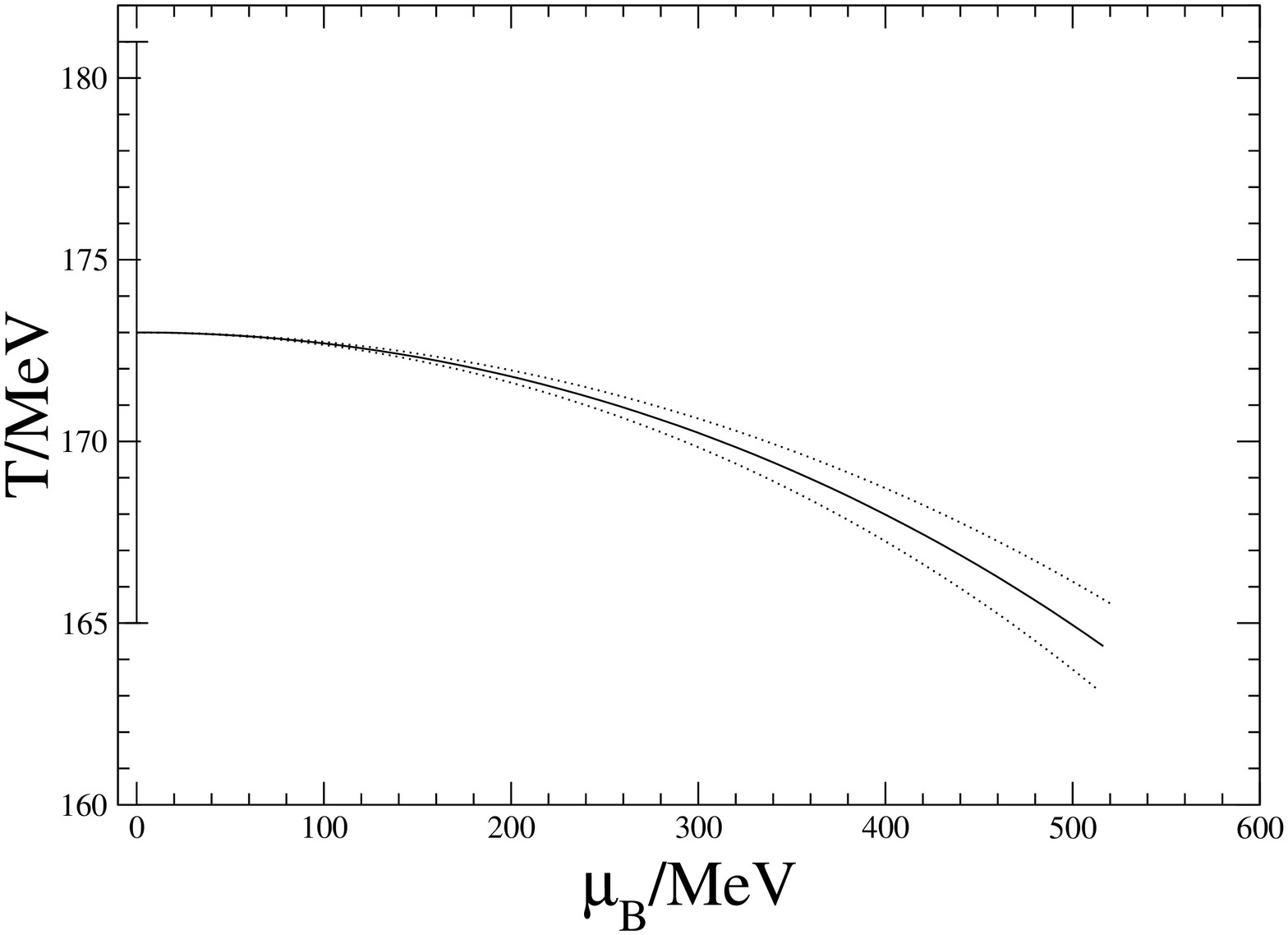}
\end{minipage}
\end{center}
\caption{
Imaginary chemical potential study by de Forcrand and Philipsen.
Critical $\beta_c$ obtained with the imaginary chemical potential
$\mu_I$ is fitted by $\beta_c(a\mu_I) = c_0 + c1(a\mu_I)^2$ (left), and
converted through ${\rm {Im}}\mu=-i {\rm {Re}}\mu$ to obtain the critical lines
in $(T,\mu)$ (right).
(Ref.~\citen{FP}).
}
\label{fig3-5}
\end{figure}

Under this constraint, we can perform the analytic continuation of
the results obtained with $\mu_I$ to the real $\mu$. 
The idea has been
known for many years, and recently de Forcrand and Philipsen, and
D'Elia and Lombardo have successfully performed simulations.
In Ref.~\citen{FP}, the critical $\beta$ is fitted as
\footnote{
On the genuine phase transition line, the partition function should
be singular, but small finite lattice has only a relic of it.
}
\begin{equation}
\beta = a_0 + a_2\mu_I^2 + \cdots
\end{equation}
and transformed to the real chemical potential,
\begin{equation}
\beta = a_0 - a_2 \mu^2 + \cdots
\end{equation}
The result is shown in Fig.\ref{fig3-5}.

Imaginary chemical potential may be considered as a special boundary
condition.

\section{Two Color}\label{sec-2color}

Since SU(2) matrix $U$ has the following property,
\begin{equation}
U_\mu^{*} = \sigma_2 U_\mu \sigma_2,
\end{equation}
\begin{equation}
\Delta(U, \gamma_\mu)^{*} = \Delta(U^{*}, \gamma_mu^{*}) 
\sigma_2 \Delta(U, \gamma_mu^{*})  \sigma_2,
\label{Wfermion-SU2}
\end{equation}
then $\{ \det \Delta(x,x'; \gamma_\mu) \}^{*}
= \det  \Delta(x,x'; \gamma_\mu^{*})$.
$\gamma_\mu^{*}$ are a representation which satisfies the anti-commutation
relations as $\gamma_\mu$
and $\det \Delta$ should not depend on the representation.
\footnote{
An explicit transformation between $\gamma_\mu$ and $\gamma_\mu^{*}$ 
can be easily found: Since $C\gamma_\mu C = -\gamma_\mu^{*}$,
$\gamma_\mu^{*} = V\gamma_\mu V^{-1}$ with $V=C\gamma_5$, where
$C=i\gamma_0 \gamma_2$ is the charge conjugation matrix and we use
hermitian $\gamma$ matrices.
}
Consequently,
\begin{equation}
 \det \Delta(x,x'; \gamma_\mu)^{*} =  \det \Delta(x,x'; \gamma_\mu)
\end{equation}
i.e., the fermion determinant is real.
For reader's convenience, we show a compilation of color SU(2) finite 
density simulations in Table \ref{Tab-su2c}.

In Refs.~\citen{KSteT} and \citen{KSTVZ},
the authors have determined the lowest order, i.e., $O(E^2)$,
effective Lagrangian including the chemical potential, 
\begin{eqnarray}
{\cal L}_{\rm eff} 
&=& {F^2\over2} {\rm Tr} \partial_\nu\Sigma \partial_\nu\Sigma^\dagger
+ 2\mu F^2 {\rm Tr} B\Sigma^\dagger\partial_0\Sigma
\nonumber\\&&
\hskip 1em -F^2
\mu^2 {\rm Tr} \left(\Sigma B^T\Sigma^\dagger B + B B\right)
-F^2 m_\pi^2 {\rm Re} {\rm Tr}\left(\hat M\Sigma\right).
\label{ChPT}
\end{eqnarray}
where
\begin{equation}
B \equiv 
\left (
\begin{array}{cc}
+1&0\\0&-1
\end{array} \right )
\end{equation}
\begin{equation}
\hat M \equiv 
\left ( \begin{array}{cc}
 0 & 1 \\ 
-1 & 0 
\end{array} \right )
\end{equation}

\begin{figure}[hbt]
\begin{center}
\begin{minipage}{ 0.48\linewidth}
\includegraphics[width=1.0 \linewidth]{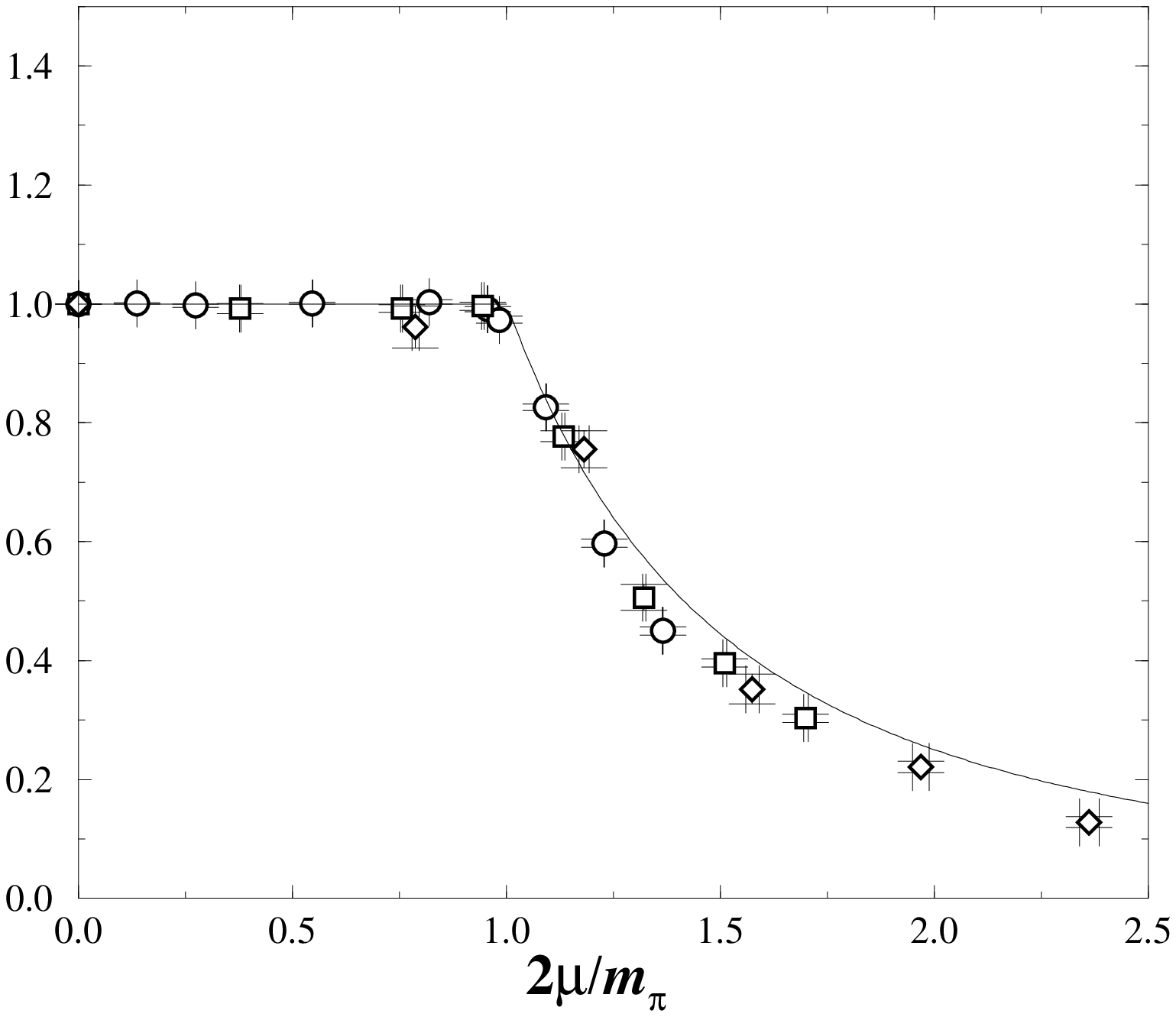}
\label{Fig-unipbp}
\end{minipage}
\hspace{1mm}
\begin{minipage}{ 0.48\linewidth}
\includegraphics[width=1.1\linewidth]{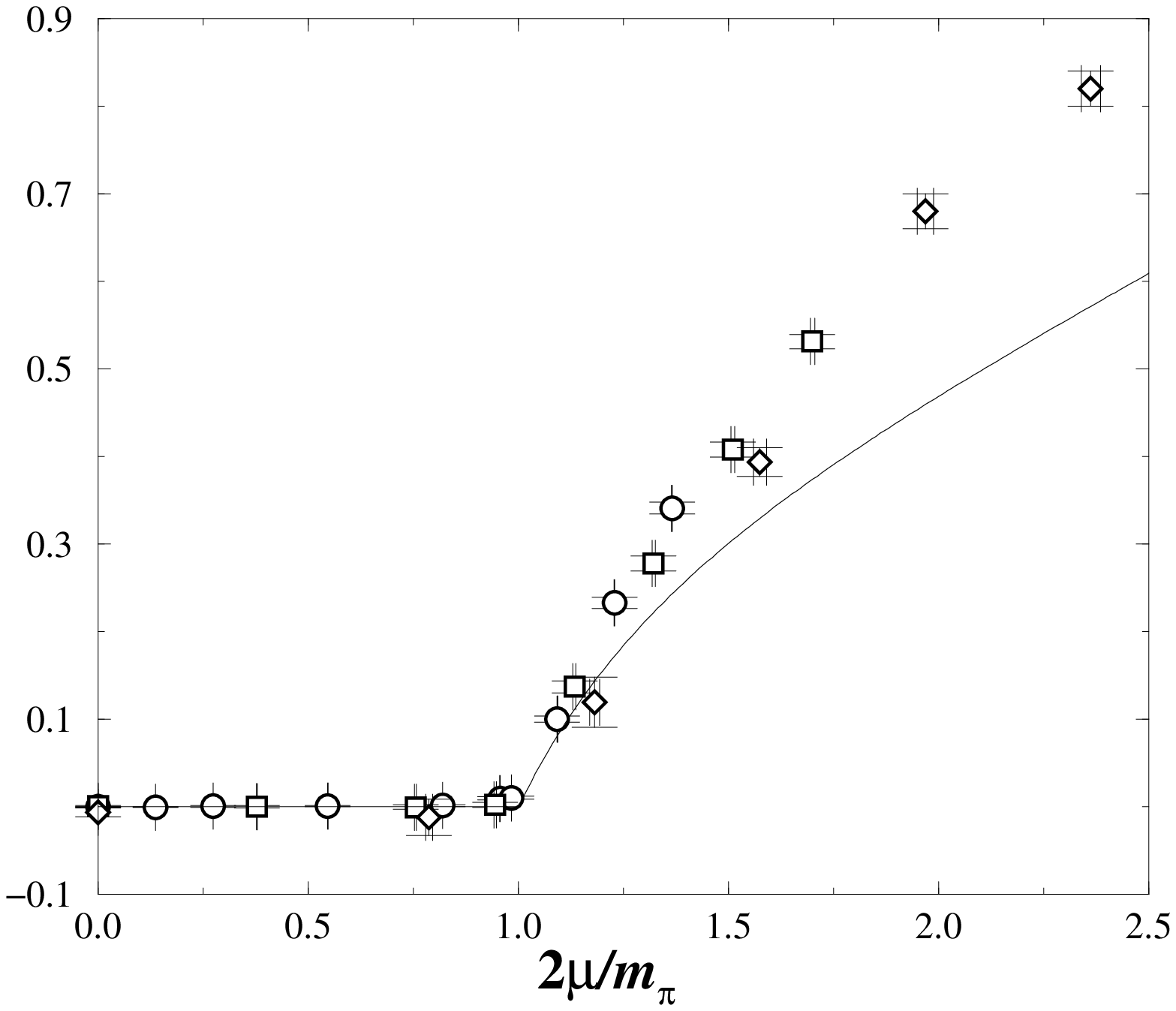}
\end{minipage}
\end{center}
\caption{
The scaled chiral condensate, $\langle \bar{\psi}\psi \rangle /
\langle \bar{\psi}\psi \rangle _0$, (left) 
and number density, $n_B$, (right) together with the chiral perturbation 
theory prediction.  (Ref.~\citen{HaMoMoOeScSk})
}
\label{fig1-11}
\end{figure}

It has been shown in Ref.~\citen{HaMoMoOeScSk} that the numerical data
are well described by the chiral perturbation theory. 
Fig.\ref{fig1-11} presents the chiral condensate and
the number density, respectively, together with the leading order
prediction,
\begin{equation}
\frac{\langle \bar{\psi}\psi \rangle }{\langle \bar{\psi}\psi \rangle _0} = \left \{ 
\begin{array}{c}
 1 \\
 \frac{1}{x^2} 
\end{array} \right. 
;
\left ( \frac{m_\pi}{8\langle \bar{\psi}\psi \rangle _0} \right ) n_B= \left \{ 
\begin{array}{cc}
 0                           & x<1 \\
 \frac{x}{4}(1-\frac{1}{x^4}) & x>1 
\end{array} \right. 
\end{equation}
where
\begin{equation}
x \equiv \frac{2\mu}{m_\pi}
\end{equation}

Using extended QCD inequality, 
Kogut, Stephanov and Toublan have shown that in SU(2) color QCD
at finite density, condensation must occur in scalar diquark channel.
\cite{KSteT}
The correlator of a meson, $\bar{\psi}\Gamma\psi$ is given by
$\langle \Tr G(x,0)\Gamma G(0,x)\Gamma \rangle $, where $G=\Delta^{-1}$ are quark
propagators and taken as a matrix in Dirac and color space.
At $\mu=0$, Eq.(\ref{det2}) leads to $\Delta^\dagger = \gamma_5
\Delta \gamma_5$. From Cauchy-Schwarz inequality, one obtains
$\Tr A B^\dagger \le \sqrt{ \Tr AA^\dagger}\sqrt{ \Tr BB^\dagger }$.
Then
\begin{eqnarray}
\Tr G(x,0) \Gamma G(0,x) \Gamma 
& = & \Tr G(x,0) \Gamma \gamma_5 G(x,0)^\dagger \gamma_5 \Gamma
\nonumber
\\ 
& \le &  
\sqrt{G(x,0)G(x,0)^\dagger} 
\sqrt{\Gamma \gamma_5 G(x,0)^\dagger \gamma_5 \Gamma
\left (
\Gamma \gamma_5 G(x,0)^\dagger \gamma_5 \Gamma
\right )^\dagger \nonumber} \\
&  = &  \Tr G(x,0) G(x,0)^\dagger
\label{QCDineq0}
\end{eqnarray}

The right-hand side corresponds to the pion propagator which behaves as
$C_\pi \exp(-m_\pi |x| )$ at large $x$.
Therefore
\begin{equation}
C e^{-m|x|} \le C_\pi e^{-m_\pi |x| }
\end{equation}
This relationship holds at arbitrary large $x$ only if $m \ge m_\pi$, i.e.,
the pion should be the lightest meson. This is the essence of QCD inequality.

The above QCD inequality does not hold at finite density. For SU(2), however,
there is another relationship, $V \sigma_2 G \sigma_2 V^{-1}$, and using this
the authors of Ref.~\citen{KSteT} show that the mass of the scalar ($0^+, I=0$)
diquark, $\psi_C  \sigma_2 \gamma_5 \psi$, is lowest, where
$\psi_C \equiv \psi^T C$.
Therefore if diquark condensation occurs, it appears in this channel.

\begin{table}[htbp]
	\begin{center}
	\caption{Two Color}
		\begin{tabular}{|l|p{50mm}|p{45mm}|l|} \hline \hline 
		Action & parameters & comments & Ref. \\ \hline
     Plaq. + Wilson & 
     $N_f=2$, $\beta=1.4 \sim 2.0$,  
     $8^3 \times 2$ 
     &  
     first calculation,  
     pseudo-Fermion Method
     &
    \citen{Nakamura84} \\ \hline
    Plaq. + KS &
     $\beta=1.3$, $m_q=0.05, 0.07$,  
    $6^3 \times 12$
    &
     symmetries and spectrum  
     of lattice gauge theory 
     & 
    \citen{HKLM99}\\ \hline
       Plaq. +  KS    &
     ($\beta=1.3$, $m_q=0.07, 0.05$),  
     ($\beta=1.5$, $m_q=0.1, 0.07, 0.05$), 
     ($\beta=1.3 \sim 2.3$, $m_q=0.1$),  
     $6^4$
     &
      interquark potential&
     \citen{MPL} \\ \hline

    Plaq. + adjoint KS & 
    $N_f=4$, $\beta=2.0$, 
    $m_q=0.01 \sim 0.1$, 
     $4^3 \times 8$ 
    &  
    adjoint dense matter,  
    Two-Step Multi-Boson algorithm
    & 
    \citen{HaMoMoOeScSk} \\ \hline
     Plaq.    + KS   & 
    $N_f=8$, $\beta=2.0$,  $m_q=0.025 \sim 0.2$, 
     $4^4$, $6^4$ 
    &  
    fermion condensates & 
    \citen{Aloisio} \\ \hline
     Plaq. + KS & 
     $N_f=4$, $m_q=0.05$,  
     $8^4$, $8^3 \times 4$, $12^3 \times 6$, $16^4$          
     &   
     diquark condensation,  
     tricritical point
     &
    \citen{KTS, KTS02} \\ \hline
     Plaq. + KS & 
     $N_f=4$, $\beta=1.5$, $m_q=0.1$,  
     $8^4$, $12^3 \times 24$
       & 
     phase diagram, 
      diquark condensation
      &
     \citen{KSHM} \\ \hline
    Plaq. + Wilson& 
    $N_f=2$, $\beta=1.6$,  
    $4^4$, $4^3\times 8$
     &
    thermodynamical quantities,  
    Link-by-Link update 
     & 
    \citen{MNN1}      \\ \hline
     Imp. + Wilson & 
    $N_f=2,3$, $\beta= 0.7$ 
    $4^4$, $4^3\times 8, 12$, 
     &
    behavior of hadrons,  
    Link-by-Link update 
    & 
    \citen{MNN2}      \\ \hline
     Plaq. + KS fermions  & 
     N$_f$=8, $12\times12\times24\times4$,
              $\beta=1.1$,  $m_q$=0.05, 0.07, 0.1
              &
              chiral condensation
      &
    \citen{LMNT}  \\ 
    \hline
    \end{tabular}
	\end{center}
	\label{Tab-su2c}
\end{table}

\subsection{Diquark condensation -- Super fluidity}

\begin{figure}[hbt]
\begin{center}
\begin{minipage}{ 0.48\linewidth}
\includegraphics[width=1.0 \linewidth]{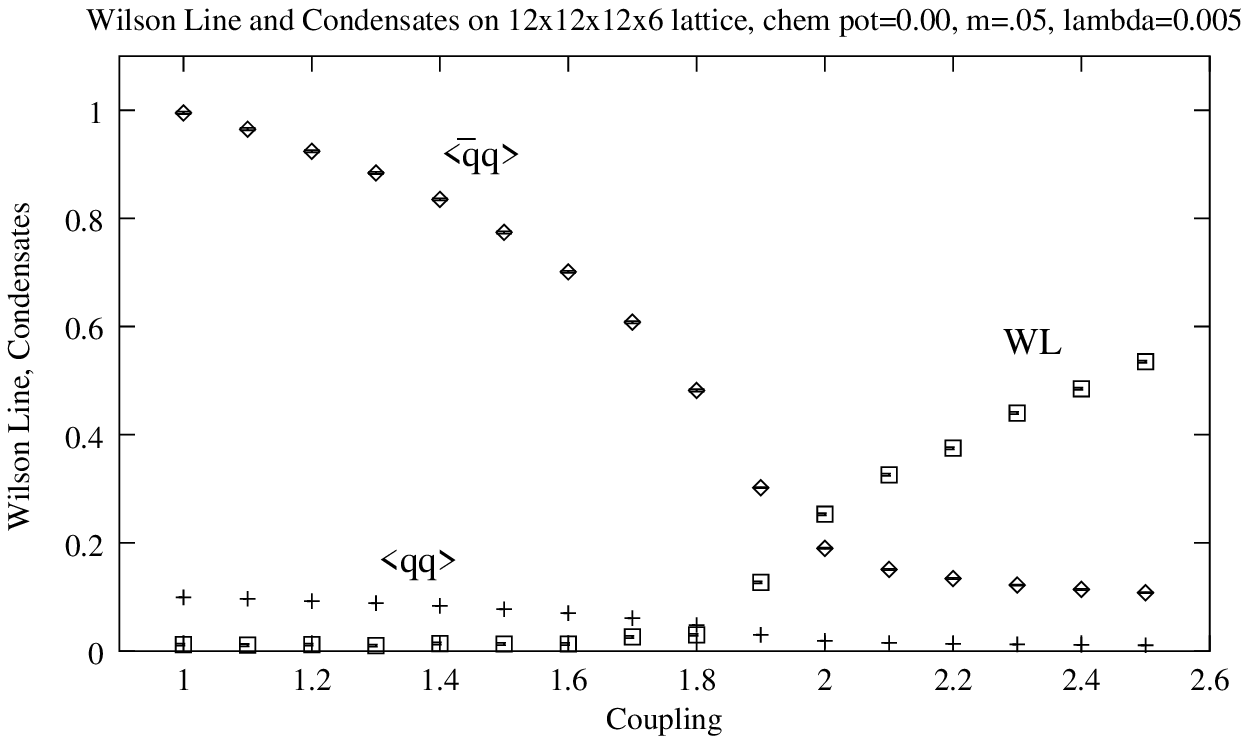}
\end{minipage}
\hspace{1mm}
\begin{minipage}{ 0.48\linewidth}
\includegraphics[width=1.1\linewidth]{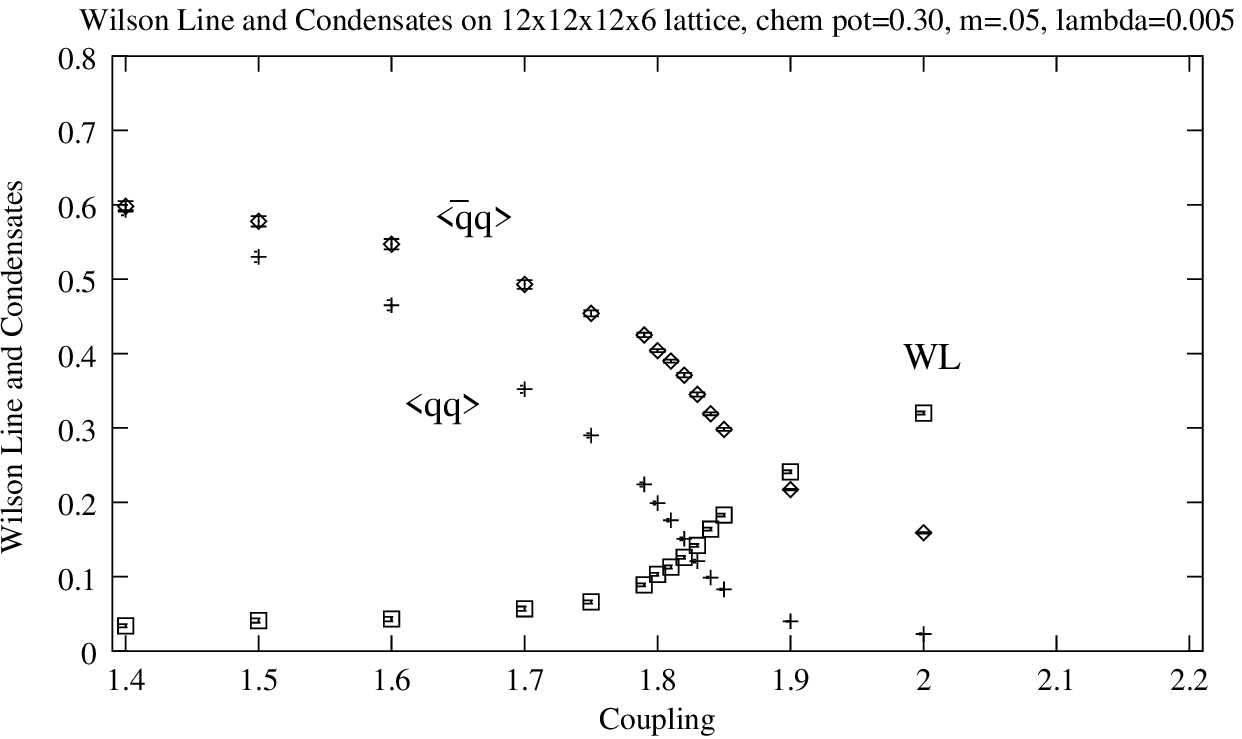}
\end{minipage}
\end{center}
\caption{
Wilson line and condensate vs. $\beta$ calculated by
Kogut, Toublan and Sinclair. Standard gauge + Staggered fermions.
$N_F=4$.
Left : $\mu=0.0$ Right : $\mu=0.3$.
(Ref.~\citen{KTS02})
}
\label{fig4-1}
\end{figure}

In Figs.\ref{fig4-1}, we show the Wilson line, the chiral condensate
$\langle \bar{\chi}\chi \rangle $ and the diquark condensate $\langle \chi\sigma_2\chi \rangle $ obtained
by Kogut, Toublan and Sinclair. \cite{KTS02} 
Here $\chi$ and $\bar{\chi}$ are staggered fermions.
Using the staggered fermions, they simulated a system with a diquark
source,\footnote{Without the source terms, the expectation values of
diquark states, $\langle \chi\Gamma\chi \rangle $, $\langle \bar{\chi}\Gamma\bar{\chi} \rangle $ vanish. }
\begin{equation}
S = \bar{\chi}\Delta\chi 
+ j \bar{\chi}\sigma_2\bar{\chi}^{T} + j^{*} \chi^{T}\sigma_2\chi
+ S_G
= (\chi,\chi^T) 
\left ( \begin{array}{cc} 
j\sigma_2              & \frac{1}{2}\Delta \\
-\frac{1}{2}\Delta^{T} & j^{*}\sigma_2
\end{array} \right )
\left ( \begin{array}{c} \bar{\chi}^{T} \\ \chi \end{array} \right )  
+ S_G
\end{equation}
After integrating out $\bar{\chi}$ and $\chi$, we obtain
\begin{equation}
Z 
 = \int \mD U \sqrt{\det\Delta\Delta^{T}+|j|^{2}} \, e^{-\beta S_G} .
\label{ZwithSource}
\end{equation}
The fermionic measure is always positive.  The source term acts as
an external field in spin systems, and $j$ should be extrapolated to
zero at the end. Due to the square root, one cannot employ Hybrid
Monte Carlo simulation, but the molecular-dynamics type algorithm
is available to generate configurations.

\begin{figure}[hbt]
\begin{center}
\includegraphics[width=.6 \linewidth]{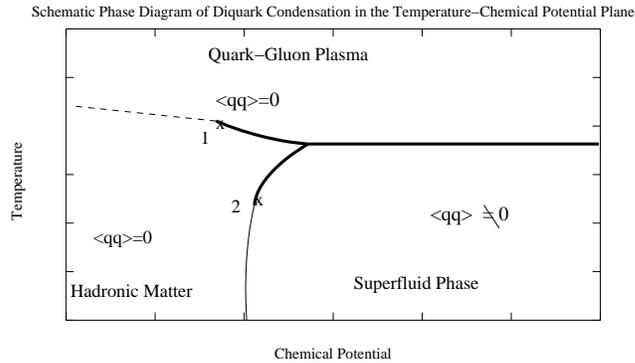}
\end{center}
\caption{ 
Schematic phase diagram in the $(T,\mu)$ plane by 
Kogut, Toublan and Sinclair. 
(Ref.~\citen{KTS02})
}
\label{fig4-2}
\end{figure}

Fig.\ref{fig4-1} (right) indicates the temperature behavior of
the phase at fixed finite $\mu$ with small $j$. When the temperature
increases (the coupling increases), the Wilson line and the chiral
condensate increase, which shows the confinement to deconfinement
transition.  The same behavior is seen at zero chemical potential
shown in Fig.\ref{fig4-1} (left).  A remarkable feature at finite $\mu$
is that $\langle \chi^{T}\sigma_{2}\chi \rangle $ has finite values at low temperature.
Therefore at finite density and low temperature, there is a region in
which the di-quark condensation occurs.
From these observations, they suggest the phase shown in Fig.\ref{fig4-2}.

The operator $\langle \chi^{T}\sigma_{2}\chi \rangle $ is a color singlet 
and cannot bring about color charge.  
Therefore this is not a color super conductivity, but it can
be super fluidity such as liquid $^{3}{\rm H}e$.  
This region belongs to the confinement phase, and this is probably
because this diquark condensation phase is revealed by the color singlet
external source.

\subsection{Hadrons at finite density -- In-medium effect ?}

In-medium hadron features are of great interest in high-energy
phenomenology.
Brown and Rho conjectured that
\begin{equation}
\frac{m^{*}}{m} = \frac{f_{\pi}^{*}}{f_{\pi}}  ,
\end{equation}
which is now known as Brown Rho scaling,
to explain the large low mass lepton pair enhancement
observed in CERES \cite{CERES}.
Here $m^*$ and $f_\pi^*$
are the mass and the pion decay constant in the medium \cite{BR-theo}.
Based on the QCD sum rule, Hatsuda and Lee predicted a decrease of
$\rho$ meson mass as a function of $\mu$ \cite{H-L}.
Harada et al. showed that the vector meson mass vanishes at the
critical density as a consequence of an effective theory
with hidden local symmetry \cite{Harada}.
Yokokawa et al. proposed simultaneous softening of $\sigma$ and
$\rho$ mesons associated with the chiral restoration
\cite{Yokokawa}.
If lattice data show such a phenomenon, it is a
new feature of QCD at finite density, and will become an
interesting experimental target in future heavy ion experiments.

\begin{figure}[hbt]
\begin{center}
\begin{minipage}{ 0.48\linewidth}
\includegraphics[width=1.0 \linewidth]{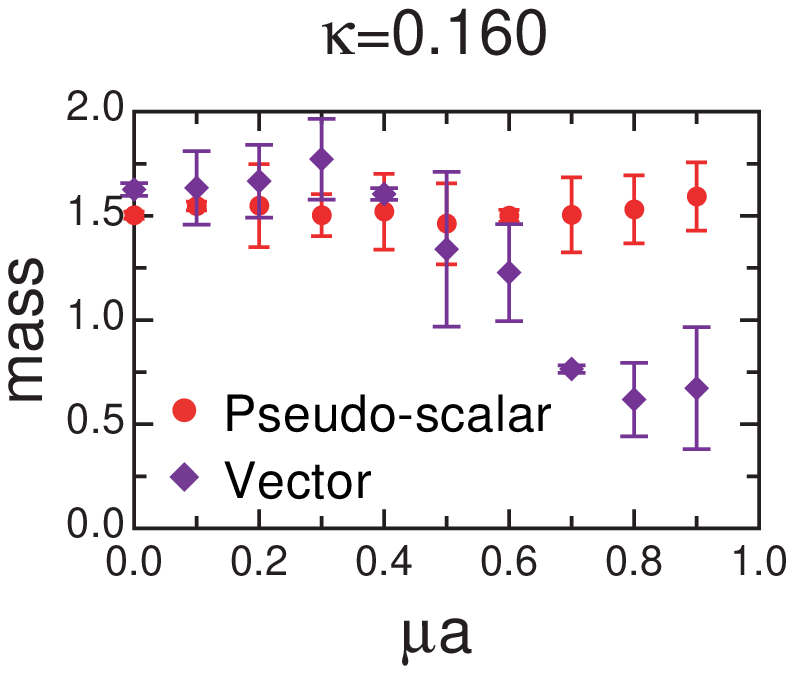}
\end{minipage}
\hspace{1mm}
\begin{minipage}{ 0.48\linewidth}
\includegraphics[width=1.1\linewidth]{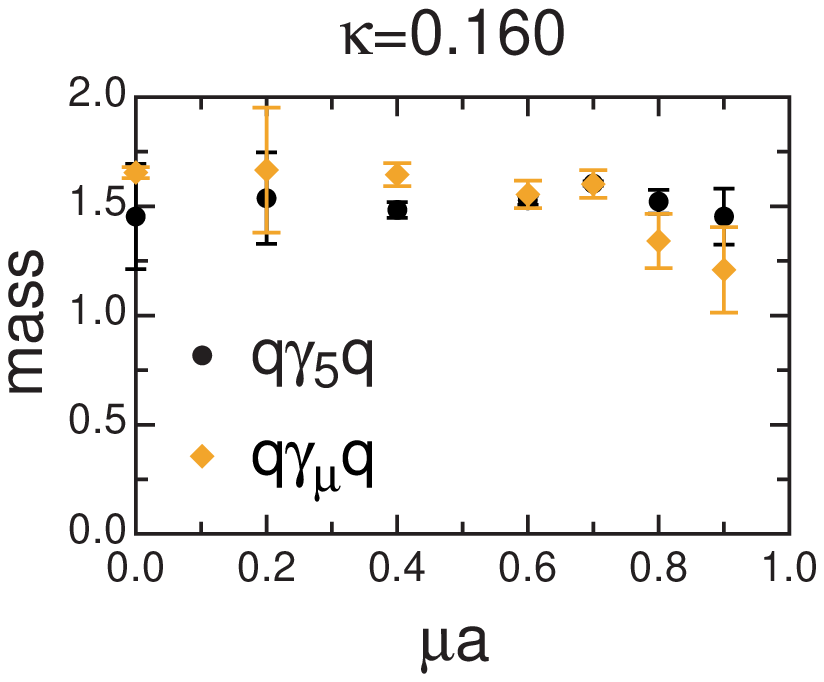}
\end{minipage}
\end{center}
\caption{
Pseudo scalar and vector meson masses (left) and
corresponding di-quark (baryon) masses (right).
(Ref.~\citen{MNN03}).
}
\label{fig-MNN03}
\end{figure}

Recently, Muroya, Nakamura and Nonaka reported a strange behavior
of the propagator of a vector meson channel with chemical 
potential \cite{MNN03}. They investigated finite density state of
color SU(2) QCD with Wilson fermion. As for the gauge part of
the lattice action, both plaquette and Iwasaki improved actions 
are adopted.
In order to make a reliable update, the ratio of fermion determinants 
is calculated exactly through the Woodbury formula.  As a result,
the lattice is limited to a small size. 
They investigated the propagators of mesons and diquarks; 
pseudo-scalar ($\pi$), scalar ($a_0$), vector ($\rho$)
mesons, and scalar diquark ($q\gamma_5 q$), pseudo-scalar 
diquark ($q1q$) and axial-vector diquark ($q\gamma_\mu q$) are
studied. 
At $\mu=0$, the propagators of the corresponding 
meson and diquark are degenerate with each other.  
With finite $\mu$, the propagators change their shapes.  
It is quite different from the finite temperature case, the 
propagator of the 
pseudo-scalar meson changes only slightly and 
the vector meson becomes lighter.  The propagators of the 
corresponding diquarks show clear asymmetry in the chronological
direction due to the charge corresponding to the chemical potential.  
However,
the behavior of the obtained mass is very mild.
Only the vector meson mass shows a marked change
around $\mu a=0.7$.  This behavior might be the expected result of 
the phenomenological models.
Figs.\ref{fig-MNN03} are typical behaviors of masses of the pseudo-scalar,
vector mesons, and di-quarks as a function of $mu$. 

The propagators of the scalar meson and pseudo scalar diquark
decay so quickly that it is very difficult to catch signal against 
noise.

Another possible explanation for the lightening of
the mass of the mesonic channel is a particle-hole pair 
around a Fermi surface.  Hands, Kogut, Strouthos and Tran
investigated the (2+1)-dimensional Gross-Neveu mode and discussed
the effect of the Fermi surface.  They evaluate the mesonic propagator
in the high density state with a Fermi surface.  In the mesonic
channel, in addition to the ordinary quark-anti-quark pair which 
contributes as a massive meson,  quark-hole gapless pair around 
Fermi surface may exist.  
In Ref.\cite{HaKoStrTr}, authors investigated propagators
with fixed momenta and found that at high density, 
a mesonic propagator whose momentum is up to some value 
shows massless behavior.  
On the other hand, propagators with larger 
momenta behave massive.  
The existence of such a particle-hole
pair is well known in the many-body theory and from direct results of the Fermi 
surface. 
In Ref.\cite{HaKoStrTr}, a momentum dependent massless mode appears in 
the scalar channel and vector channel, 
but, in Ref.~\citen{MNN03} it appears in the vector
channel only. In order to clarify the Fermi surface effect, 
we need larger lattice simulation and calculation of the $\sigma$
channel which also contains a disconnected diagram.

\section{Related works}\label{sec-related}

\subsection{Gross-Neveu model - Simple model for test}

QCD in four dimensions is highly nontrivial.
A simple model may help us to understand the theory, and
may give us a hint toward determining  a correct route.
The Gross-Neveu model has been studied for this purpose.

Let us start from a model in two dimensions
with Nambu-Jona-Lasinio type four fermion
interaction which has a `flavor' degree of freedom $N$.
The  action is given by
\begin{equation}
 S = \sum_{k=1}^{N} \left[ \bar{\psi}^{(k)} ( 
    \gamma_\mu\partial_\mu + m ) \psi^{(k)}
  - \frac{g^2}{2N} \{ (\bar{\psi}^{(k)}\psi^{(k)} )^2
                + (\bar{\psi}^{(k)} i\gamma_5 \psi^{(k)} )^2 \}
      \right],
\label{GN-0}
\end{equation}
where the Dirac matrices are
\begin{equation}
  \gamma_1 = \sigma_2,   \gamma_2 = \sigma_1, 
  \gamma_5 = i \gamma_1 \gamma_2 = \sigma_3.
\end{equation}

When its mass $m$ vanishes, Eq.(\ref{GN-0}) possesses the chiral  
symmetry,
\begin{equation}
 \psi \rightarrow e^{i\gamma_5\theta} \psi,
 \bar{\psi} \rightarrow \bar{\psi} e^{i\gamma_5\theta} .
\label{GN-sym1}
\end{equation}
We introduce two auxiliary fields, $\sigma(x)$  and $\pi(x)$, which
correspond to 
$-(g^2/N) \bar{\psi}\psi$ and $-(g^2/N) \bar{\psi}i\gamma_5\psi$,
respectively, in the standard manner, 
\begin{eqnarray}
 S &=&  \bar{\psi} (
      \gamma_\mu\partial_\mu + m ) \psi
   -  \frac{g^2}{2N} \{ (\bar{\psi}\psi )^2
                + (\bar{\psi} i\gamma_5 \psi )^2 \}
\nonumber \\
&& 
  + \frac{N}{2g^2} (\sigma + \frac{g^2}{N}\bar{\psi}\psi )^2
  + \frac{N}{2g^2} (\pi    + \frac{g^2}{N}\bar{\psi} i\gamma_5 \psi )^2 ,
\nonumber \\
   &=& \bar{\psi} (
    \gamma_\mu\partial_\mu + m + \sigma +  i\gamma_5\pi) \psi
    + \frac{N}{2g^2} (\sigma^2 + \pi^2).
\label{GN-1}
\end{eqnarray}
Here we suppress the flavor index $k$.
It is easy to verify that Eq.(\ref{GN-1}) is invariant under the
following rotation together with Eq.(\ref{GN-sym1})
in a massless limit,
\begin{equation}
\left(
\begin{array}{c}
\sigma \\
\pi    
\end{array}
\right)
\rightarrow
\left(
\begin{array}{cc}
\cos 2\theta & \sin 2\theta \\
-\sin 2\theta & \cos 2\theta 
\end{array}
\right)
\left(
\begin{array}{c}
\sigma \\
\pi    
\end{array}
\right)
\label{GN-sym2}
\end{equation}

In a large $N$ limit, the fields $\sigma$ and $\pi$ can be
replaced by their mean fields, and we may rotate them so that
$\pi=0$.
Then we have a model with only the first four-fermion interaction
term in Eq.(\ref{GN-0}), or
\begin{equation}
   S = \bar{\psi} (
    \gamma_\mu\partial_\mu + \sigma ) \psi
    + \frac{N}{2g^2} \sigma^2 . 
\label{GN-2}
\end{equation}

Although the model has no confinement,
it has the following features common to QCD:
\begin{itemize}
\item
it is an asymptotically free theory,
\item
it shows spontaneous breakdown of a (discrete) chiral symmetry,
i.e., $\psi \rightarrow \gamma_5\psi$.
\end{itemize}
The Gross-Neveu model is well studied in the continuum,
\begin{itemize}
\item
the phase structure at finite temperature and density
is obtained in $N\rightarrow \infty$ limit by a mean field
approach.
\item
Finite $N$ case can be treated by the factorized S-matrix method \cite{ZaZa79}.
\item
There is a conjecture that a generalization of the Gross-Neveu
model from $U(N)$ to $O(N)$ flavor symmetry leads to the
appearance of a pairing condensate at high density. \cite{Chodos}
\end{itemize}

\begin{figure}[hbt]
\begin{center}
\begin{minipage}{ 0.48\linewidth}
\includegraphics[width=1.0 \linewidth]{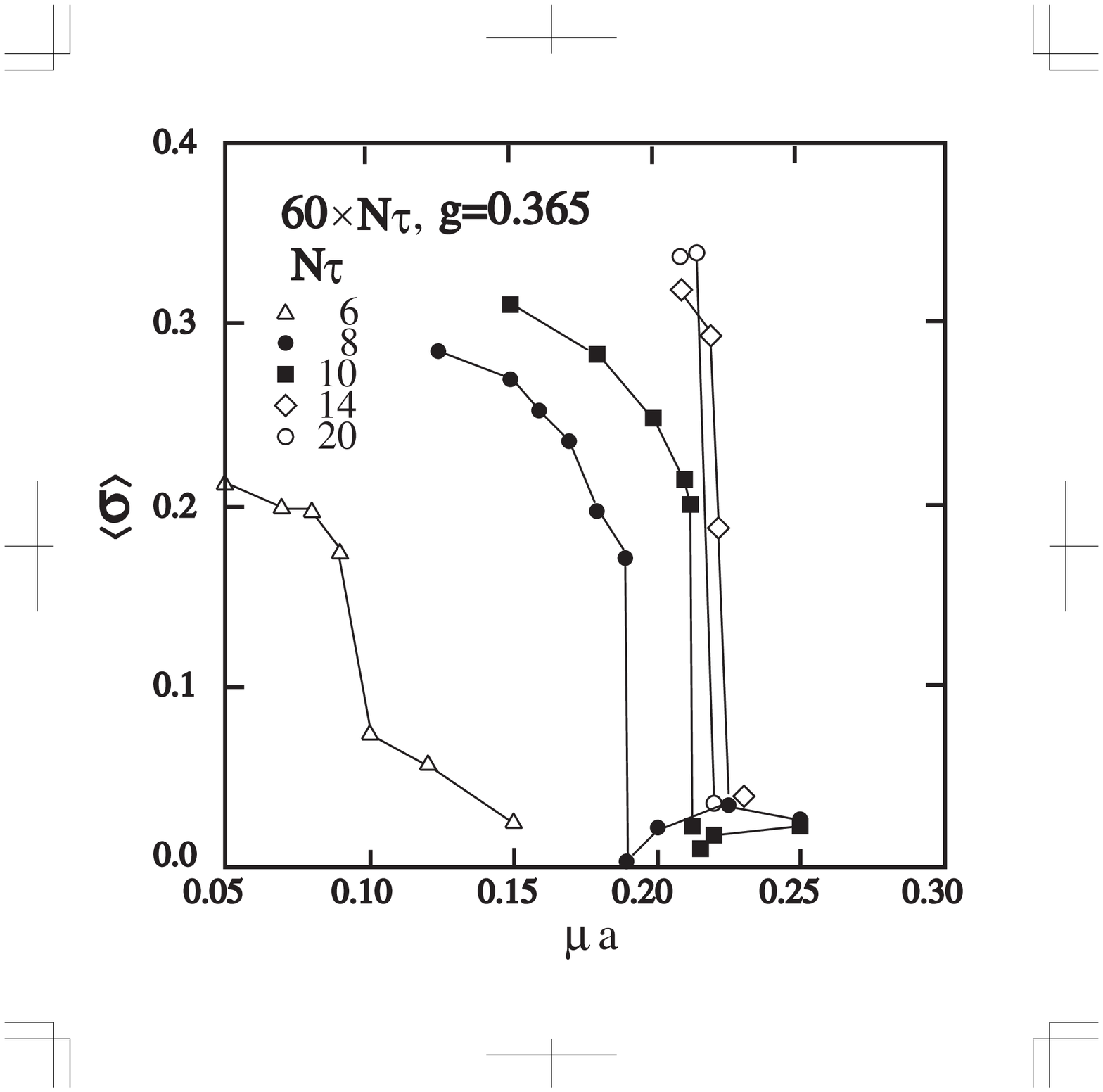}
\end{minipage}
\hspace{1mm}
\begin{minipage}{ 0.48\linewidth}
\includegraphics[width=1.1\linewidth]{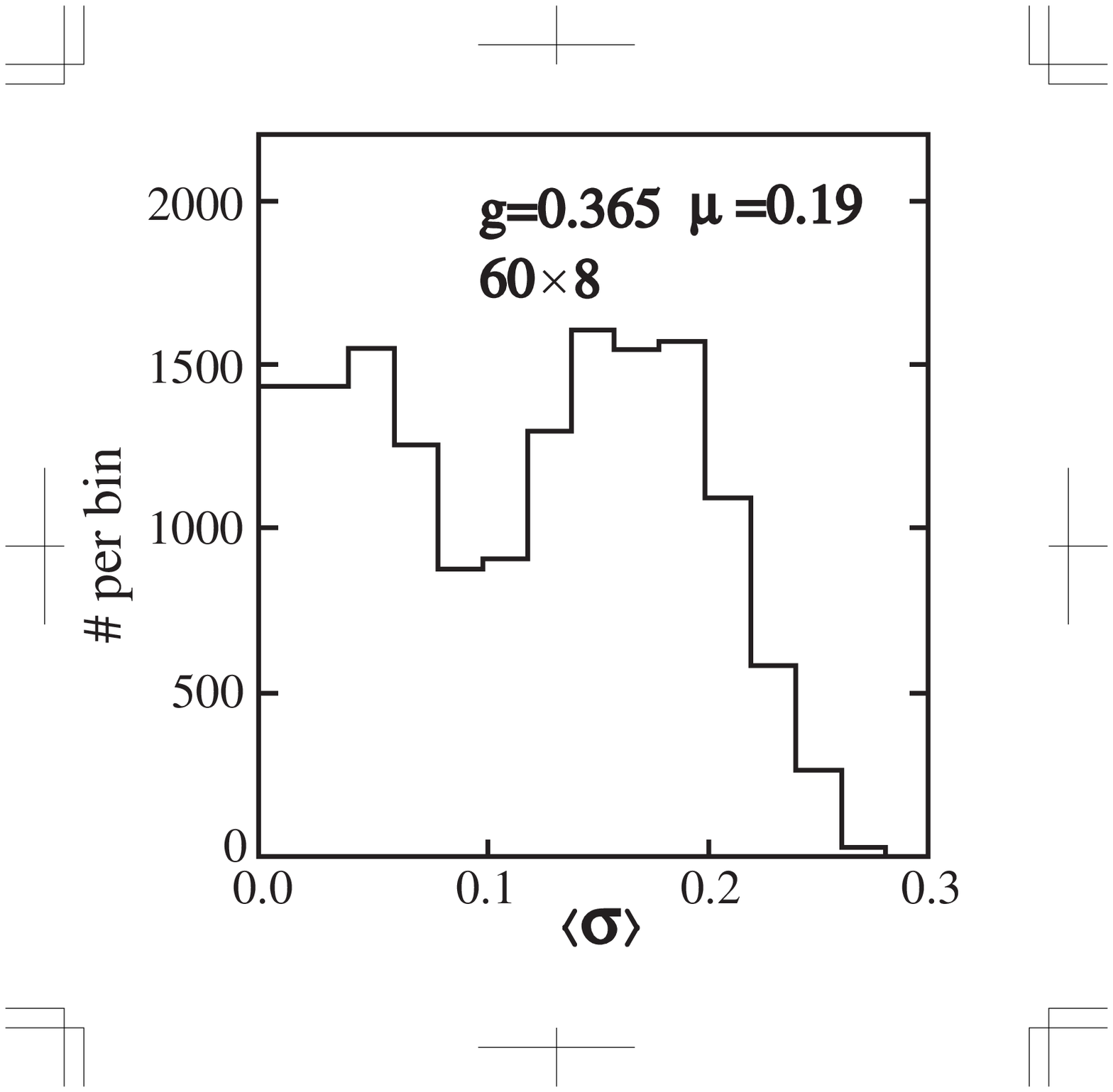}
\end{minipage}
\end{center}
\caption{
Left : $\langle\sigma\rangle$ as a function of $\mu$ on lattices of 
size $60\times N_\tau$,$N_\tau=6, 8, 10, 14, 20$. 
Right : Coexisting states at $\mu_c$.  Ref.\citen{GN-lat1}.
}
\label{fig5-1}
\end{figure}

There are two lattice studies of the Gross-Neveu model in the
literature. 
Karsch, Kogut and Wyld analysed the staggered fermion version
of Eq.(\ref{GN-1}) \cite{GN-lat1},
\begin{equation}
S = \sum_{k=1}^{N} \bar{\chi}^{k}\Delta\chi^{k} + \frac{1}{2g^2}\sigma^2,
\end{equation}
where the fermion matrix is given by
\begin{equation}
\Delta_{ij} = \frac{1}{4} (\sigma_j + \sigma_{j-\hat{0}}
   +  \sigma_{j-\hat{1}} + \sigma_{j-\hat{0}-\hat{1}} ) \delta_{ij}
+ \frac{1}{2} (e^\mu \delta_{j,i+\hat{0}} 
             - e^{-\mu}\delta_{j,i-\hat{0}} )
+  \frac{1}{2} (-1)^t
( \delta_{j,i+\hat{1}} - \delta_{j,i-\hat{1}} )
\end{equation}
All elements of the fermion matrix are real, therefore
its determinant is also real.
In Ref.~\citen{GN-lat1}, the Langevin algorithm was used to simulate
the system with $N=12$.  
Fig.\ref{fig5-1} shows typical simulation results.

Izubuchi, Noaki and Ukawa investigated a lattice version of
Eq.\ref{GN-0} using Wilson fermions at finite temperature and
density, \cite{GN-lat2}
\begin{eqnarray}
S = &-& \frac{1}{2a}[ \bar{\psi}(x) (r-\gamma_1) \psi(x+\hat{1})
   \bar{\psi}(x) (r+\gamma_1) \psi(x-\hat{1}) ]
\nonumber \\
    &-& \frac{1}{2a}[ \bar{\psi}(x) (r-\gamma_2) e^{-\mu a}\psi(x+\hat{2})
   \bar{\psi}(x) (r+\gamma_1) e^{\mu a}\psi(x-\hat{1}) ]
\nonumber \\
   &+& \frac{1}{a} (2r+\delta m) \bar{\psi}(x)\psi(x) 
   - \frac{g_\sigma^2}{2N}(\bar{\psi}(x)\psi(x))^2
   - \frac{g_\pi^2}{2N}(\bar{\psi}(x)i\gamma_5\psi(x))^2 .
\end{eqnarray}
Here they introduce two couplings $g_\sigma$ and $g_\pi$ to
controle the continuum limit.
The lattice effective potential is constructed with two auxiliary fields,
$\sigma$ and $\pi$ as in the continuum, 
\begin{eqnarray}
V_L = -\left( \frac{\delta m_L}{g_\sigma^2} + 2C_1 \right)\sigma_L
     + \left( \frac{1}{g_\pi^2} - 2C_0 \right)\pi_L^2
     + \left( \frac{1}{g_\sigma^2} - C_0 + 2C_2 \right)\sigma_L^2
\nonumber \\
     + \frac{1}{4\pi}(\sigma_L^2 + \pi_L^2) 
      \log\frac{\sigma_L^2 + \pi_L^2}{e} + O(a^3) ,
\end{eqnarray}
where $C_0$, $C_1$ and $C_2$ are constant. 
In order to get the correct chiral symmetric theory in the continuum,
we tune three parameters, $g_\pi$, $g_\sigma$ and $\delta m_L$
so that
\begin{eqnarray}
\frac{1}{g_\pi^2} = \frac{1}{g_\sigma^2} + 4C_2 + O(a) ,
\nonumber \\
\frac{\delta m_L}{g_\sigma^2} = -2C_1 + O(a^2) .
\end{eqnarray}

\subsection{NJL model}
At present, there are no reliable methods for evaluating
low temperature and
high density regions in nonperturbative QCD calculation.
Therefore effective
theories such as Nambu-Jona-Lashino
(NJL) model play an important role
and the comparison of the results of
lattice QCD for a simple case to
the results of effective theories is a meaningful approach in order to
understand the superconducting behavior at high density.
Hands and Walters apply the lattice analyses to
the NJL model in (3+1)-dimension, focusing on BCS diquark
condensation.\cite{njl-3d}
\footnote{The (2+1)-dimensional lattice NJL model at finite density
has been simulated, but BCS condensation is not observed.\cite{njl-2d}}
The action of this model is written by
\begin{equation}
S = \psi^{\rm{tr}} {\cal{A}} \psi + \frac{2}{g^2} \sum_{\tilde{x}}(\sigma^2 +
\vec{\pi}\cdot \vec{\pi}),
\end{equation}
where the bispinor $\psi$ is written in terms of staggered isospinor
fermions fields, $\psi^{\rm{tr}}=(\bar{\chi}, \chi^{\rm {tr}})$
 and $\sigma$ and the triplet $\vec{\pi}$ are real auxiliary fields defined on
 the dual site $x$. In the fermion matrix ${\cal{A}}$ they introduce the
 diquark source term $j$,
 \begin{equation}
 {\cal{A}}= \frac{1}{2} \left (
 \begin{array}{cc}
 \bar{j}\tau_2 &  M  \\
 M^{\rm {tr}} & j \tau_2
 \end{array}
 \right ),
 \end{equation}
 where the matrix $M$ is defined by
 \begin{eqnarray}
 M^{pq}_{xy}[\sigma, \vec{\pi}] & = &  \frac{1}{2} \delta^{pq}
 \left [
 ( {\rm e}^\mu \delta_{yx+\hat0}-{\rm e}^{-\mu} \delta_{yx-\hat0}) \right .
 \nonumber \\
  &  & + \sum_{\nu = 1,2} \eta_\nu(x)(\delta_{yx+\hat{\nu}}
  -\delta_{yx-\hat{\nu}}) + 2m_0 \delta_{xy}
 \left . \right ] \nonumber \\
  & & + \frac{1}{16} \delta{xy}\sum_{\langle \tilde{x}, x \rangle}
  ( \sigma (\tilde{x}) \delta^{pq}+ i \epsilon (x)
  \vec{\pi}(\tilde{x}) \cdot \vec{\tau}^{pq}).
 \end{eqnarray}
 Here $m_0$ is the bare mass and $\eta_\nu(x)$, $\epsilon (x)$ are the
 phases $(-1)^{x_0+\cdots x_{\nu-1}}$ and $(-1)^{x_0+x_1+x_2+x_3}$,
 respectively. The $\vec{\tau}$ are Pauli matrices and
 $\langle \tilde{x},x \rangle$ represents the set of 16 dual lattice sites
 neighboring $x$.
\begin{figure}[hbt]
\begin{center}
\begin{minipage}{ 0.48\linewidth}
\includegraphics[width=1.0 \linewidth]{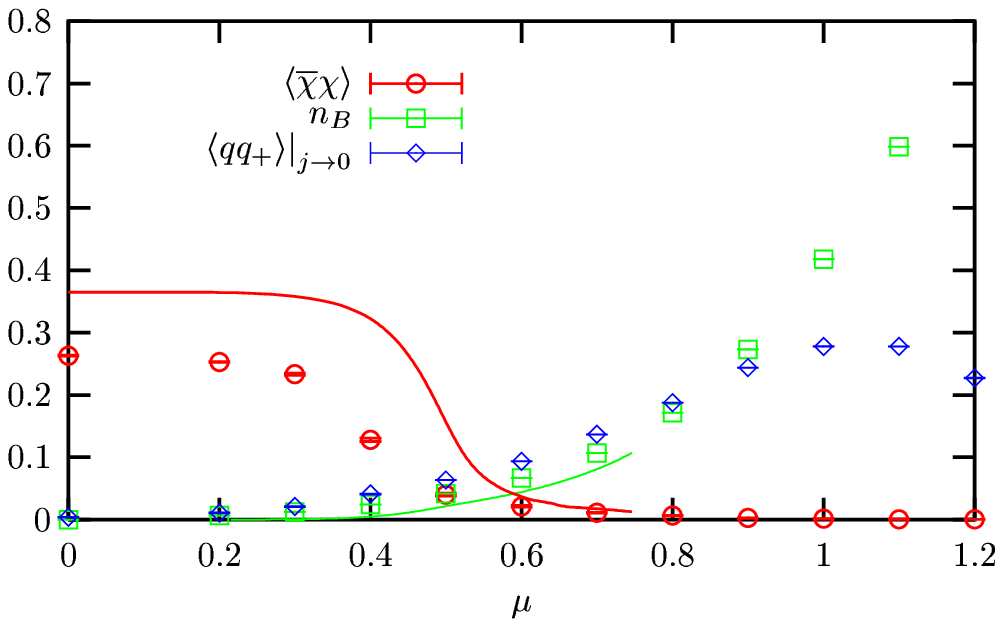}
\end{minipage}
\hspace{1mm}
\begin{minipage}{ 0.48\linewidth}
\includegraphics[width=1.1\linewidth]{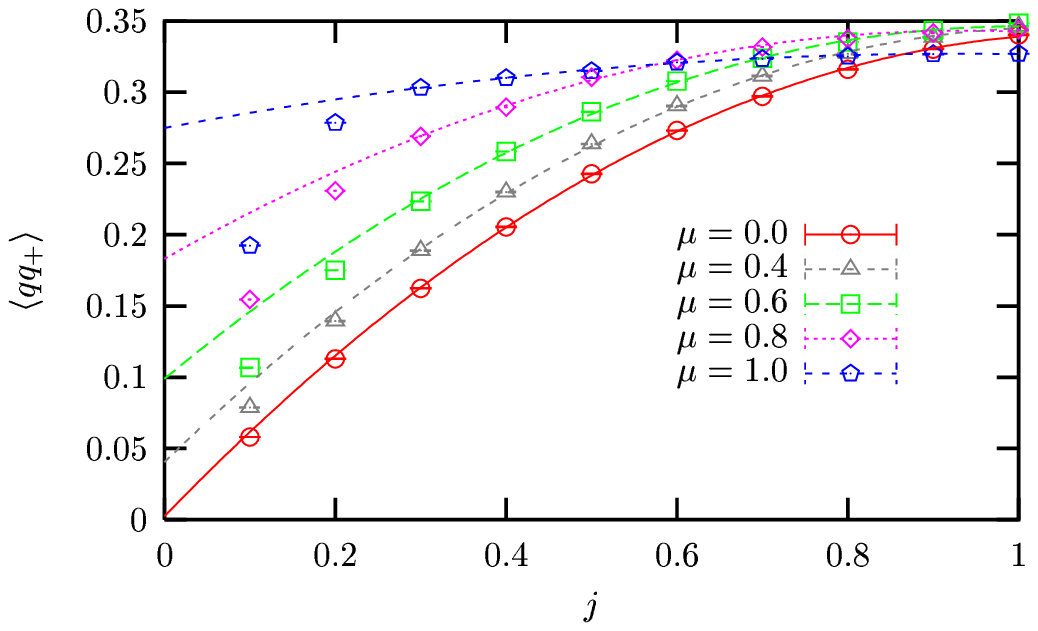}
\end{minipage}
\end{center}
\caption{The diquark condensate, $\langle q q \rangle$
as a function of $\mu$(left).
$\langle q q \rangle$ as a function of $j$ (right).
}
\label{fig5-1.1}
\end{figure}
They measure diquark condensate $\langle qq \rangle$
as a function of chemical potential, extrapolating to $j \rightarrow 0$.
From this analysis, they determine the existence of a nonzero BCS condensate
at high baryon density (Fig. \ref{fig5-1.1} (left)) .
However, the behavior of $\langle qq \rangle$ as a function of $\mu$
depends on the evaluation of the data at $j \rightarrow 0 $.
They ignore the values of $\langle qq \rangle$ at small $j$,
because at high $\mu$,  $\langle qq \rangle$ at small $j$ falls rapidly
away from the tendency of $\langle qq \rangle$ at large $j$
(Fig. \ref{fig5-1.1}(right)).
They discuss that the low-$j$ discrepancy comes from the finite volume
effect and the partially quenched approximation and
some issues concerning thermodynamic limit need to be
resolved for a definite conclusion.

\subsection{Strong coupling calculation}

It is comfortable if we have any analytical or semi-analytical
method to study QCD at finite baryon density in non-perturbative
region.
Strong coupling approximation of lattice QCD is one of 
such possibilities. 
Indeed the lattice formulation was introduced to allow
the strong coupling expansion and the confinement was
shown there.
We can also compare our numerical data to the analytical
calculations and take a deep view of lattice results. 

When the gauge coupling $g$ is very large, $\beta S_G = 6/g^2 S_G$
may drop,
\begin{equation}
Z  = \int \mD U \mD\psibar \mD\psi
    \, e^{- \psibar \Delta \psi} .
\label{StrongCouplingEq}
\end{equation}
In this case, the integrations over time like link variables, $U_4(x)$,
and space-like ones, $U_i(x)$, are decouples, and one can integrate 
over the gauge fields. 

The first strong coupling calculation was performed by Ilgenfritz
and Kripfganz \cite{IK} and Damgaard, Hochberg and Kawamoto, \cite{DHK}
together with the mean field approximation.
Bili\'c, Demeterfi and Petersson extended the calculation to the next
order of $1/g^2$. \cite{BDP}
Recently Nishida, Fukushima and Hatsuda applied the strong coupling
method to two-color QCD to study the dynamics of the theory. \cite{NFH}

For warming-up, let us first study a very simple case in a pedagogical
way:
\begin{equation}
Z = \int dU d\bar{\psi_1}d\psi_1 d\bar{\psi_2}d\psi_2 
e^{\bar{\psi}_1 U \psi_2 - \bar{\psi}_2 U^{\dagger} \psi_1},
\end{equation}
i.e., two fermion fields, $\psi_1^a, \psi_2^a$ are connected
by a link filed $U$, where $a=1,2,\cdots, N_c$ are color index.
Series expansion of the exponential is finie due to the nil
potency of the Grassmann variable. 
With the help of group integration properties,
\begin{eqnarray}
\int dU &=& 1 \nonumber \\
\int dU U_{a,b} {U^{\dagger}}_{c,d} = \frac{1}{3}\delta_{a,d} \delta_{b,c} 
=
\mbox{
\includegraphics[width=2.2cm,height=2.0cm,keepaspectratio]{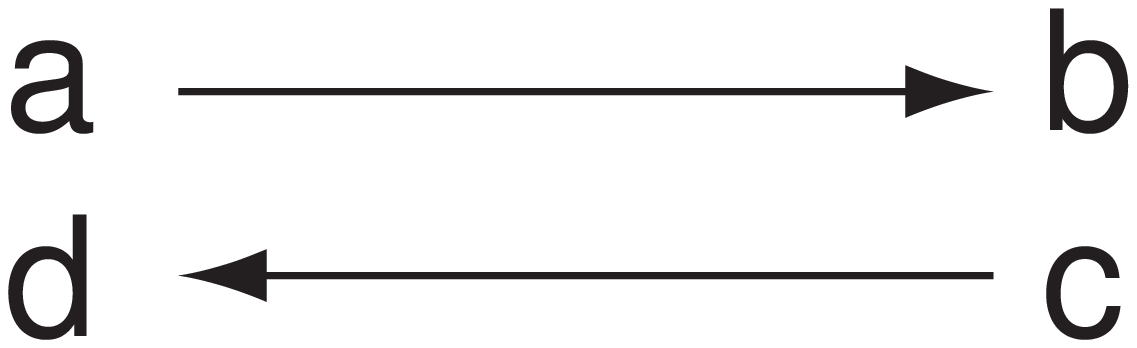}
}
\nonumber \\
\int dU U_{a_1,b_1} U_{a_2,b_2} U_{a_3,b_3} 
&=& \frac{1}{3!} \epsilon_{a_1 a_2 a_3} \epsilon_{b_1 b_2 b_3}
=
\mbox{
\includegraphics[width=2.2cm,height=2.0cm,keepaspectratio]{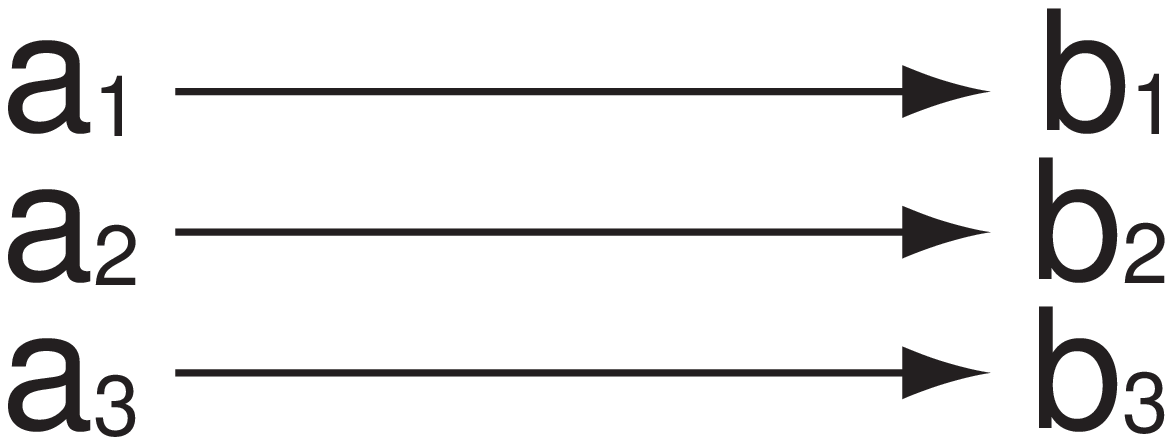}
}
\nonumber \\
\int dU {U^{\dagger}}_{a_1,b_1} {U^{\dagger}}_{a_2,b_2} {U^{\dagger}}_{a_3,b_3} 
&=& \frac{1}{3!} \epsilon_{a_1 a_2 a_3} \epsilon_{b_1 b_2 b_3}
\end{eqnarray}
where $U_{a,b}$ is $(a,b)$ element of the matrix $U$,
we obtain
\begin{eqnarray}
Z &=& 1 + \frac{1}{3}M_1 M_2 + (-\bar{B}_1 B_2 + \bar{B}_2 B_1) + \cdots
\nonumber \\
&\sim& e^{\frac{1}{3}M_1 M_2 + (-\bar{B}_1 B_2 + \bar{B}_2 B_1)},
\end{eqnarray}
where
\begin{eqnarray}
M_i &\equiv& {\bar{\psi}_i}^a {\psi_i}^a ,
\nonumber \\
B_i &\equiv& \frac{1}{3!} \epsilon_{a_1 a_2 a_3}
 {\psi_i}^{a_1}{\psi_i}^{a_2}{\psi_i}^{a_3} .
\end{eqnarray}
More systematic and precise formulae are given in Ref.\cite{RossiWolff84}. 
Using the identities with auxiliary bosonic field $\phi$ and fermionic field
$b$,
\begin{eqnarray}
e^{\frac{1}{2} ^t{M}\mathcal{V}^{-1}{M}} 
= \int d\phi_1 d\phi_2 \, e^{-\frac{1}{2}^t{\phi}\mathcal{V}{\phi}
   - ^t{\phi}{M}},
\\
e^{\bar{B}\mathcal{W}^{-1}B} =  
\int d\bar{b} db  \, e^{-\bar{b}\mathcal{W}b + \bar{b}B + \bar{B}b},
\end{eqnarray}
we linearize $M$ and $B$ fields,
where
$M = ^t ( M_1 M_2 )$.

Now we come back to the realistic case.
We integrate first the spatial link variables for the staggered fermions,
\begin{eqnarray}
Z = \int \mD U_4 \mD \bar{\chi} \mD\chi
\exp\left\{-\sum_{x,y} \bar{\chi}(x) \left[ m\delta_{x,y} 
+ \frac{1}{2r}e^{\mu r} U_4(x)\delta_{x,y-\hat{4}}
\right. 
\right. 
\nonumber \\
\left.  
- \frac{1}{2r}e^{-\mu r} U_4(x)^{\dagger}\delta_{x,y+\hat{4}}
\right] \chi(y)
- \frac{1}{4N_c}\sum_{\langle x,y \rangle} 
	\bar{\chi}(x)\chi(x)\bar{\chi}(y)\chi(y)
\nonumber \\
\left.  
- \frac{1}{32N_c^2 (N_c -1)}\sum_{\langle x,y \rangle} 
	[\bar{\chi}(x)\chi(x)]^2 [\bar{\chi}(y)\chi(y)]^2
\right\}
\end{eqnarray}
where an anisotropy parameter of time-like and space-like
lattice spacing, $r\equiv a_t/a_s$, is introduced to controle
the temperature.
Introducing auxiliary fields, $\lambda$, we can linearize the exponential,
\begin{eqnarray}
Z = \int \mD U_4 \mD \bar{\chi} \mD\chi \mD \lambda
 \exp\left\{-\frac{N_c}{d} \sum_{x,y} 
 (\lambda(x)-m)V^{-1}(x,y) (\lambda(y)-m)
\right .
\nonumber \\
\left .
- \sum_{x,y} \bar{\chi}(x) \left[ \lambda(x) \delta_{x,y}
 + \frac{1}{2r}e^{\mu r} U_4(x)\delta_{x,y-\hat{4}}
- \frac{1}{2r}e^{-\mu r} U_4(x)^{\dagger}\delta_{x,y+\hat{4}}
\right] \chi(y)
\right\}
\end{eqnarray}
where $d$ is the spatial dimension ($=3$), and
the matrix $V(x,y)$ is defined by
\begin{equation}
V(x,y) = \frac{1}{2d}
\sum_{k=1}^{d} (\delta_{y,x+\hat{k}} + \delta_{y,x-\hat{k}} )
\end{equation}

The integration over $\bar{\chi}$ and $\chi$ can be performed.
Then we can further integrate over $U_4$.  When $\lambda$ is constant
(mean field approximation), then
\begin{equation}
Z = 2 \cosh(N_t N_c \mu r) 
+ \frac{\sinh(N_c+1)N_t\lambda'}{\sinh(N_t\lambda')}
\end{equation}
where $\lambda'=\sinh^{-1}[(\lambda+m)r]$.
$\lambda$ is determined by minimizing the free energy 
$F=N_t N_c \lambda^2/d - \log Z$, 
i.e., $\partial F/\partial \lambda = 0$.
The chiral condensation is given by the $\lambda$ by
$\langle\bar{\chi}\chi\rangle = 2N_c\lambda/d$.
In the infinite coupling limit, the transition value of the chemical
potential is given as
\begin{equation}
\mu_0 = \frac{1}{r} \sinh^{-1}(\lambda_0 r) 
- \frac{\lambda^2_0}{dr},
\end{equation}
where 
\begin{equation}
\lambda_0 = \frac{1}{r\sqrt{2}} (\sqrt{1+d^2 r^4} - 1)^{1/2}.
\end{equation}
The transition point with $1/g^2$ correction can be found in Ref.\cite{BDP}.

\begin{figure}[hbt]
\begin{center}
\begin{minipage}{ 0.45\linewidth}
\includegraphics[width=1.2 \linewidth]{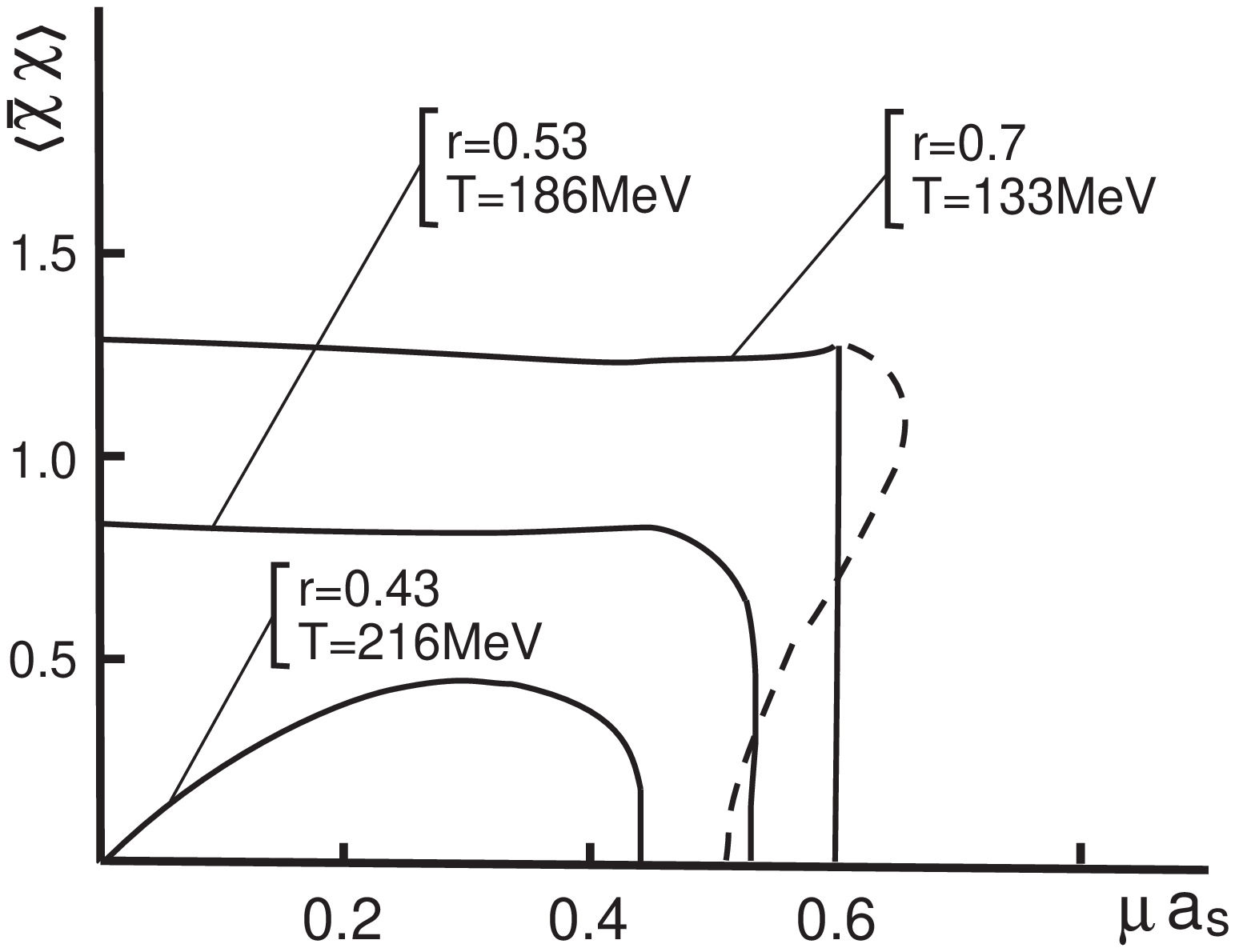}
\end{minipage}
\hspace{1mm}
\begin{minipage}{ 0.45\linewidth}
\includegraphics[width=1.2 \linewidth]{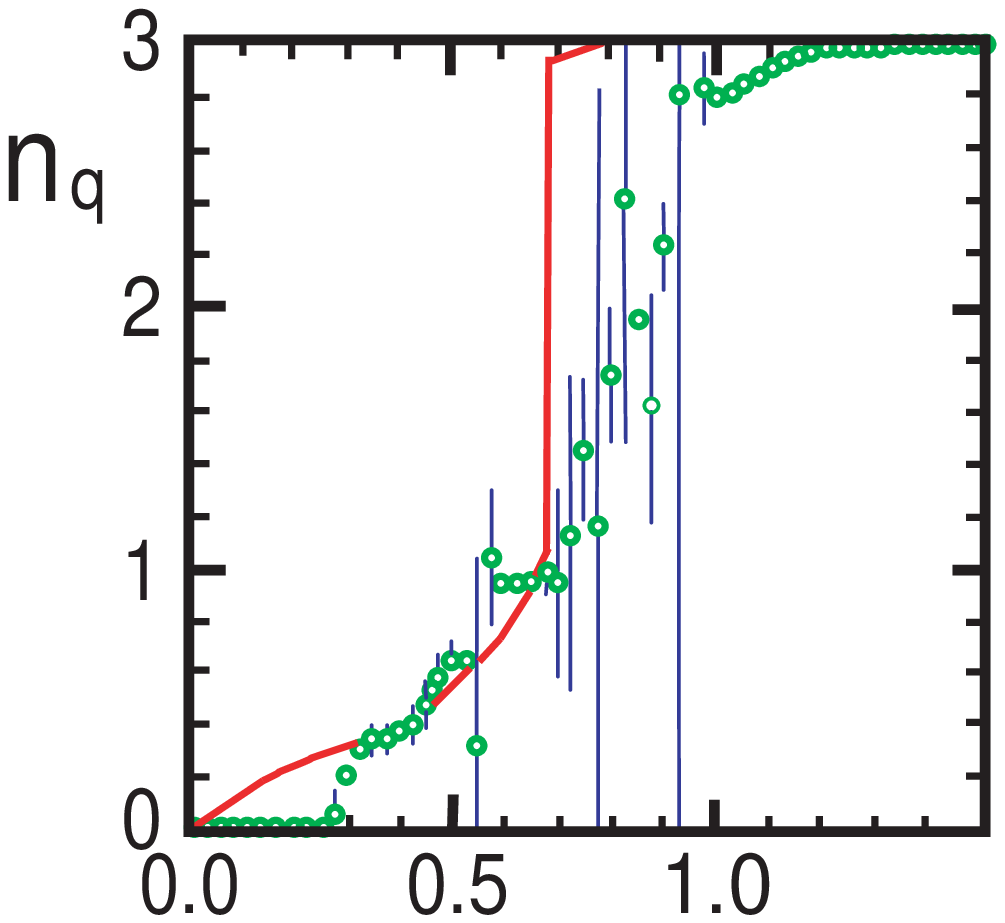}
\end{minipage}
\end{center}
\caption{
Strong coupling calculations
by  Bili\'c, Demeterfi and Petersson
(Ref.~\citen{BDP}).
Chiral condensate as a function of $\mu$ for various 
temperatures at $6/g^2=2$.
The dashed line represents the instability region. (left) 
Comparison of the strong coupling calculation of the quark
number density with numerical data for $N_t=4$ and $6/g^2=4.8$. 
The points are data of Ref.\cite{Hasen-To}.
(right)
}
\label{fig5-2x}
\end{figure}

In Fig.\ref{fig5-2x}, we show the chiral condensate as a function
of the chemical potential for various temperatures at $6/g^2=2$
taken from Ref.\cite{BDP} where the $1/g^2$ correction is included.
They calculated also the quark number density as a function of the chemical
potential and found good agreement with the numerical calculation given
by Hasenfratz and Toussaint \cite{Hasen-To}.

Karsch and M\"{u}tter have given an elegant form, `monomer-dimer-polymer'
system. \cite{KaMu89} 
For the fermion matrix of the form,
\begin{equation}
 \Delta = 2ma + \sum_{\langle xy \rangle} U(x,y)\zeta(x,y),
\end{equation}
\begin{equation}
Z = \int \mD\psibar\mD\psi \, e^{2ma\sum_x M(x)} 
\Pi_{\langle xy \rangle} F(x,y) .
\end{equation}
where
\begin{eqnarray}
F(x,y) &=& \int \mD U \, 
e^{\psibar(x)U\zeta(x,y)\psi(y) + \psibar(y)U^{\dagger}\zeta(y,x)\psi(x) }
\nonumber \\
&=& 1 - \frac{1}{3}\rho(x,y)M(x)M(y)+\frac{1}{12}(\rho(x,y)M(x)M(y))^2
 -\frac{1}{36}(\rho(x,y)M(x)M(y))^3
\nonumber \\
&& +\zeta(x,y)^3\bar{B}(x)B(y) + \zeta(y,x)^3\bar{B}(y)B(x).
\end{eqnarray}
Here $M(x)$, $B(x)$ and $\bar{B}(x)$ are `meson', `baryon' and `antibaryon'
fields.  For staggered fermions, the quark field $\psi(x)$ has only color
degrees of freedom, not Dirac indices. Meson, baryon and antibaryon 
fields are given as, 
\begin{equation}
M(x) \equiv \psibar(x)\psi(x) = \sum_{a=1,2,3}\psibar^a(x)\psi^a(x)
= \sum_{a=1,2,3} M_a(x),
\end{equation}
\begin{eqnarray}
B(x) &\equiv& \psi^1(x)\psi^2(x)\psi^3(x), \nonumber \\ 
\bar{B}(x) & \equiv& \psibar^3(x)\psibar^2(x)\psibar^1(x),
\end{eqnarray}
respectively, and
\begin{eqnarray}
e^{2maM(x)} = 1 &+& 2ma\left( M_1(x)+M_2(x)+M_3(x) \right)
\nonumber \\
 &+& (2ma)^2\left( M_1(x)M_2(x) +  M_2(x)M_3(x) + M_3(x)M_1(x) \right)
\nonumber \\
 &+& (2ma)^3\left( M_1(x)M_2(x)M_3(x) \right).
\end{eqnarray}
The last equation gives `monomer' contributions.
$\rho(x,y)$ is the `mesonic' link factor,
\begin{equation}
\rho(x,y) = \zeta(x,y)\zeta(y,x) ,
\end{equation}
from which three types of ``dimers'', $\rho(x,y)$, $\rho(x,y)^2$ 
and $\rho(x,y)^3$, are constructed.
For staggered fermions,
\begin{equation}
\zeta(x,y) = \eta(x,y) 
\left\{
\begin{array}{ll}
\pm \frac{a_s}{a_t}\, e^{\pm\mu a_t}, & \mbox{for } x = y+\hat{4}, \\
\pm 1,                                & \mbox{for } x = y+\hat{k}, (k=1,2,3)
\end{array}
\right.
\end{equation}
with $\eta$ being the staggered sign factor. 

Integration over Grassmann variables, $\psi$ and $\bar{\psi}$, is straightforward.
Nonvanishing contributions appear only if each site of the lattice is
occupied either by three mesons, $M_1(x)M_2(x)M_3(x)$ or by a 
baryon-antibaryon pair, $\bar{B}(x)B(x)$.
Finally, the partition function is written as a sum over monomer-dimer
loop configurations $K$,
\begin{equation}
Z(2ma,a_s/a_t,\mu a_t) = \sum_K w_K,
\end{equation}
where
\begin{eqnarray}
w_K &=& (2ma)^{(\mbox{\# of Monomer})} 
\left( \frac{a_s}{a_t}\right)^{2(\mbox{\# of Dimer lines in t direction})} 
\nonumber \\
&\times& \left(\frac{1}{3}\right)^{(
\mbox{\# of the first and second type Dimers})}
\Pi_x w(x) \Pi_C w(C) .
\label{MDP}
\end{eqnarray}
$w(x)$ and $w(C)$ are site and loop weights associated with the configuration.

\begin{figure}[hbt]
\begin{center}
\begin{minipage}{ 0.48\linewidth}
\includegraphics[width=1.0 \linewidth]{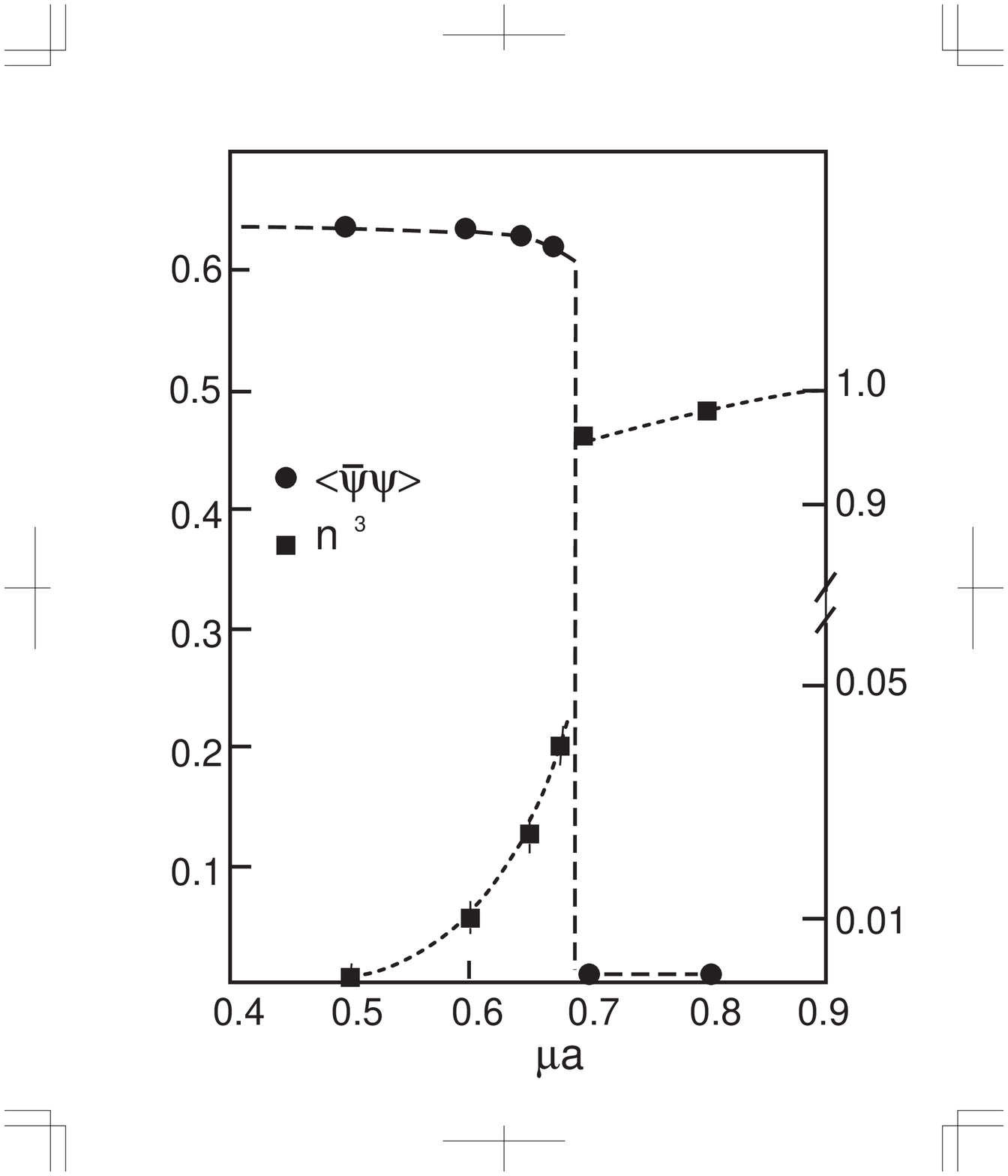}
\end{minipage}
\hspace{1mm}
\begin{minipage}{ 0.48\linewidth}
\includegraphics[width=1.1\linewidth]{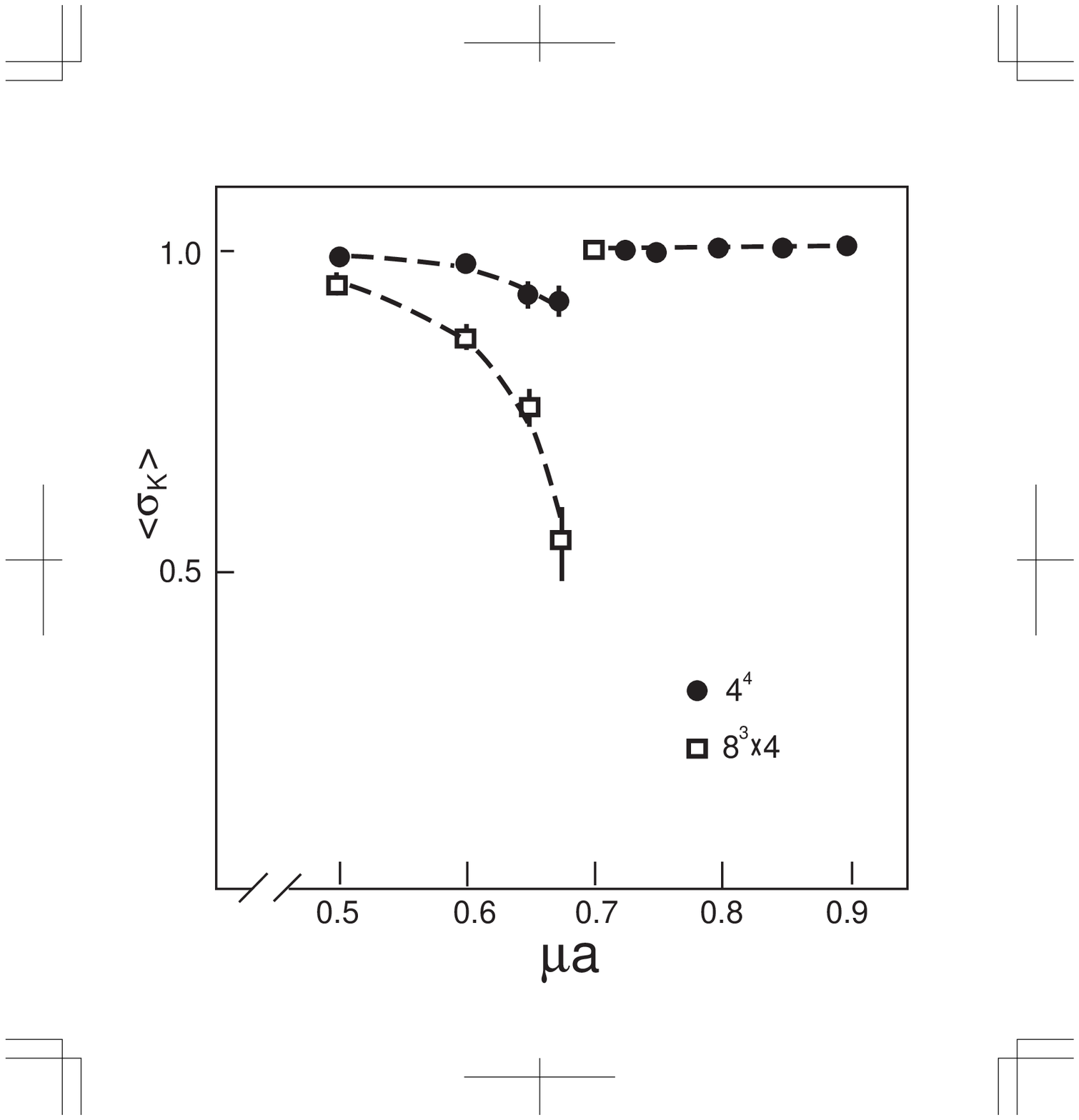}
\end{minipage}
\end{center}
\caption{
(a) The chiral condensate 
$\langle\psibar\psi\rangle =V^{-1}\partial log Z/\partial (2ma)$ 
and baryon number density
$na^3= (3VN_t)^{-1}\partial log Z/\partial (\mu a_t)$.
(b) The average sign of the Boltzmann weights.
(Ref.~\citen{KaMu89})
}
\label{fig5-2}
\end{figure}

In Eq.(\ref{MDP}), $w_K$ is not always positive.  A sign $\sigma_K$ for
each monomer-dimer loop configuration is introduced as
\begin{equation}
w_K = \sigma_K |w_K|,
\end{equation}
and the simulation was carried out with the positive Boltzmann weight $|w_K|$,
\begin{equation}
\langle O \rangle = \frac{\langle \sigma_K O \rangle_{+}}
{\langle \sigma_K \rangle_{+}} .
\end{equation}
Fig.\ref{fig5-2} shows the chiral condensate and baryon number density
as a function of $\mu a$ for $ma=0.1$ on $8^3\times 4$ lattice
obtained by Karsch and M\"utter.
They determined that $\mu_c a = 0.69 \pm 0.015$.  
The average of the sign factor 
drops at the critical $\mu$ shown in Fig.\ref{fig5-2}-(b).

\subsection{Canonical ensemble }

A thermodynamical system can be studied either as a grand canonical
formulation or as a canonical one.  When the grand canonical
partition function $Z(\mu)$ is expanded in terms of the fugacity, 
$\zeta\equiv\exp(\mu/T)$,
each coefficient is a canonical partition function $Z_N$ with a fixed 
quark baryon number $N$,
\begin{equation}
 Z(\mu/T) = \sum_{N=-\infty}^{+\infty} \left( e^{\frac{\mu}{T} } \right)^N
  Z_N.
\label{Canonical1}
\end{equation}
The canonical partition function can be formally obtained
from  the grand canonical partition function with imaginary chemical
potential, \cite{RobWei, MillRed}
\begin{equation}
Z_N = \frac{1}{2\pi} \int_0^{2\pi} d\phi Z(i\phi\equiv \mu/T)
 = \sum_{n,\bar{n}} C(n,\bar{n}) \zeta^n \zeta^{* \bar{n}},
\end{equation}
where $\zeta = e^{i\phi},\zeta^* = e^{-i\phi}$.
The canonical partition function with the quark baryon number $N=n-\bar{n}$ 
has contributions from $\zeta^n \zeta^{* \bar{n}}$ terms in the
grand partition function.
Miller and Redlich developed a general lattice canonical formulation, 
and investigated it using the hopping parameter expansion.

To date, only one numerical simulation was performed by Engels et al. 
\cite{EnKacKarLa99} 
They used Wilson fermions, Eq.(\ref{Wfermion}), and the expression 
(\ref{muatEdge2}).  Then the coefficient of $\zeta^n \zeta^{* \bar{n}}$
should include $\kappa^{n+\bar{n}}$.  Therefore for a heavy quark
system, i.e., for small $\kappa$, the leading contribution comes from
the $\bar{n}=0$ sector.  An explicit expression of the expansion is very
complicated, and yet they succeeded in calculating up to
$N = 12$ in the quenched approximation.
The baryon number density is given as
\begin{equation}
n_B = 
\frac{B/3}{N_s a^3}.
\end{equation}
$N=12$ on $8^3\times 2$ lattice at $T \sim T_c = 270$ MeV corresponds 
to $n_B \sim 0.15$ fm$^{-3}$ , i.e., approximately nuclear matter density.

\begin{figure}[hbt]
\begin{center}
\includegraphics[width=.6 \linewidth]{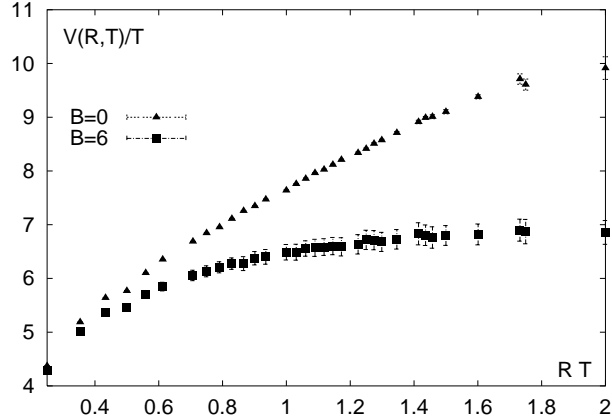}
\end{center}
\caption{Heavy quark potential for $T\simeq 0.86 T_c$ and $B=0$ and 6.
on $16^3\times 4$.
}

\label{fig5-3}
\end{figure}

The heavy quark potential shows finite
values for the large distance shown in Fig.\ref{fig5-3}.
$N=6$ on this lattice at $T \sim 0.86$ $T_c$ corresponds to
$n_B = 0.049$ fm$^{-3}$.  The heavy quark potential is screened even
below the nuclear matter density.

\subsection{Density of states method}

In lattice QCD, the average value of an operator $\langle O \rangle $ is given by 
\begin{equation}
\langle O \rangle =\frac1{Z}\int \mD U O[U]\det \Delta (U) e^{-\beta S_g(U)} ,
\end{equation}
with 
\begin{equation}
Z=\int \mD U \det \Delta (U) e^{-\beta S_g(U)} .
\end{equation}
It is impractical to directly perform this integration of very high dimensions. 
Instead, we can use a importance sampling
where configurations $U_i$ are generated with the probability distribution:
\begin{equation}
P(U)\propto \det \Delta (U) e^{-\beta S_g(U)}
\end{equation}
and $\langle O \rangle $ is given by $\displaystyle \lim_{N \rightarrow \infty }\frac1{N}\sum_i^N O[U_i]$ 
where $N$ is the number of configurations.

If one knows the density of states of some parameters, then 
the integration can be reduced to an integral of few dimensions.
Let us assume a one-parameter case and define the density of states as
\begin{equation}
\rho(E)=\int dU\delta(E-x(U))g(U)
\label{dos1}
\end{equation}
where $E$ is a certain parameter associated with $x(U)$ for which we construct the density 
of states $\rho(E)$,
and $g(U)$ is any function of $U$.
Then using the density of states $\rho(E)$,  $\langle O \rangle $ is given by
\begin{equation}
\langle O \rangle =\frac1{Z_{\rho} }\int dE \rho(E) 
\langle O \det \Delta  e^{-\beta S_g}/g(U) \rangle _E ,
\label{dos2}
\end{equation}
\begin{equation}
Z_{\rho} =\int dE\rho(E)\langle  \det \Delta  
e^{-\beta S_g}/g(U) \rangle _E ,
\label{dos3}
\end{equation}
where $\langle O \rangle _E$ stands for  the microcanonical average 
with fixed $E$, defined as
\begin{equation}
\langle O \rangle _E=\frac1{\rho(E)}\int dU \delta(E-x(U))g(U)O(U) .
\label{dos4}
\end{equation}
In this way one can define any $\rho (E)$ which depends on $x(U)$ and $g(U)$, 
and $\langle O \rangle $ is rewritten by using the density of states $\rho(E)$.
In the following, we show three examples of $x(U)$ and $g(U)$ which have been  
actually studied in the literature.

\subsubsection{Density of states of energy}
The simplest choice of the parameter for the density of states 
may be the energy of the gauge action.
In this case we set 
\begin{equation}
x(U)=S_g(U) ,
\end{equation}
\begin{equation}
g(U)=1.
\end{equation}
Substituting these equations to Eqs.(\ref{dos1})-(\ref{dos4}) we obtain
\begin{equation}
\rho(E)=\int dU\delta(E- S_g) ,
\end{equation}
\begin{equation}
\langle O \rangle =\frac1{Z_{\rho}}\int dE \rho(E) e^{-\beta E} 
\langle O \det \Delta  \rangle _E ,
\label{dos6}
\end{equation}
\begin{equation}
Z_{\rho}=\int dE\rho(E) e^{-\beta E} \langle  \det \Delta  \rangle _E .
\label{dos7}
\end{equation}

\begin{figure}[hbt]
\begin{center}
\rotatebox[origin=c]{-90}{
\includegraphics[width=.6 \linewidth]{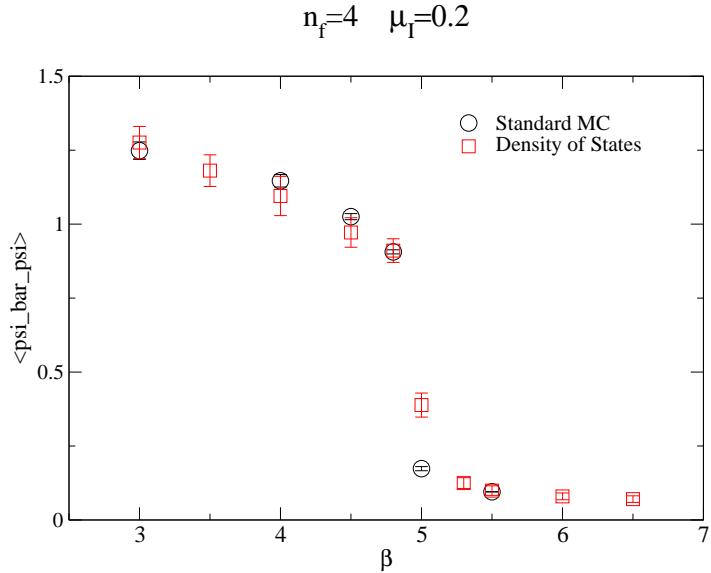}
}
\end{center}
\vspace{-2cm}
\caption{$<\bar{\psi}\psi>$ for $n_f=4$ on a $4^4$ lattice at $\mu_I=0.2$ and $m_q=0.025$.
The standard Monte Carlo results were obtained by the R-algorithm\cite{R-algo}. Ref.\citen{TAKAISHI} 
}
\label{ISOSPIN}
\end{figure}

This method including dynamical fermion was applied in
compact QED\cite{Azcoiti1990} and then in QCD\cite{LUO}. 
Both were simulated at zero density. 
Considering the complex phase in the above equations, one may apply 
this method for the finite density case. If we explicitly introduce 
the complex phase in the equations, Eqs.(\ref{dos6}) and (\ref{dos7}) are 
rewritten as
\begin{equation}
\langle O \rangle =\frac1{Z_{\rho}}\int dE \rho(E) e^{-\beta E} 
\langle O |\det \Delta|e^{i\theta}  \rangle _E ,
\end{equation}
\begin{equation}
Z_{\rho}=\int dE\rho(E) e^{-\beta E} \langle  | \det \Delta |e^{i\theta} \rangle _E .
\end{equation}
Since there is still the sign problem, e.g. $e^{i\theta}$ fluctuates,   
it is not obvious that this works better than the standard simulation.
If we omit the phase factor, resulting in simulating the finite isospin model, 
the method should work.
Fig.\ref{ISOSPIN} shows a comparison of $<\bar{\psi}\psi>$ 
between results from the density of states method and those from the standard 
Monte Carlo simulation(e.g. importance sampling)\cite{TAKAISHI}. The density of states method reproduce 
the results from the standard Monte Carlo simulation well.

\subsubsection{Density of  states of phase}
Gocksch\cite{Gocksch} used the complex phase of the determinant as
the parameter  for the density of states
and studied the chiral condensate $\langle \bar{\Psi}\Psi \rangle $ and the phase diagram of QCD 
at infinite gauge coupling.
In Ref.~\citen{Gocksch} he sets 
\begin{equation}
x(U)=\theta(U), 
\end{equation}
\begin{equation}
g(U)=|\det\Delta (\mu) | e^{-\beta S_g}.
\end{equation}
Then we obtain 
\begin{equation}
\rho(E)=\int dU\delta(E- \theta(U)) |\det\Delta (\mu) | e^{-\beta S_g},
\end{equation}
\begin{equation}
\langle O \rangle =\frac1{Z_{\rho}}\int dE \rho(E) e^{i E} \langle O  \rangle _E,
\end{equation}
\begin{equation}
Z_{\rho}=\int dE\rho(E) e^{iE}. 
\end{equation}
In order to obtain $\rho(E)$, simulations were carried out 
according to the probability 
$\displaystyle P(U)dU \sim |\det\Delta (\mu) | e^{-\beta S_g}dU$
and the histogram of $E$, which forms the density of state $\rho(E)$ 
in the end, was constructed. 
In the actual simulations, in order to speed up the process, 
the period of $E$ was divided into 
bins and the simulations were performed for each bin. Each bin has overlaps
with the neighboring bins to match the adjacent density of states.
The updating process is constrained in the range of each bin.
Therefore the configuration leading out of the range is not accepted.
Nevertheless the event has to be recorded in order to form the correct histogram\cite{HISTOGRAM}.

\subsubsection{Density of states of observable}
One may use observables to form the density of states.
In Refs.~\citen{AmAnaNishiVer} and  \cite{Ana-Nishi} the random matrix theory for finite density QCD was studied by
making use of the density of states for the quark number density.
Originally they call the method "factorization method" which is considered as  a variety of the density of states method.

Setting 
\begin{equation}
x(U)=O(U) ,
\end{equation}
and
\begin{equation}
g(U)=|\det\Delta (\mu)| e^{-\beta S_g},
\end{equation}
we obtain
\begin{equation}
\rho(E)=\int dU\delta(E- O(U)) |\det \Delta (\mu)| e^{-\beta S_g},
\end{equation}
\begin{equation}
\langle O \rangle =\frac1{Z_\rho}\int dE \rho(E) E \langle  e^{i\theta} \rangle _E ,
\end{equation}
\begin{equation}
Z_\rho = \int dE \rho(E) \langle  e^{i\theta} \rangle _E .
\end{equation}


If the operator $O$ takes a complex number, we may evaluate real and complex terms independently.
First we rewrite $\langle O \rangle $ as
\begin{equation}    
\langle O \rangle =\langle O_R \rangle +i\langle O_I \rangle 
\end{equation}
where $O_R$ and $O_I$ stand for ${\rm {Re}}(O)$ and ${\rm Im}(O)$ respectively.
Then, defining the density of states $\rho_R(E)$ and $\rho_I(E)$ as
\begin{equation}
\rho_R(E)=\int dU\delta(E- O_R)|\det\Delta (\mu)| e^{-\beta S_g},
\label{dosR}
\end{equation}
\begin{equation}
\rho_I(E)=\int dU\delta(E- O_I)|\det\Delta (\mu)| e^{-\beta S_g},
\label{dosI}
\end{equation}
we obtain $\langle O_R  \rangle $ and $\langle O_I  \rangle $ 
as given bellow.
For $i=R,I$,
\begin{equation}
\langle O_i  \rangle =\frac1{Z_i}\int dE \rho_i(E) E \langle e^{i\theta} \rangle _{i,E}.
\end{equation}
\begin{equation}
Z_i=\int dE\rho_i(E)\langle  e^{i\theta} \rangle _{i,E},
\end{equation}
where 
\begin{equation}
\langle  e^{i\theta} \rangle _{i,E} =\frac1{\rho_i(E)}\int dU \delta(E- O_i) e^{i\theta}|\det\Delta (\mu)| e^{-\beta S_g}.
\end{equation}
Finally the real part of $\langle O \rangle $ is given by
\begin{equation}
{\rm {Re}} \langle O \rangle = \langle \tilde{O_R} \rangle -\langle \tilde {O_I} \rangle 
\end{equation}
where
\begin{equation}
\langle \tilde{O_R} \rangle =\frac1{C}\int dE \rho_R(E) E \langle \cos\theta \rangle _{R,E}
\end{equation}
\begin{equation}
\langle \tilde{O_I} \rangle =\frac1{C}\int dE \rho_I(E) E \langle \sin\theta \rangle _{I,E}
\end{equation}
and $C=Z_R=Z_I$.
In Ref.\cite{AmAnaNishiVer}, to obtain 
$\langle \cos\theta \rangle _{R,E}$ and $\langle \sin\theta \rangle _{I,E}$,
they performed simulations with the constrained partition functions 
which are actually the density of states as given 
in Eq.(\ref{dosR}) and (\ref{dosI}).
The density of states itself was also evaluated through 
those simulations. They argue that since the simulations are performed with the constrained partition function
which forces the simulations to sample the important region, the results are obtained accurately.

\subsection{Complex Langevin}

Stochastic quantization by Parisi and Wu does not rely on the
path integral, and its simulation is realized by the Langevin 
equation\cite{Namiki}.
Parisi \cite{Parisi83} and Klauder \cite{Klauder84} proposed independently
an extension of the Langevin algorithm to a system with a complex action.

For the dynamical variable $x$ which is subject to the (Euclidean) action 
$S(x)$, let us consider a stochastic process along the ``fictitious" 
time $\tau$,
\begin{equation}
\frac{d x(\tau)}{d\tau} = - \frac{d S}{d x} + \eta(\tau) , 
\end{equation}
where $\eta(\tau)$ is a Gaussian white noise characterized by 
$\langle \eta(\tau) \rangle =0$ and $\langle \eta(\tau)\eta(\tau') \rangle =2\delta(\tau-\tau')$. 
From the Langevin equation, we can derive an equivalent 
Fokker-Planck equation,
\begin{equation}
\frac{\partial}{\partial \tau} P(x;\tau)=\frac{\partial }{\partial x}
\left (\frac{\partial}{\partial x}+\frac{d S}{d  x} \right )P(x;\tau). 
\end{equation}
By virtue  of the positive semi-definite property of 
the Fokker-Planck Hamiltonian, $H_{FP}=-\frac{\partial }{\partial x}
(\frac{\partial}{\partial x}+\frac{d S}{d  x} )$,
we can obtain a stationary solution as,
\begin{equation}
P(x,\tau) 
\maprightb{t \to \infty}
P_{eq}(x) \propto {\rm e}^{-S(x)}.
\end{equation}
Therefore, putting the solution of the Langevin equation into 
the quantities $O(x)$ and taking the average 
over the ensemble of the random noise $\eta$, 
we can obtain the expectation value $\langle O(x) \rangle $ with probability 
measure $P_{eq}(x) \propto {\rm e}^{-S(x)}$. This is a rough 
sketch of the basic procedure of the stochastic quantization. 
In practical use, instead of an ensemble average, the long time average 
is taken over the simulation time $\tau$ after thermalization,
where we assume ergodicity.

Once the action $S(x)$ becomes complex, the situation is not trivial. 
Putting a complex action into the above Langevin equation leads to 
the complex-valued extended version, 
\begin{equation}
\frac{d z(\tau)}{d\tau} =
 - \frac{d S(z)}{d z} + \eta(\tau)  
\end{equation}
with a complex variable $z(\tau) = x(\tau) +iy(\tau)$  
and {\it real} noise $\eta(\tau)$.  
The corresponding Fokker-Planck equation is defined for 
a {\it complex} distribution function $P_{\rm c}(x)$ of the {\it real} 
variable $x$ which satisfies,
\begin{equation}
\langle O(z(\tau)) \rangle _{\eta}=\int dx P_{\rm c}(x) O(x).
\end{equation}

In addition to the problem of the interpretation of the complex 
generalized distribution function $P_{\rm c}(x)$, 
in this case, Fokker-Planck Hamiltonian $H_{FP}$ 
loses the positive semi-definite 
property and there exists no general proof for the convergence.

However, if the spectrum of the Fokker-Planck Hamiltonian 
is composed of very small imaginary part and positive semi-definite
real part, it is expected that a complete set ${z_{n}}$ with 
positive semi-definite eigenvalue ${E_n}$ and 
stationary solution ${\rm e}^{-S(z)}$ exist.
For a simple case, this conjecture has been checked 
explicitly by Klauder and Petersen\cite{Klauder85}.  

Karsch and Wyld applied the algorithm to a three dimensional 
SU(3) spin model with the chemical potential\cite{KaWy}.  
The Langevin simulations converge even for large 
values of the chemical potential, and are in agreement
with exact solutions at the strong coupling limit.

Unfortunately the complex Langevin method sometimes gives
incorrect results.  
For some simple noncompact variable case, the Parisi and
Klauder algorithm can be proved to work well if the partition 
function satisfies certain conditions of the analytic behavior 
\cite{MatuiNaka},
but for the compact variable case, it is difficult to judge if
the algorithm gives correct results or not. 

Extensive use of the fictitious time as a redundant variable 
introducing a kernel in the Langevin
equation has been discussed by the Waseda and Siegen group 
intensively\cite{PTP-Suppl}.
A kernel  $K(z)$ is introduced as,
\begin{equation}
\frac{d z(\tau)}{d\tau} = -K(z)\frac{d S}{d z} +
\frac{d K}{d z}+ \xi(\tau) ,
\end{equation}
where $\xi(\tau)$ is a noise satisfying 
$\langle \xi(\tau) \rangle =0$ and $\langle \xi(\tau)\xi(\tau') \rangle =2K(z)\delta(\tau-\tau')$.
The corresponding Fokker-Planck equation becomes,
\begin{equation}
\frac{\partial}{\partial \tau} P_{\rm c}(x;\tau)=
- H_{FP}P_{\rm c}(x;\tau),
\end{equation}
where
\begin{equation}
-H_{FP} = \frac{\partial }{\partial x}K(x)
\left( \frac{\partial}{\partial x}+\frac{d S}{d  x} \right ).
\end{equation}
Therefore, the role of the kernel is to change relaxation process only 
keeping the same asymptotic behavior,
\begin{equation}
P_{\rm c}(x,\tau) 
\maprightb{t \to \infty}
P_{eq}(x) \propto {\rm e}^{-S(x)}.
\end{equation}
Thereby, a choice of the appropriate kernel may provide a 
better relaxation process without changing the physical result.
Indeed in some cases, the use of a kernel improves the relaxation 
process and converges the random walk. 
For example, Okamoto et al. applied a constant kernel method to the
polynomial model and succeeded in extending Klauder and Petersen's 
convergent region in the complex eigenvalue space\cite{Okamoto}.
Though a pole in the second Rieman sheet sometimes leads 
the constant kernel method to an incorrect result, 
an appropriate choice of the field dependent kernel solves the
problem in the case of the polynomial model\cite{Okano1}.
For the simple models, the origin of the problems in the complex
Langevin method is well analyzed and some answers with kerneled
Langevin equation exist(see more details in Ref.~\citen{Okano2}).
However, the general recipes for the complex system are 
still unclear and how to manage the finite density QCD is 
still open.


\subsection{Finite isospin model}\label{sec-FiniteIsospin}

Imagine a system in which the chemical potentials of $u$ and $d$ quarks
have the opposite signs to each other,
\begin{equation}
\mu_u = -\mu_d \equiv \mu
\end{equation}
The system is called a finite isospin model \cite{SONST} or 
the isovector model \cite{QCD-Taro02}.

The fermion determinant of this system is
\begin{equation}
\det\Delta(\mu_u) \det\Delta(\mu_d)
= \det\Delta(\mu) \det\Delta(-\mu)
= | \det\Delta(\mu) |^2
\label{detFI}
\end{equation}
where we use the relationship (\ref{det2}).
Therefore the finite isospin model is considered as a two-flavor
system where one drops the phase of the fermion determinant,
and of course Monte Carlo calculation is possible.

\begin{figure}[hbt]
\begin{center}
\includegraphics[width=.6 \linewidth]{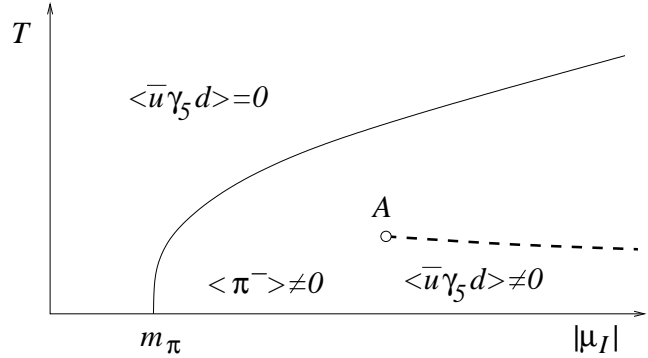}
\end{center}
\caption{
Finite Iso-spin phase diagram calculated by Son and Stephanov.
(Ref.~\citen{SONST}).
}
\label{fig5-4}
\end{figure}

\begin{figure}[hbt]
\begin{center}
\begin{minipage}{ 0.48\linewidth}
\includegraphics[width=1.0 \linewidth]{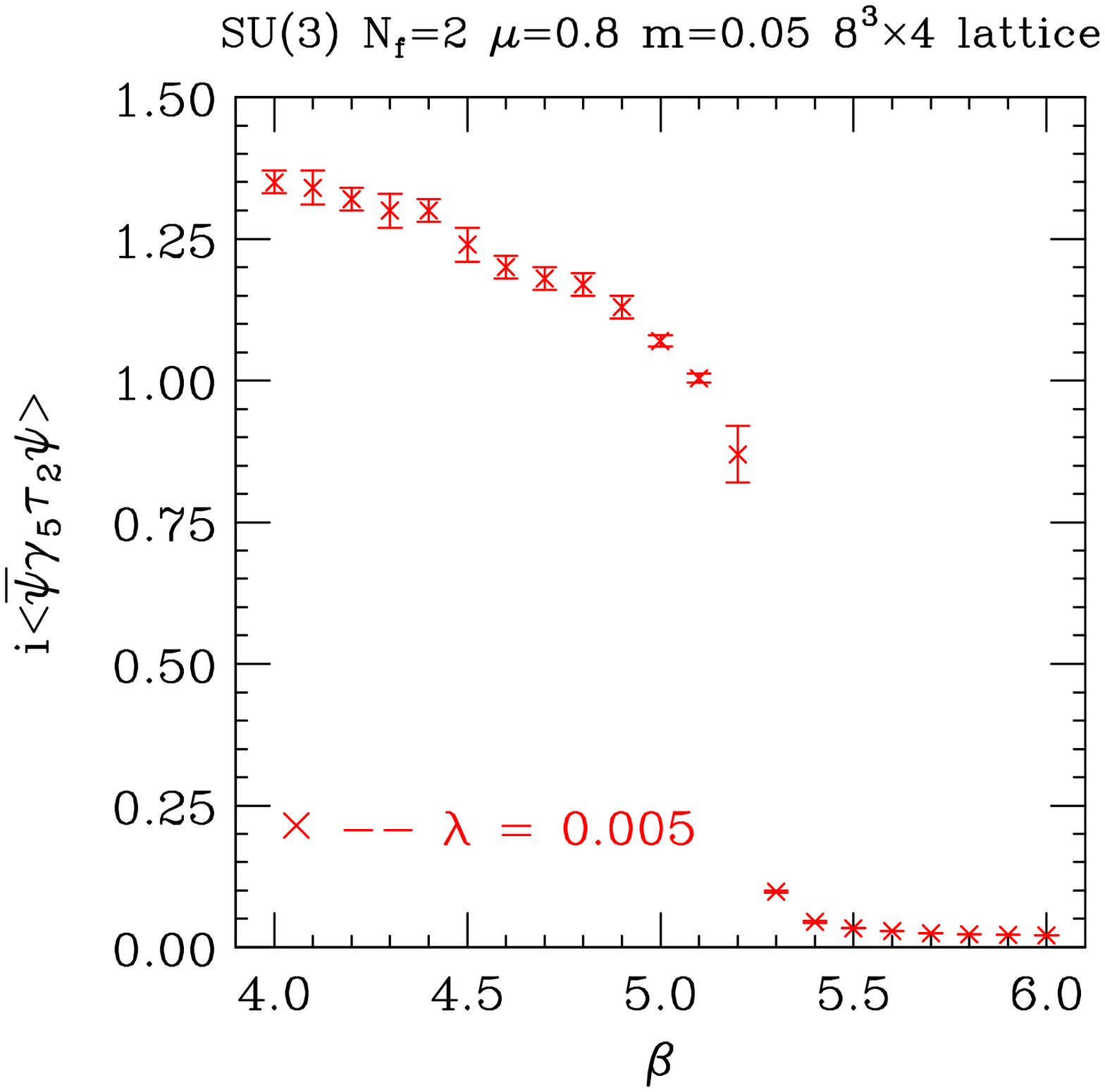}
\end{minipage}
\hspace{1mm}
\begin{minipage}{ 0.48\linewidth}
\includegraphics[width=1.1\linewidth]{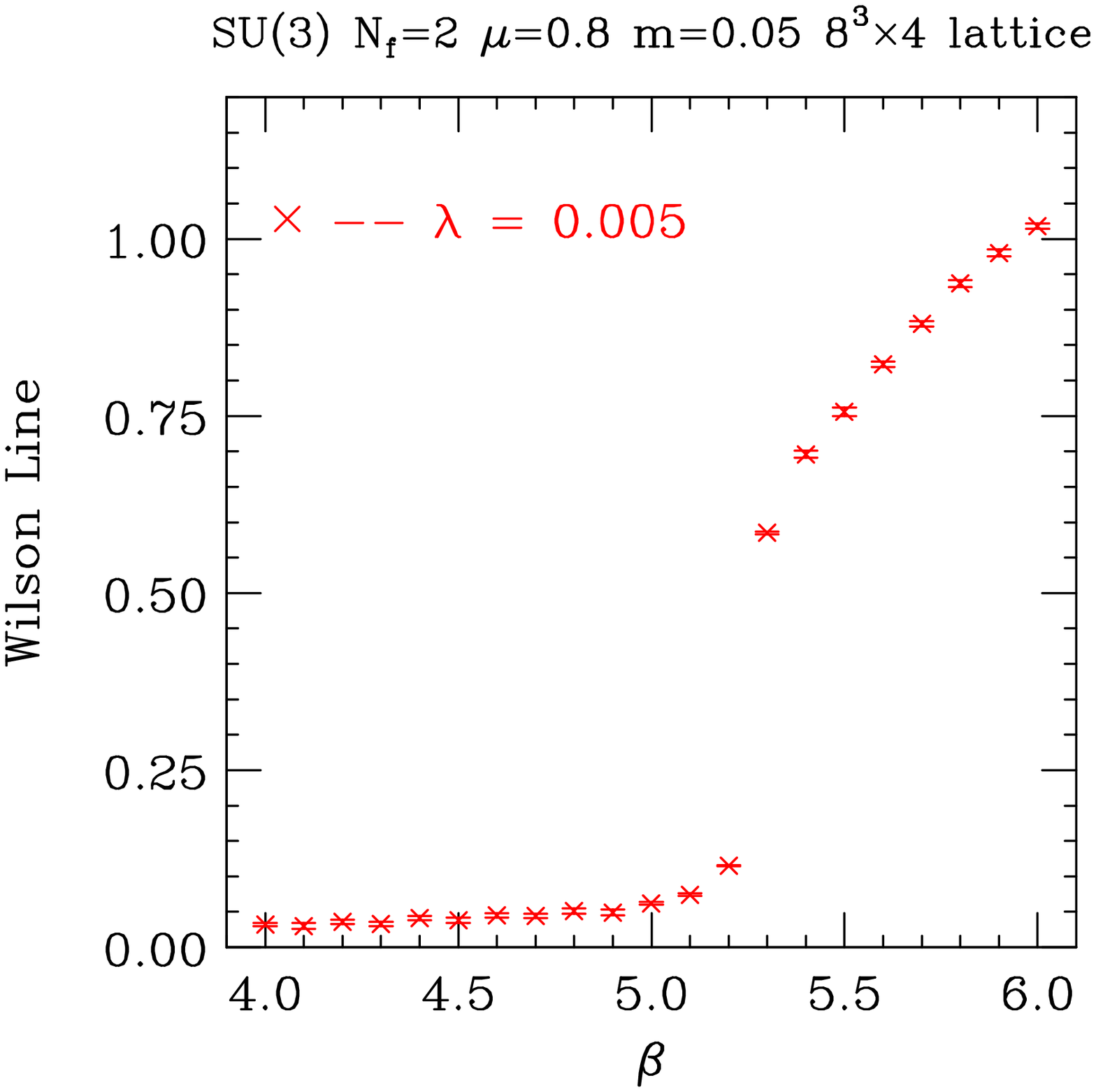}
\end{minipage}
\end{center}
\caption{
Numerical results of finite iso-spin model by Kogut and Sinclair.
Left : $i\langle \bar{psi}\gamma_5\tau_2\psi \rangle $ at $\mu_I=0.8$ has finite value
when the temperature is low, and decreases as $T$ increases.
Right : Wilson line at $\mu_I=0.8$ at low $T$ is zero and increases
when $T$ increases. (Ref.~\citen{KoSi2002}).
}
\label{fig5-5}
\end{figure}

Son and Stephanov have obtained the phase of the system shown in 
Fig.\ref{fig5-4}\cite{SONST}, 
and the qualitative features of the diagram were 
confirmed by numerical simulation by Kogut and Sinclair \cite{KoSi2002}.
See Figs. \ref{fig5-5}.

\section{Summary}

After the brief introduction of the lattice QCD actions with
the chemical potential and their characters, we have seen
many activities in the field.  Recent progress in small $\mu$
and finite $T$ regions is very promising. Indeed until several
years ago, we did not expect that any physically relevant
data were provided by lattice QCD.
This is the area
which recent high energy heavy ion experiments are investigating.
We anticipate now lattice data in the next several years
which can be compared with experimental results.
For that algorithm development is important.

Still it is difficult to study finite density QCD in large baryon 
density and low temperature regions,
where many interesting phenomena such as color super condictivity
and partial chiral restoration may occur.
Two color QCD provides us hints on what may occur.
Lattice simulations discussed in Sec.\ref{sec-2color} show us 
many interesting phenomena.  Some of them are relevant for QCD and some not.
In any case, they are interesting features in field theories and
it is desirable to investigate their mechanisms.

We discussed many attempts in Sec.\ref{sec-related} related with 
lattice field theory at finite density, i.e., 2-dimensional model, 
lattice NJL model, application of the strong coupling technique, 
density of state method, complex Langevin and finite isospin model.
Surely there are more in the literature.  We have made a survey of
these works by hoping that a new idea appears from these studies,
i.e., taking a lesson from the past.

\noindent
{\bf Acknowledgement}
It is our pleasure to thank Prof. T. Kunihiro for providing us the chance 
to write this review and for continuous encouragement.
We would like to thank Dr. S. Ejiri for many discussions, from which we have
learned lots.
We appreciate kind help by Prof. Y. Sasai for preparing the appendix.
We are much indebted to Profs. T. Hatsuda, T. Inagaki, K. Okano 
and I. O. Stamatescu for many valuable discussions.
We wish to thank Drs. K. Iida, S. Sasaki and M. Tachibana for providing us
many facts and wisdoms on color super conductivity. 
This work is supported by DOE grants DE-FG02-96ER40495, 
Grant-in-Aide for Scientific Research by
Monbu-Kagaku-sho (No.11440080, No. 12554008 and No. 13135216),
and ERI, Tokuyama Univ..
Our simulations in this paper were performed on SR8000 at IMC, Hiroshima
Univ., SX5 at RCNP, Osaka Univ., SR8000 at KEK..

\appendix
\def\appendixsymbol{Appendix}

\section{Gibbs formula to condense the fermion determinant}

In lattice QCD simulations of finite baryon density system, we
often need to calculate the fermion determinant.
Gibbs has obtained a useful expression for this purpose. 
Here we describe a brief derivation of the formula 
following \citen{Gibbs,Hasen-To}. 
Our proof given below is not very smart, but we hope that it helps
the readers who need the formula.

We consider KS fermions,(\ref{KSfermion}), and
employ Eqs.(\ref{muatEdge1}),(\ref{muatEdge2}).
\footnote{We renormalize a factor two in (\ref{KSfermion}).}
The fermion matrix $\Delta$ is 
$(3 \times N_{x} N_{y} N_{z} N_{t})^{2}$ complex matrix.
We take the axial gauge as $U_t(x)=1$ except at $t=N_{t}-1$,
the fermion matrix can be written in time slices as follows:

\begin{equation}
\Delta
=
\left(
\begin{array}{c|c|ccc|c|c}
   B_0 & \1 & 0 & \cdots& & 0 &U_{N_t-1}^{\dagger}{\rm e}^{-N_t \mu}
\\ \hline
   -\1  &B_1 & \1 &\cdots  &  & 0& 0 
\\ \hline
  0 & -\1 & B_{2} & \cdots & &  & 
\\ 
 \cdots  &\cdots  &\cdots & \cdots& \cdots &\cdots  &\cdots
\\ 
   &  & &\cdots &   & \1 &0
\\ \hline
    0    & 0 & &  \cdots & -\1 &B_{N_t-2}& \1 
\\ \hline
-U_{N_t-1}{\rm e}^{N_t \mu}& 0 &  &\cdots &0& -\1 &B_{N_t-1} \\
\end{array}
\right) 
\end{equation}
where each column and row corresponds to the time slice
$t = 0, 1, 2, \cdots N_{t}-1$ and $U_{N_t-1}$ is the abbriviation
of a $(N_cV_s) \times (N_cV_s)$ diagonal matrix whose elemets 
are $U_t(\vec{x},{N_t-1})\delta_{\vec{x},\vec{x}'}$.
Here we use the anti-periodic boundary condition along $t$ direction.
The element of the above matrix is a $(N_cV_s) \times (N_cV_s)$ block matrix,  
where $ V_s \equiv N_x N_y N_z$. 
$B_i$ stands for the spatial derivative and quark mass parts of the matrix
at $i$-th time slice ($ i = 0,\cdots,N_t-1$).
The chemical potential dependence has been summarized to 
the last time slice only.  
Let us denote $U_{N_t-1}{\rm e}^{N_t \mu}$ as $\Xi$ for the abbreviation.

Let us define $\tilde{\Delta}$ by  multiplying $  \Xi $ only 
to the last column of $\Delta$,
\begin{equation}
\tilde{\Delta} =
\left(
\begin{array}{ccccccc}
 B_0&\1  & &       & &      &\1\\
  -\1&B_1& &       & &     &0\\ 
    &-\1 & &       & &     & \\
    &   &     \mbox{\quad } \ddots  & &     &         & \\
    &   &     &       & &  \1   & \\
    &   &     &       & &B_{N_{t}-2} & \Xi \\
-\Xi &  0 &  & &  &-1       & B_{N_{t}-1} \Xi \\
\end{array}
\right).
\end{equation}
\noindent
We assume that $N_t$ is even.
Since
\begin{equation}
\det(\Xi) =  \det(U_{N_t-1}{\rm e}^{N_t\mu} )  
                = {\rm e}^{3V_s N_t \mu}, 
\end{equation}
the determinant of $\tilde{\Delta}$ is related to $\det\Delta$ as
$\det\tilde{\Delta} = \det\Delta {\rm e}^{3V_s N_t \mu}$. 
Multiplying the following matrix whose determinant is one,
\begin{equation}
{\cal B}=\left(
\begin{array}{cccccccc}
   \1&B_0&0&   &&      &         &         \\
   0&  \1&0&   &&      &         &         \\ 
    &   &\1&B_2&&      &         &         \\
    &   & &\1  &&      &         &         \\
    &   & &   &\ddots&&         &         \\
    &   & &   &&      &0        &    0    \\
    &   & &   &&      &\1        &B_{N_{t}-2}\\
    &   & &   &&      &         &\1        \\
\end{array}
\right), \nonumber
\end{equation}
to the fermion matrix $\tilde{\Delta}$ from the left,
we obtain,
\begin{eqnarray}
&&
{\cal B} \times \tilde{\Delta} \nonumber \\
&=&\left(
\begin{array}{ccccccccc}
   0&1+B_0 B_1&B_0& 0 &      &         & 1        \\
  -1&  B_1    & 1 & 0 &      &         & 0        \\ 
   0& -1      & 0 &1+B_2 B_3& B_2 &         &         \\
    &  0      &-1 & B_3 &  1    &         &         \\
    &   & &   &\ddots&         &         \\
 -B_{N_{t}-2}\Xi &&&&&0&(1+B_{N_{t}-2}B_{N_{t}-1})\Xi\\
-\Xi &&&&&-1&B_{N_{t}-1} \Xi \\
\end{array}
\right). \nonumber
\end{eqnarray}
\noindent
Permuting the first column to the last and multiplying -1 to the last column, $\det{\tilde{\Delta}} $ 
is rewritten as,
\begin{equation}
\det\left(
\begin{array}{ccccccccc}
1+B_0 B_1&B_0& 0        &      &     & 1 &   0 \\
  B_1    & 1 & 0        &      &     & 0 &  1 \\ 
  -1     & 0 &1+B_2 B_3 & B_2  &     &   &   0 \\
   0     &-1 & B_3      &  1   &     &   &     \\
         &   &          &\ddots&     &   &     \\
&&&&0&(1+B_{N_t-2}B_{N_t-1}) \Xi &B_{N_t-2}\Xi \\
&&&&-1&B_{N_t-1} \Xi & \Xi \\
\end{array}
\right) 
. \nonumber
\end{equation}
Denoting as,
\begin{equation}
\Omega_{j,j+1} = \left(
                   \begin{array}{cc}
                      B_j B_{j+1}+1&B_j\\
                      B_{j+1}&1
                   \end{array}
                   \right) 
               = \left(
                   \begin{array}{cc}
                      B_j & 1\\
                      1   & 0
                   \end{array}
                   \right)
                   \left(
                   \begin{array}{cc}
                      B_{j+1} & 1\\
                      1   & 0
                   \end{array}
                   \right),
\end{equation}
and
\begin{equation}
1_{2} = \left( 
\begin{array}{cc}
1 & 0 \\
        0 & 1 
\end{array}
\right)
\end{equation}
the determinant of the fermion matrix can be written as,
\begin{equation}
\det(\tilde{\Delta})=\det\left(
\begin{array}{ccccccc}
\Omega_{01}&           &           &  &   1_2  \\
  -1_{2}       &\Omega_{23}&           &  &     \\ 
           & -1_{2}        &\Omega_{45}&  &     \\
           &           & \ldots    &  &     \\
           &           &           &-1_{2}&\Omega_{N_t-2,N_t-1}\Xi \\
\end{array}
\right)  
 .
\end{equation}
Note that $\Omega_{N_t-2,N_t-1}\Xi$ is $(3V_s \times 2)$ times 
$(3V_s \times 2)$ matrix.

We multiply a matrix of which determinant is 1  from the right.
\begin{eqnarray}
\det(\tilde{\Delta})& = & 
\det\left(
\begin{array}{ccccccc}
\Omega_{01}&           &           &  &   1_2  \\
  -1_{2}       &\Omega_{23}&           &  &     \\ 
           & -1_{2}        &\Omega_{45}&  &     \\
           &           & \ldots    &  &     \\
           &           &           &-1_{2}&\Omega_{N_t-2,N_t-1}\Xi \\
\end{array}
\right)\left(
\begin{array}{ccccccc}
1_2 &           &           &  &  -\Omega_{01}^{-1}  \\
         &1_2&           &  &     \\ 
           &         &1_2&  &     \\
           &           &     & \ldots &     \\
           &           &           & &1_2 \\
\end{array} 
\right) \nonumber \\
& = &  
\det\left( 
\begin{array}{ccccccc}
\Omega_{01}&   0        &   \ldots        & 0  &   0  \\
  -1_{2}       &\Omega_{23}&           &  & \Omega_{01}^{-1}    \\ 
           & -1_{2}        &\Omega_{45}&  &     \\
           &           & \ldots    &  &     \\
           &           &           &-1_{2}&\Omega_{N_t-2,N_t-1}\Xi \\
\end{array}
\right) \nonumber \\
& =&  
\det(\Omega_{01})\det\left(
\begin{array}{ccccccc}
\Omega_{23}     &           &  & \Omega_{01}^{-1}    \\ 
        -1_{2}  &\Omega_{45}&  &     \\
                & \ldots    &  &     \\
                &           &-1_{2}&\Omega_{N_t-2,N_t-1}\Xi \\
\end{array}
\right)
\nonumber \\
&=& \cdots
\nonumber \\
&=& \det\left(\Omega_{01} \Omega_{23}\cdots \Omega_{{N_t-6},{N_t-5}}\right)
\nonumber \\
& \times &
\begin{array}{|cc|}
\Omega_{N_t-4,N_t-3}&     0 \\
 -1                 & \Omega_{N_t-2,N_t-1}\Xi + (\Omega_{01}\cdots\Omega_{Nt-3,N_t-2})^{-1}\\
\end{array}
\end{eqnarray}
where we used a formula for blocked matrixes $A_{ij}$:
\begin{equation}
\begin{array}{|cccc|}
A_{11}&     0&\ldots&     0\\
A_{21}&A_{22}&\ldots&A_{2n}\\
\vdots&\vdots&      &\vdots\\
A_{n1}&A_{n2}&\ldots&A_{nn}\\
\end{array}
=
\left|A_{11}\right|
\begin{array}{|ccc|}
A_{22} & \ldots & A_{2n} \\
\vdots &        & \vdots \\
A_{n2} & \ldots & A_{nn} \\
\end{array}.
\end{equation}

Taking the similar step recursively, we can reduced the matrix 
up to two-time slices.
We obtain finally, 
\begin{equation}
\begin{array}{cll}
\det(\Delta) 
&=& \det\tilde{\Delta}  \times e^{-3V_sN_t\mu}
\\
&=& \det ( 1+ e^{N_t\mu} P) 
\times e^{-3V_sN_t\mu}  \\
 \\
&=& \det \left\{ e^{N_t\mu} ( e^{-N_t\mu}+P )\right\}
    \times e^{-3V_sN_t\mu}  \\ \\
&=& \det \left(P+e^{-N_t\mu} \right)\times ( e^{3V_s N_t \mu} )^2
    \times e^{-3V_s N_t \mu}  \\ \\
&=& \det \left( P+e^{-N_t\mu} \right) \times e^{3V_s N_t \mu}  ,
\end{array}
\end{equation}
where
\begin{equation}
P  =  \displaystyle \left( \Pi_{j=0,2,4}^{N_t-2} \Omega_{j,j+1} 
\right) U_{N_t-1} \\ 
   =  \displaystyle \left( \Pi_{j=0}^{N_t-1} \left(
\begin{array}{cc}
B_j & 1 \\
1 & 0 \\
\end{array}
\right) \right) U_{N_t-1} .
\end{equation}
 
The rank of the matrix $P+e^{-N_t\mu}$ is $3\times(2V_s)$.
Therefore if $N_t>2$, we obtain a condensed smaller matrix.
To estimate the determinant, $LU$ decomposition is fastest
for small size lattices
on many computers if there is enough memory.
If we have eigenvalues of the matrix $P$, $\{\lambda_i\}$, 
which do not depend on the chemical potential $\mu$, 
we obtain Eq.(\ref{Eq-Gibbs}), which gives us values of the
fermion determinant at any $\mu$.

\section{Meson and baryon propagators in color SU(2) space}

A $\bar{q}-{q}$ bound state belonging to {\bf 1} representation in
$\bar{\bf2} \otimes {\bf 2}$  is called a meson.    
With a quark field $\psi(x)$, the meson operators are constructed as
$$\bar{\psi} \Gamma \psi,$$
with $\Gamma $ being Dirac's $\gamma$ matrices.
$\Gamma = 1$ corresponds to a scalar meson, 
$\Gamma = \gamma_{5}$ corresponds  to a pseudoscalar meson, 
$\Gamma =\gamma_{\mu}$ corresponds to 
a vector meson, and $\Gamma = \gamma_{5}\gamma_{\mu}$ to an 
axial vector meson.

In SU(2), {\bf 1} representation also exists in ${\bf2} \otimes {\bf 2}$ 
and  $q-q$ bound state can be realized  as a color singlet state.
It means that a baryon in SU(2) theory is a color 
singlet {\it di-quark} state.  
On this point, color-SU(2) and color-SU(3) are essentially different.
The di-quark operator is given as,
$$ \varepsilon^{ab} {\psi_{a} }^{T} C \Gamma \hat{\tau} \psi_{b} ,$$
with $a$ and $b$ being color indices.
$\hat{\tau}$ stands for Pauli matrix  $\tau_2$ or $ \vec{\tau} \tau_2$ in the 
flavor SU(2) space.
In order to satisfy the fermionic property of the quark,  the 
di-quark state must be totally antisymmetric  utilizing flavor 
degrees of freedom. 
We must take iso-singlet channel, 
$\hat{\tau}= \tau_2$, for $\Gamma = 1$, $\gamma_5$ and 
$\gamma_{\mu}\gamma_{5}$.  Iso-vector di-quarks are 
realized for the vector channel, $\Gamma = \gamma_{\mu}$.
Because of the charge conjugation $C$, we must turn our
attention to the parity of the di-quark states;
$\Gamma = 1$ corresponds to a 
pseudoscalar di-quark, $\Gamma = \gamma_{5}$ corresponds to a 
scalar di-quark, $\Gamma =\gamma_{\mu}$ corresponds to 
an axial vector  di-quark, and $\Gamma = \gamma_{5}\gamma_{\mu}$ to a 
vector di-quark.
 
By virtue  of the simple structure of SU(2), propagators of meson 
and di-quark are degenerate in a vacuum state, i.e. ,  propagators of 
pseudoscalar meson and scalar di-quark, and vector meson and 
axial vector di-quark are degenerate.

\begin{figure}[htb]
\begin{center}
\begin{minipage}{ 0.48\linewidth}
\includegraphics[width=1.0 \linewidth]{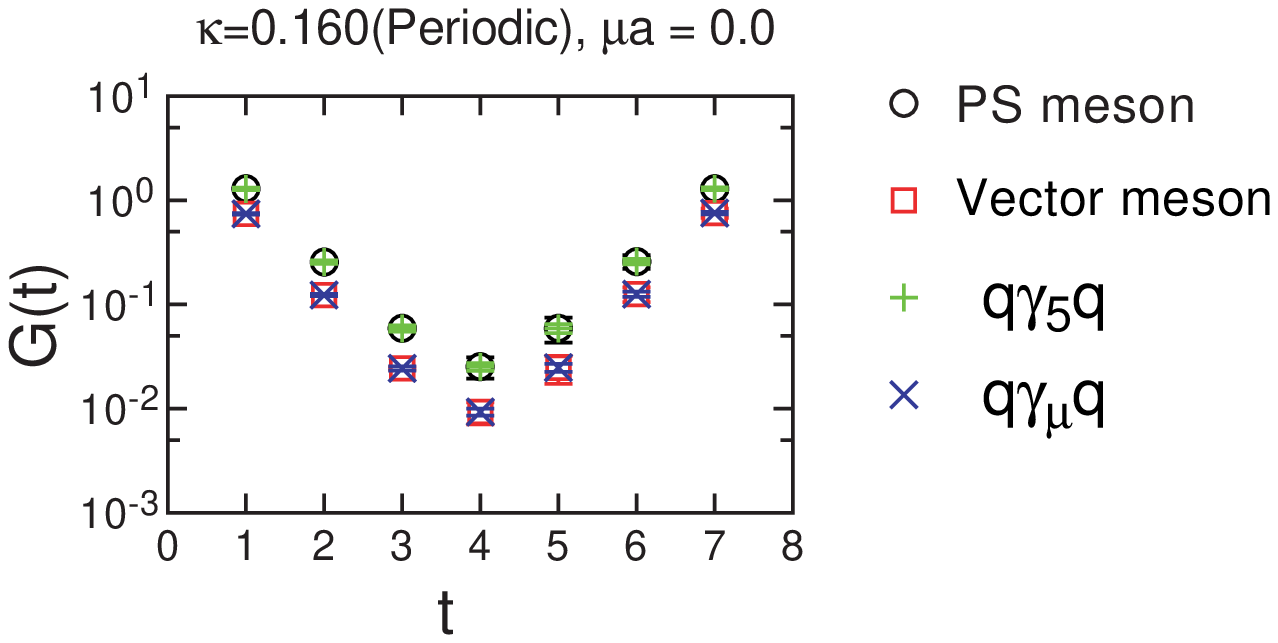}
\end{minipage}
\begin{minipage}{ 0.48\linewidth}
\includegraphics[width=1.\linewidth]{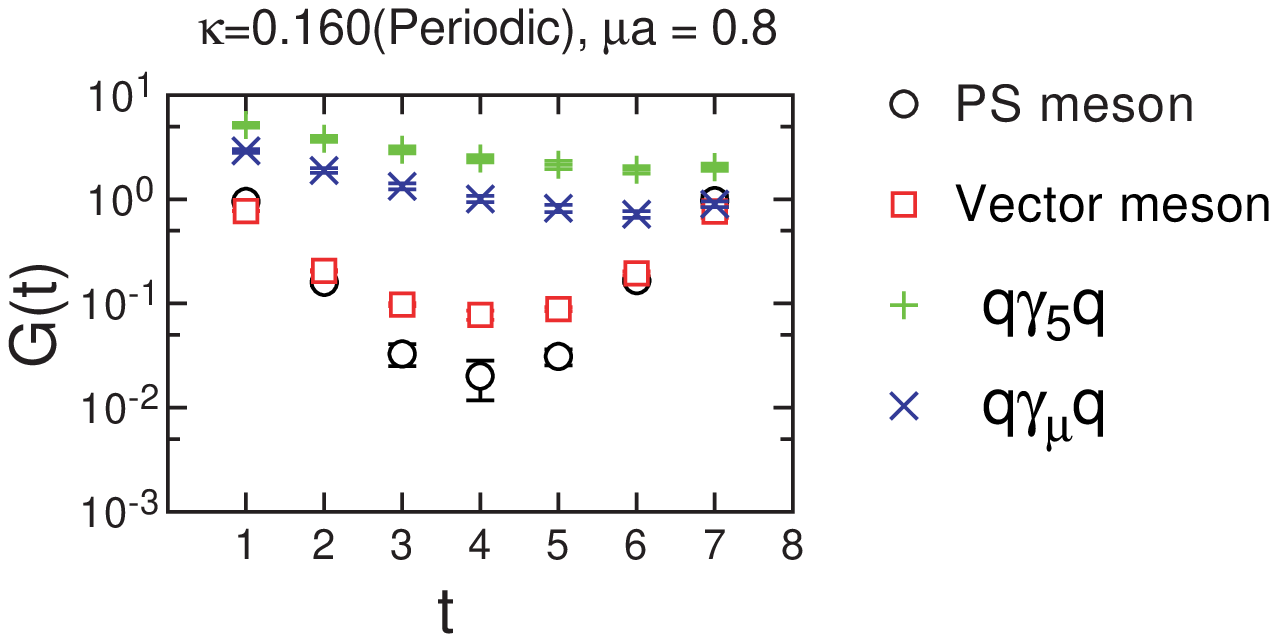}
\end{minipage}

\end{center}
\caption{
Hadron propagators at $\mu a=0$ (left) and $\mu a=0.8$ (right).
Pseudoscalar and vector mesons and scalar and axial vector
baryons are shown. 
}
\label{figb-1}
\end{figure}

Figure (\ref{figb-1}) shows the propagators on a $4^3 \times 8$
lattice with 
spatially 
periodic boundary condition at $\beta=0.7$ and $\kappa=0.150$.  
At $\mu =0$ (\ref{figb-1}a), pseudoscalar meson and scalar di-quark, 
and vector boson and axial vector di-quark are degenerate. 
With the finite $\mu$,  the $\mu B$ term in the $W$ makes the di-quark 
propagators  asymmetric in the time-direction. On the other hand,
the meson propagators keep the symmetric shape. 
Therefore, degeneracy between the meson 
propagator and di-quark propagators is solved with finite $\mu$.

The asymmetry of the di-quark propagator is caused by the difference of the 
charge of "di-quark" and "anti-di-quark". The
chronological propagation is provided by a particle with 
charge $Q$ and  of which the chemical potential is $\mu Q$. 
On the other hand,  the anti-chronological 
propagation is governed by an anti-particle with charge $-Q$ and  
which is affected by the chemical potential as $-\mu Q$.
Taking this asymmetry into account, we adopt 
$$
G_{d}(t) = C_{1}{\rm e}^{-(m-2\mu)t} + C_{2}{\rm e}^{-(m+2\mu)(N_{t}-t)}
$$
as a fitting function to the di-quark propagator, where 
$C_{1}$ and $C_{2}$  must 
keep the constraint on the bosonic boundary condition of di-quark.
For the mesons, even in finite density we can use a usual form as,
$$
G_{m}{t} = C_{1}{\rm e}^{-mt} + C_{2}{\rm e}^{-m(N_{t}-t)}.
$$

\section{Thermodynamical variables on the lattice}

The path integral form of the partition function is given as \cite{G-P-Y} ,
\begin{eqnarray}
Z &=& {\rm{Tr}} ({\rm e }^{-\beta (H - \mu N)})
\nonumber \\
 &=& \int \mathcal{D}U \mathcal{D}
\bar{\psi}\mathcal{D}\psi{\rm e}^{-\int_0^\beta d \tau \int d^3x
(L + \mu n)}
\nonumber \\
& = & {\rm e}^{-\beta(\epsilon -\mu n)V}, \nonumber \\
\label{PartFunc}
\end{eqnarray}
The energy density $\epsilon$, and the number density $n$ are given by 
\begin{eqnarray}
\epsilon & = & \frac{1}{V}
\left (
- \frac{\partial}{\partial \beta}
+ \frac{\mu}{\beta}\frac{\partial}{\partial \mu}
\right )
\log Z|_V, \nonumber \\
n & = &
\frac{1}{\beta V}
\frac{\partial}{\partial \mu} \log Z|_V,
\end{eqnarray}

On the lattice, $\beta = N_t a_t$, and
\begin{equation}
\frac{\partial}{\partial \beta}
=
\frac{1}{N_t} \frac{\partial}{\partial a_t}
\end{equation}
We must be careful that the temperature, $T$, changes when we 
vary the lattice spacing $a_t$ which is a function of the coupling
constant.

A simple and safe way is to write down a formula in terms of 
non-dimensional quantities and to start to calculate.
For example, the number density is given in the continuum world as
\begin{equation}
n = \frac{\partial p}{\partial \mu}  
= \frac{\partial}{\partial \mu} \left(\frac{T}{V} \log Z \right)
\end{equation}
We may write a formula 
\begin{equation}
\frac{n}{T^3} = \frac{N_t^3}{N_s^3} \frac{\partial}{\partial\hat{\mu}}
\log Z ,
\end{equation}
where
\begin{equation}
\hat{\mu} \equiv \frac{\mu}{T} .
\end{equation}

\end{document}